\documentclass[sigconf,nonacm,10pt]{acmart}

\hypersetup{hidelinks}

\usepackage{tabularx}
\usepackage{amsmath}
\usepackage{siunitx}
\usepackage{enumitem}
\usepackage{algorithm}
\usepackage[noend]{algpseudocode}
\usepackage{multirow}
\usepackage{xspace}

\usepackage[font=small]{caption}
\DeclareSIUnit{\TECU}{TECU}
\DeclareSIUnit{\electron}{e^-}

\newlist{params}{description}{1}
\setlist[params]{font=\normalfont\ttfamily,labelsep=0.8em,leftmargin=2.6cm,style=nextline,itemsep=0.2em,topsep=0.2em}

\usepackage{longtable}
\usepackage{subcaption}

\newcommand{\sysname}{\textsc{Gnomon}\xspace}

\newlength{\figurecaptionaboveskip}
\newlength{\figurecaptionbelowskip}
\begin{document}

\title[GNOMON: Reading the Sky to Forecast the Ground]{\textit{\Large Reading the Sky to Forecast the Ground:}\texorpdfstring{\\}{ }
{\LARGE Physics-Informed Link-State Forecasting for LEO Networks at Any Location}
}

\author{Yunxiang Chi}
\affiliation{%
  \institution{Princeton University}
  \city{Princeton}
  \state{New Jersey}
  \country{USA}}
\email{yc3926@princeton.edu}

\author{Zhenlin An}
\affiliation{%
  \institution{University of Georgia}
  \city{Athens}
  \state{Georgia}
  \country{USA}}
\email{zhenlin.an@uga.edu}

\author{Longfei Shangguan}
\affiliation{%
  \institution{University of Pittsburgh}
  \city{Pittsburgh}
  \state{Pennsylvania}
  \country{USA}}
\email{longfei@pitt.edu}

\author{Kyle Jamieson}
\affiliation{%
  \institution{Princeton University}
  \city{Princeton}
  \state{New Jersey}
  \country{USA}}
\email{kylej@princeton.edu}

\begin{abstract}
In this paper, we introduce \sysname, a physics-informed system that forecasts user-perceived low-Earth-orbit (LEO) downlink throughput, uplink throughput, and round-trip time (RTT) under different levels of trace availability. \sysname's physics layer reconstructs the serving geometry and four-leg bent-pipe attenuation from public weather, orbital, routing, and licensing data. Based on what is available, \sysname conditions on the target terminal's own history (Mode~1), measurements from nearby publicly reachable dishes (Mode~2), or the physical covariates alone (Mode~3) to predict the link state: Modes~1 and~2 share a fine-tuned time-series foundation model, while Mode~3 uses a compact boosted-tree estimator. All three modes expose a common output interface and can be selected without retraining.
We evaluate \sysname using $8{,}260$ minutes of $1$~Hz measurements collected at nine sites across five states in the U.S. We train on three sites and hold out the remaining six sites and their serving beams. On these unseen sites, the own-trace mode reduces downlink-throughput and RTT prediction error by $17\%$ and $11\%$ relative to the strongest published baseline and, to our knowledge, provides the first LEO uplink forecasts. The neighbor-trace mode requires no on-site hardware, while the covariate-only mode reduces downlink-throughput and RTT error by $24.6\%$ and $78.8\%$ relative to the only prior covariate-only forecaster. Moreover, experiments show that \sysname provides calibrated quantile bands and improves adaptive-bitrate streaming driven over real TCP flows on replayed Starlink links.

\end{abstract}

\maketitle

\section{Introduction}
\label{sec:intro}

\begin{figure}[t]
  \centering
  \includegraphics[width=\columnwidth]{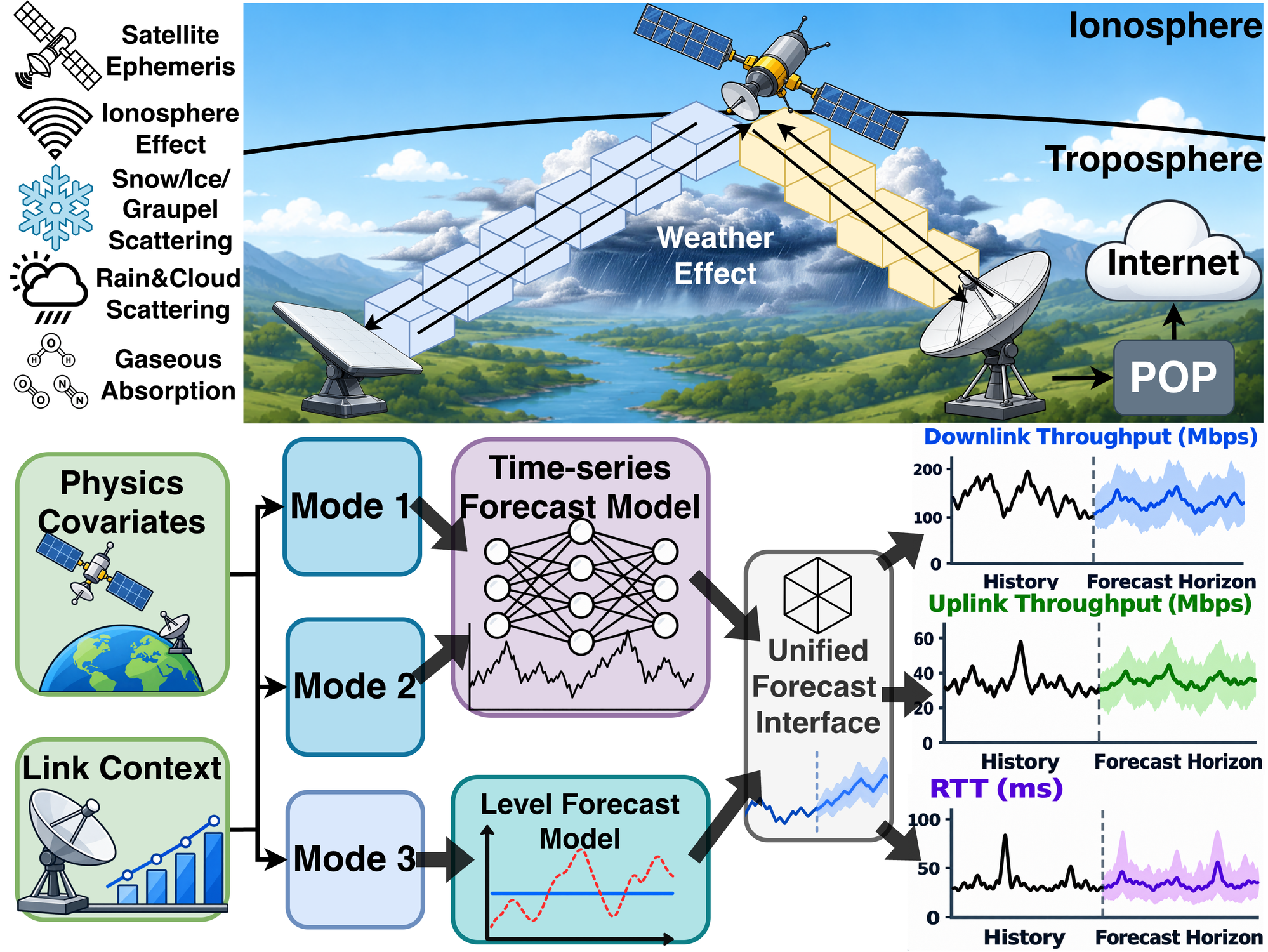}
\caption{\textbf{\sysname{} overview.} \textnormal{\sysname{} models how weather, atmospheric conditions, satellite geometry, and other physical factors affect LEO links, and combines them with the available link measurements. Its forecasting interface is powered by a large time-series model that predicts future downlink throughput, uplink throughput, and RTT under different levels of measurement availability. These forecasts allow network systems to anticipate link dynamics and adapt before performance changes occur.}}
  \label{fig:teaser}
  \Description{Two-panel system overview. Top: a user terminal reaches the Internet through a satellite, a ground station and a PoP, with the four bent-pipe legs crossing cloud, rain and the ionosphere, and a legend of the atmospheric effects modeled. Bottom: physics covariates and link context feed three operating modes, which share a unified interface emitting downlink, uplink and RTT bands over a forecast horizon.}
  \vspace{-2mm}
\end{figure}

Low-Earth-orbit (LEO) broadband is expanding rapidly, with Starlink alone serving millions of terminals across dozens of countries~\cite{mohan2024, t3p}. However, LEO links are highly dynamic. Their performance depends on atmospheric attenuation along the propagation path, contention among users sharing a beam, and the rapidly changing geometry of the serving satellite. As a result, downlink throughput, uplink throughput, and latency can vary substantially over short time scales. Prior studies show that rain can reduce throughput significantly while sometimes leaving latency unchanged~\cite{ullah2025starlink, lottermoser2026weather, wetlink}; under heavier storms, round-trip time (RTT) and packet loss degrade as well~\cite{ehsani2026storm}. Starlink also reassigns satellites and beams roughly every 15 seconds, and these reconfigurations cause second-scale disruptions in throughput and latency~\cite{mohan2024, starnet}. Forecasting these changes would help applications and network controllers make proactive decisions.


LEO link forecasting relies on two types of information: \emph{physical covariates} and \emph{measured link traces}.

\begin{itemize}[leftmargin=*]
    \item Physical covariates describe atmospheric conditions along the user and ground-station links, as well as satellite visibility, elevation, and hardware generation. These can be obtained from public weather products, orbital ephemerides, and regulatory filings, which exist for any location.
    \item Measured link traces capture factors that public data cannot reveal, including cell contention, demand patterns, and the operator's scheduling policies. However, such traces are available only at locations where a terminal (e.g., a Starlink user terminal) is already deployed and measured.
\end{itemize}

At an uninstrumented location, the physical covariates remain available, but the local link history does not.
Even at instrumented locations, the available trace information can change over time. For instance, during site planning, a forecaster may have only the physical covariates. A nearby publicly reachable dish may provide a neighbor trace, and after deployment, the site's own measurement history becomes available. Nearby dishes may also appear or disappear from the public Internet. A forecaster tied to a fixed input configuration would require a separate model, or repeated retraining and calibration, whenever the available observations change.
The central challenge is therefore to forecast LEO link performance under different levels of trace availability without changing the interface exposed to downstream applications.

Existing work does not fully address this challenge. Most fine-grained LEO link-state predictors, including StarNet~\cite{starnet}, T3P~\cite{t3p}, and BG-CFQS~\cite{bgcfqs}, require recent measurements from the target terminal. These systems therefore cannot operate at a location without an on-site trace. Horizon~\cite{horizon} expands the spatial coverage by learning from crowdsourced speed-test records and forecasting from geographic, temporal, and weather covariates. However, it forecasts only at coarse spatial and temporal granularity, and a short study likewise predicts Starlink link quality from weather data alone over the WetLinks dataset~\cite{lanfer2024weather}. Moreover, these methods return point predictions---the sole exception, BG-CFQS~\cite{bgcfqs}, emits a single conservative lower quantile for admission control---so none reports a predictive quantile band whose coverage is calibrated and measured, and none forecasts uplink throughput. Prior measurement studies have characterized Starlink either from a small number of instrumented terminals~\cite{firstlook, ullah2025starlink} or from sporadic crowdsourced speed tests~\cite{mohan2024}. However, the former cannot be extended to a new location without deploying hardware there, and the latter cannot resolve the per-second link state a planner acts on, nor can new measurements be collected on demand at a chosen location.


This paper presents \sysname, a physics-informed system that forecasts downlink throughput, uplink throughput, and RTT for LEO links.
It combines public physical data with an on-site link trace or with a trace from a nearby publicly reachable dish, and still forecasts when no measured trace is available at all.

At the core of \sysname is a physics layer that constructs link-quality covariates for any coordinate and time. As shown at the top of Fig.~\ref{fig:teaser}, a bidirectional bent-pipe connection contains two atmospheric paths: a user link between the terminal and the satellite, and a feeder link between the ground station and the satellite. Each path carries an uplink and a downlink at different frequencies. For each of these four links, \sysname estimates atmospheric attenuation using its carrier frequency, slant geometry, and public weather data. It also derives satellite visibility, elevation, hardware generation, and demand-related features from public orbital, routing, licensing, population, and calendar data. The trace-based forecaster combines these covariates with the available link history to predict per-step downlink throughput, uplink throughput, and RTT. Moreover, \sysname returns multiple quantiles rather than only a point estimate, allowing downstream policies to select a prediction according to their risk tolerance.

\sysname supports three operating modes, shown at the bottom of Fig.~\ref{fig:teaser}. 
\textit{(1) Mode~1} uses the query terminal's own recent trace for channel forecasting. 
\textit{(2) Mode~2} introduces a new forecasting capability absent from prior work: when no local trace is available, it uses as the history context the RTT, downlink, and uplink measurements that \sysname obtains from nearby publicly reachable dishes through low-rate, bounded active probes.
Modes~1 and~2 use the same fine-tuned time-series foundation model and differ only in how their history channels are populated. \textit{(3) Mode~3} applies when neither trace is available: it estimates the link-state level from the physical covariates alone using a lightweight boosted-tree model. All three modes output downlink, uplink, and RTT forecasts in the same format. \sysname can therefore switch modes as trace availability changes, without retraining or modifying the downstream application.

\begin{figure}[t]
  \centering
  \includegraphics[width=\columnwidth]{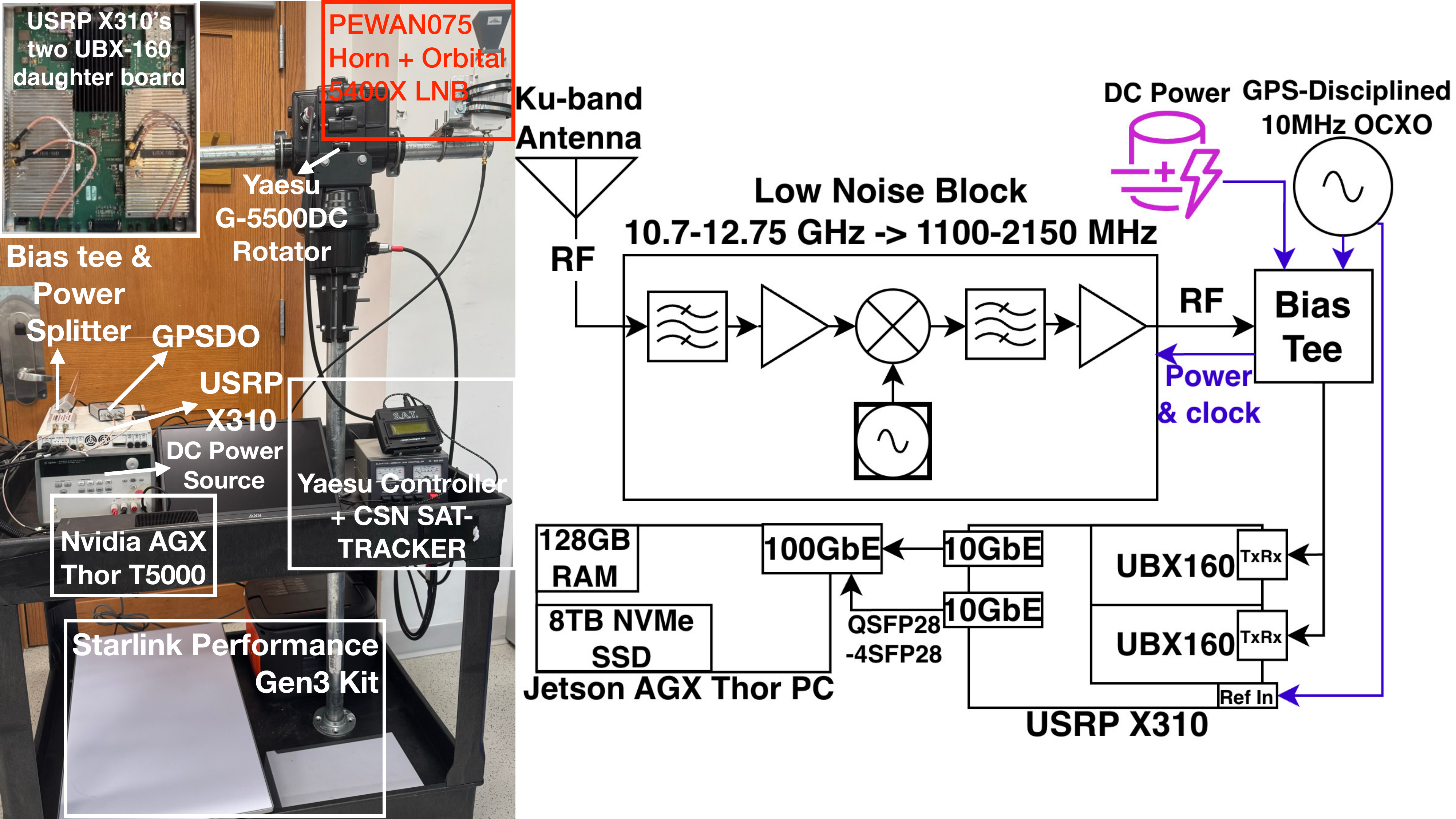}
  \caption{\textbf{The LEO measurement platform.} \textnormal{\emph{Left:} the deployed hardware setup used to collect direct Ku-band satellite signal measurements. \emph{Right:} the circuit diagram of the measurement platform, illustrating the RF signal path, clocking and synchronization, software-defined radio (SDR), and host interface.}}
  \label{fig:hardware-real}
  \Description{Left: photograph of the deployed platform, with the Starlink terminal, horn and downconverter on a rotator, software-defined radio, reference clock and host annotated. Right: block diagram of the same chain from antenna through low-noise block, bias tee and radio to the host.}
\end{figure}

To validate \sysname's physics layer over the air, we build a customized LEO measurement platform that directly captures the Ku-band Starlink satellite downlink, as shown in Fig.~\ref{fig:hardware-real}. We design the receiver-chain hardware and a custom Starlink sniffer algorithm that reliably detects the Starlink frame's synchronization sequence in the raw Ku-band capture, enabling direct link-level channel measurements. It records the received satellite signal power at five sites in the eastern United States over three days. Separately, we deploy a commercial Starlink user terminal~\cite{starlink_user_terminal} to record downlink and uplink throughput, RTT, packet loss, and dish state at 1~Hz at nine sites across five states in the eastern and midwestern U.S. Over 16 measurement days, we collect $8{,}260$ minutes of continuous link-state traces under clear skies, cloud cover, and rain of varying intensity.

We train \sysname on three sites and reserve the other six sites for evaluation, so the forecasting model is tested entirely on locations not observed during training.
On the held-out sites, \sysname in Mode~1 reduces downlink-throughput error by $17\%$ and RTT error by $11\%$ relative to the state-of-the-art StarNet baseline. More importantly, \sysname forecasts uplink throughput, a capability that, to our knowledge, no prior LEO forecaster provides. In Mode~3, with no trace at all, \sysname reduces downlink-throughput and RTT error by $24.6\%$ and $78.8\%$ relative to Horizon, the only published covariate-only LEO forecaster.

\noindent\textbf{Contributions.} We make the following contributions:

\noindent\textbf{(i)} A physics layer that derives four frequency-specific bent-pipe attenuation estimates from public weather, orbital, and Federal Communications Commission (FCC) licensing data at any coordinate. We validate its user-downlink leg over the air using direct Starlink Ku-band measurements and show that its covariates reduce StarNet's downlink forecasting error by 18\%.

\noindent\textbf{(ii)} A unified forecasting system with three modes based on trace availability: the target terminal's own trace, a nearby public terminal's trace, or physical covariates alone. It generalizes to unseen sites and beams, provides the first LEO \emph{uplink} forecasts and the first calibrated two-sided \emph{quantile bands} for a LEO link, and outperforms the existing downlink and RTT forecasters of each mode's input class.

\noindent\textbf{(iii)} A zero-on-site-deployment measurement method that combines public data with low-rate, bounded probes of nearby publicly reachable dishes. Extending the HitchHiking public-dish measurement methodology~\cite{hitchhiking}, it obtains downlink, uplink, and RTT context for locations that have not been instrumented.


\section{System Design}
\label{sec:design}
\begin{figure}[t]
  \centering
  \includegraphics[width=\linewidth]{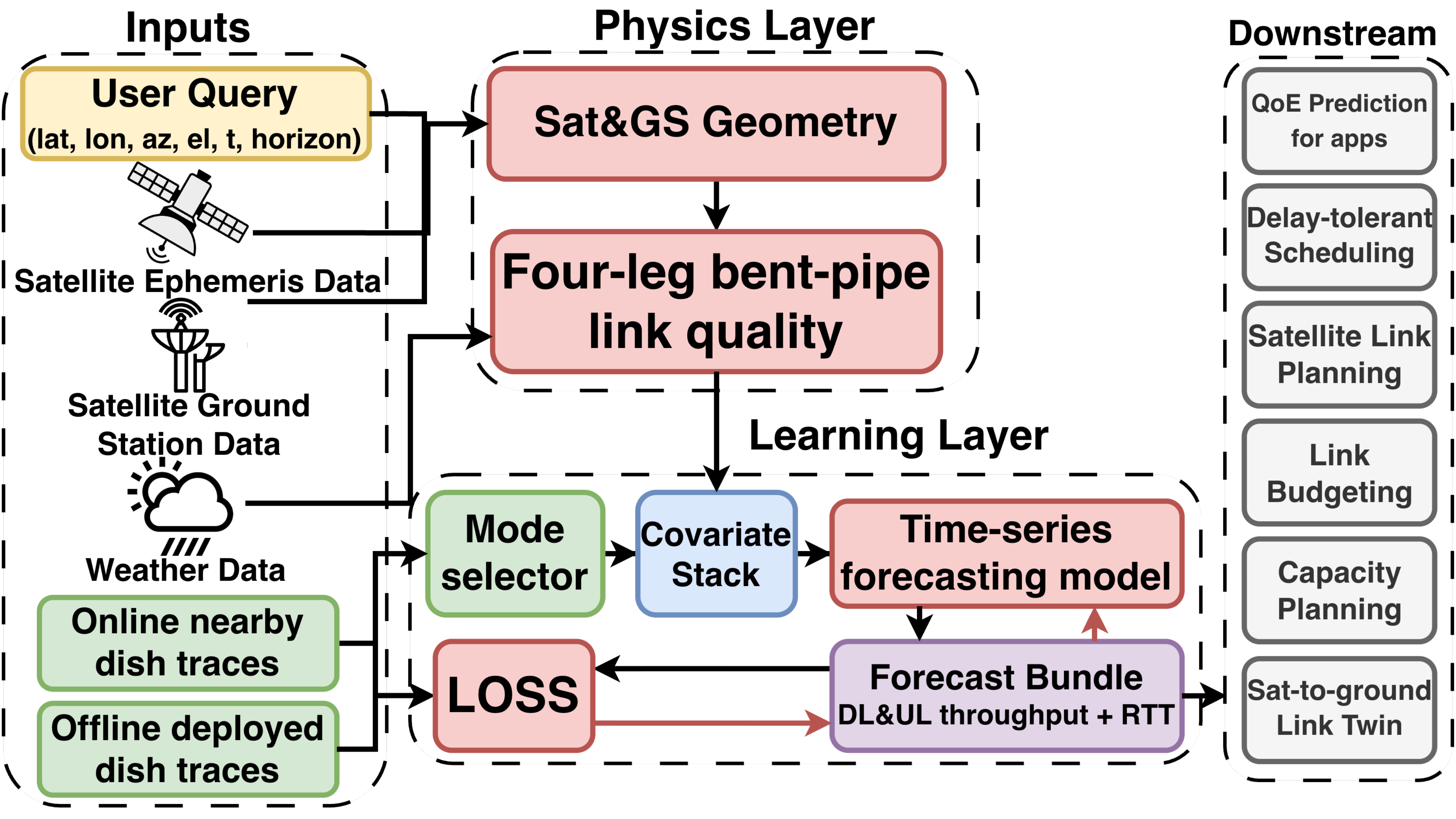}
  \caption{\textbf{\sysname system architecture.} \textnormal{The \emph{physics layer} turns the query, weather, and satellite ephemeris into attenuation and geometry covariates; the \emph{learning layer} forecasts link state from a context plus the covariates.}}
  \label{fig:e2earch}
  \Description{Architecture: the physics layer produces covariates, and the learning layer forecasts link state from them plus a context.}
\end{figure}

\sysname forecasts user-perceived downlink throughput, uplink throughput, and RTT for a given location, antenna direction, start time, and forecast horizon (e.g., the next $H$ seconds, from minutes to hours). It adapts to whether recent link histories are available from the query-site terminal, from a nearby public dish, or from neither.


\subsection{Architecture Overview}
\label{sec:arch}

As shown in Fig.~\ref{fig:e2earch}, \sysname consists of a physics layer (\S\ref{sec:physics}) and a learning layer (\S\ref{sec:learning}). The physics layer combines the user query with public weather, orbital, and ground-station data to recover the serving geometry and estimate the quality of the four directional satellite links. The learning layer combines these physics-derived covariates with the available link history to produce a forecast bundle containing downlink throughput, uplink throughput, RTT, and their predictive bands. The learning-layer input is organized into a \emph{covariate stack} and a \emph{time-series stack}. The covariate stack contains weather, satellite geometry, and demand-related features over both the context and forecast horizon.
The time-series stack contains recent downlink, uplink, and RTT traces over the context window. The availability of histories determines which operating mode \sysname uses, as elaborated below.


\noindent $\bullet$ \textbf{Mode~1: Own history.} When the query terminal has already been deployed for a while, \sysname uses its recent measurements as the time-series context. This mode supports real-time forecasting for an active terminal.

\noindent $\bullet$ \textbf{Mode~2: Neighbor trace.} When the query terminal has no local history (e.g., when evaluating Starlink performance at a prospective deployment site), \sysname uses the recent measurements from nearby public dishes. \sysname prefers dishes within the same beam cell~\cite{pan2023measuring}, since they are more likely to share the serving beam and propagation conditions. Otherwise, \sysname uses the closest available dishes. This mode supports forecasting at uninstrumented locations without installing hardware on site.

\noindent $\bullet$ \textbf{Mode~3: Covariates only.} When no trace is available, e.g., at a remote site with no public dish nearby, \sysname predicts downlink, uplink, and RTT directly from the covariate stack using a compact gradient-boosted regression-tree ensemble~\cite{friedman2001}, which yields one link-state level per channel over the horizon rather than the per-step dynamics captured by Modes~1 and~2 (\S\ref{sec:forecast}).

Modes~1 and~2 use the same time-series model and differ only in the source of their context measurements. Mode~3 uses a separate, smaller model because it has no time-series input. All three modes use the same output interface, allowing downstream tasks to switch between them without modification.

\subsection{Physics Layer: Link Quality Prediction}
\label{sec:physics}

\begin{figure}[t]
  \centering
  \includegraphics[width=\linewidth]{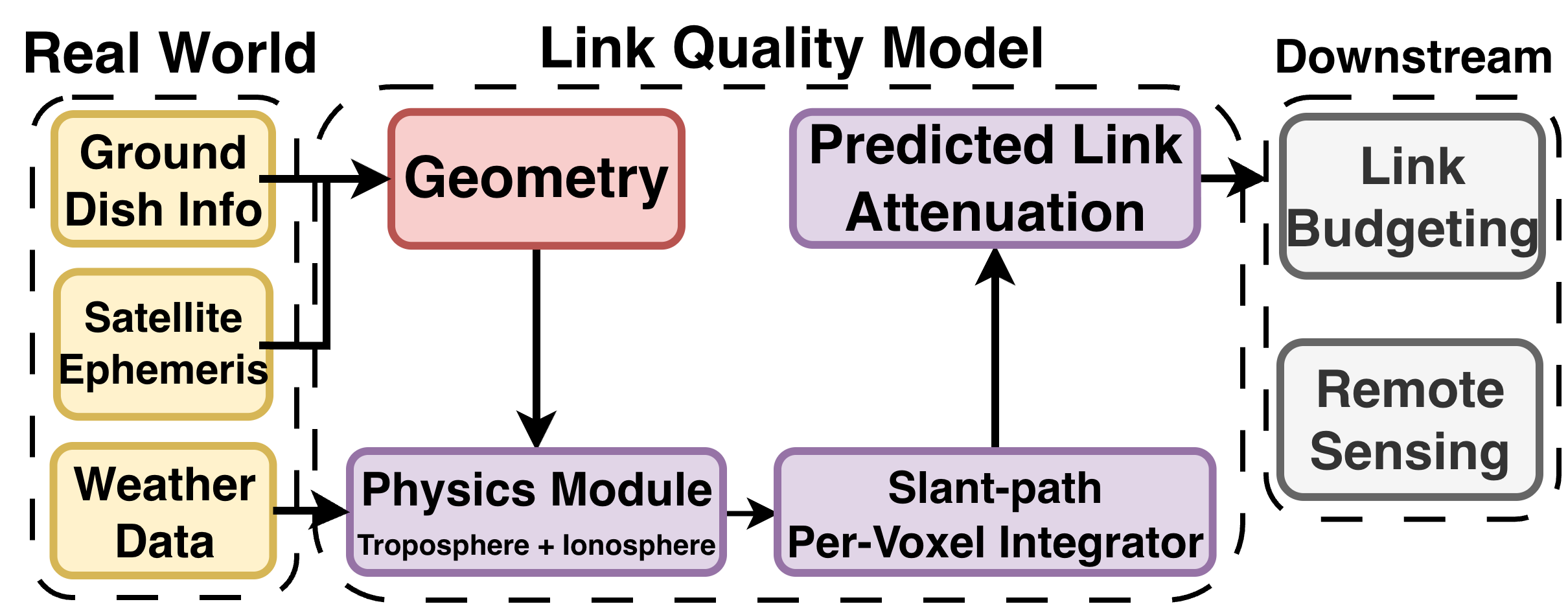}
  \caption{\textbf{Physics-layer architecture.} \textnormal{\sysname's link-quality predictor models the four-leg bent-pipe path.}}
  \label{fig:lqmodel}
  \Description{The physics-layer link-quality model, evaluated along the four legs of the bent pipe.}
\end{figure}

All three operating modes use the same physics-derived covariates; they differ only in the trace context available to the learning layer. Given a user query, the physics layer constructs these covariates in two steps (Fig.~\ref{fig:lqmodel}). It first recovers the likely serving satellite and ground station (\S\ref{sec:geometry}), and then estimates propagation attenuation along the four links of the resulting bent-pipe path (\S\ref{sec:fourleg}). Both steps use public data and require no prior link history at the query site.

LEO link propagation attenuation depends on the carrier frequency, elevation angle, and atmospheric conditions along each slant path. Its main components include gaseous absorption, cloud and hydrometeor attenuation, and ionospheric fading~\cite{ITURP676, Chen1975RAND_R1694, Zhao2025RadioWavePropagationSatelliteSystems}. Rather than represent weather using a single value at either endpoint, \sysname integrates attenuation through the weather-grid cells intersected by the signal path. \S\ref{app:background} provides additional background on the propagation mechanisms and LEO network architecture.



\subsubsection{Serving Geometry Recovery}
\label{sec:geometry}

Before computing weather-induced attenuation, \sysname must determine the bent-pipe path serving the query. This path connects the user terminal to a ground station through a serving satellite; its two segments are therefore the user--satellite path and the satellite--ground-station path, as illustrated at the top of Fig.~\ref{fig:teaser}.
Given the query location and terminal boresight, \sysname first identifies both the ground station and the satellite likely carrying this connection. But Starlink does not expose these identifiers through its user-facing telemetry interfaces~\cite{starlinkgrpcgolang, grpcissue102}. We therefore reconstruct them from satellite ephemerides, public routing records, and ground-station filings.
Specifically, inspired by prior measurement studies~\cite{tanveer2023making, hitchhiking}, we first identify the likely ground station and then determine which satellites are simultaneously visible to the terminal and that station.


\noindent $\bullet$ \textbf{Ground Station.} A terminal is statically assigned to a Point of Presence (PoP), and each PoP is fed by a set of ground stations, so recovering the PoP narrows the candidates. We read the PoP from Starlink's own routing records and take the nearest footprint-feasible station to it as the serving ground station, following the routing-anchored station selection of prior measurement work~\cite{hitchhiking}; the records themselves and their alternatives are in \S\ref{sec:data_geometry}. At a bare query coordinate with no terminal of its own, the same anchor comes from a nearby publicly exposed dish, since a PoP serves an entire country or sub-region rather than an individual dish~\cite{mohan2024,hitchhiking}.
Recovering the ground station matters because the end-to-end link is a bent pipe that crosses the atmosphere \emph{twice}, and the two crossings may see entirely different weather.

\noindent$\bullet$ \textbf{Satellite.} The potential serving satellite is recovered by propagation and masking. Every satellite of the serving constellation is propagated to time $t$ with SGP4~\cite{Vallado2006SGP4} from public ephemeris (\S\ref{sec:data_geometry}), and its elevation is evaluated at both the dish and the ground station. A satellite is a co-visibility candidate only if it clears a $25^\circ$ elevation mask at \emph{both} endpoints and falls within the user terminal's field of view~\cite{fcc2021spacexmod, tanveer2023making, ahangarpour2024fov}.
Rather than commit to a single candidate---the terminal reselects several times a minute---we keep the whole set and assign each member a serving probability by a softmax over its dish elevation with a $5^\circ$ temperature, so higher, more direct passes dominate while lower ones retain mass. The effective number of candidates, $1/\sum_i p_i^2$, then lands near one to four. The attenuation legs of \S\ref{sec:fourleg} are then evaluated for the highest-probability candidate, while the geometry consumed downstream summarizes the whole weighted set: the number of co-visible satellites $N$, the probability-weighted mean elevation $\bar\theta_{\mathrm{el}}$, and the serving candidate's hardware generation. These enter the learning layer as the covariates of \S\ref{sec:forecast}.

\subsubsection{Modeling the Four-Leg Bent-Pipe Attenuation}
\label{sec:fourleg}

The four attenuation legs are the coupling between the physics layer and the learning layer. Each leg is an end-to-end slant-path attenuation in decibels for one carrier, evaluated over the serving geometry of \S\ref{sec:geometry}, and the leg set spans both atmospheric crossings of the bent pipe: the user Ku downlink and uplink over the dish ($11.575$ and $14.25$\,GHz) and the Ka downlink and uplink over the ground station ($19.0$ and $28.75$\,GHz). A few ground stations now also operate in E band ($71$--$76$ and $81$--$86$\,GHz), which is authorized for gateway feeder links rather than user service links~\cite{fcc2026gen2, fcc2024eband}.

\noindent$\bullet$ \textbf{Slant-path attenuation model.} Along a slant ray we model total attenuation as an additive sum of independent loss mechanisms,
\begin{equation}
A_{\mathrm{atm}}(f,\theta_{\mathrm{el}},t)=
A_{\mathrm{gas}}+A_{\mathrm{cloud}}+A_{\mathrm{rain}}+A_{\mathrm{snow/ice}}+A_{\mathrm{iono}},
\label{eq:atm_sum}
\end{equation}
where each tropospheric term integrates a local specific attenuation $\gamma$ (dB/km) along the ray. Discretizing the line of sight into samples $n$ of slant length $\Delta s_n=\Delta h_n/\sin\theta_{\mathrm{el}}$, each intersecting one weather grid cell,
\begin{equation}
\label{eq:raymarch}
\begin{split}
A_{\mathrm{atm}}(f,\theta_{\mathrm{el}},t)\approx \sum_n\!\bigl(&\gamma_{\mathrm{gas},n}+\gamma_{\mathrm{cloud},n}+\gamma_{\mathrm{rain},n}\\
&+\gamma_{\mathrm{snow/ice},n}\bigr)\Delta s_n+A_{\mathrm{iono}}(t).
\end{split}
\end{equation}
\begin{figure*}[t]
  \centering
  \includegraphics[width=\textwidth]{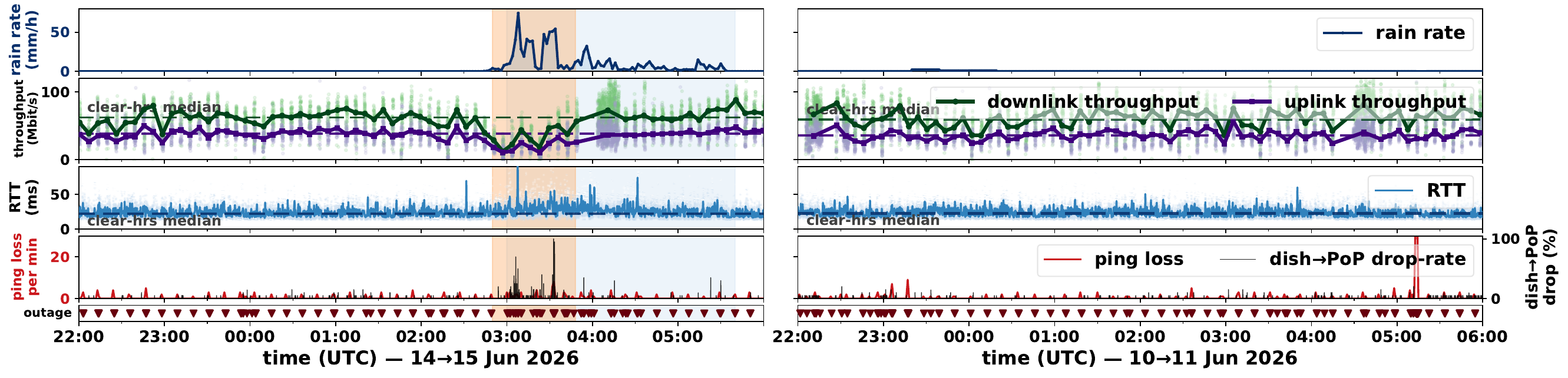}
  \caption{\textbf{Weather impact on the link.} \textnormal{Storm night (left) vs.\ clear
  night (right) at the same Starlink user terminal. During the rain (yellow range), downlink and uplink
  throughput roughly halve while RTT and loss rise; the clear night barely changes.}}
  \label{fig:weatherqos}
  \Description{Throughput, round-trip time and loss at one terminal across a stormy night and a clear night.}
\end{figure*}

The tropospheric mechanisms follow the ITU-R P-series models wherever one exists: gaseous absorption by oxygen and water vapor (ITU-R~P.676~\cite{ITURP676}), attenuation by non-precipitating cloud and fog liquid (ITU-R~P.840~\cite{ITURP840}), and rain attenuation (ITU-R~P.838~\cite{ITURP838}); frozen-hydrometeor extinction for snow, graupel, and hail has no ITU-R counterpart, so we derive it from Rayleigh and Mie scattering with an effective-medium dielectric for melting particles~\cite{Chen1975RAND_R1694, Leinonen2014PyTMatrix, Kim2007IceScattering, Bruggeman1935}; the ionosphere brings dispersive delay and scintillation fade (ITU-R~P.531~\cite{ITURP531}). The atmospheric state along each link is drawn from the gridded products of \S\ref{sec:dataset}: HRRR supplies thermodynamic state and cloud/frozen microphysics, MRMS supplies precipitation structure and melting-layer geometry, and SWPC supplies the space-weather indices that drive scintillation. Details of the derivations, coefficient tables, and the hydrometeor retrieval that instantiates particle phase and size along the link are in \S\ref{app:atten}.

Two properties of this model shape what the forecaster can learn. \emph{First}, precipitation dominates the legs, and its attenuation grows steeply with frequency: the same rain cell that costs the Ku-band legs a few tenths of a decibel costs the Ka-band legs several decibels (and the E band more still), so most of the weather signal rides the ground-station legs. \emph{Second}, rain is integrated only through the portion of the path that crosses it; at the serving elevations of Starlink the ray leaves the rain layer within a few kilometers horizontally, so the leg tracks the rain the beam actually crosses rather than a ground average. Finally, the scheduling and beam-sharing algorithms of commercial operators are nonpublic, and even a small attenuation may cost a user disproportionately once the scheduler reacts to it; this layer therefore predicts link quality rather than delivered rate.

\noindent$\bullet$ \textbf{Atmospheric effect on network performance.} Prior work documents the effect of the atmosphere on LEO communication~\cite{ullah2025starlink, lottermoser2026weather, ehsani2026storm, wetlink}.
At our instrumented site, we observe the same effect: during the storm, measured downlink and uplink throughput roughly halve, round-trip time rises, and loss and outages increase, whereas a clear night at the same hours shows no such drop (Fig.~\ref{fig:weatherqos}). Two loss metrics are shown: \emph{packet loss} is the end-to-end probe's drop rate, so congestion or a routing event anywhere on the path can raise it, whereas the \emph{dish-to-PoP drop rate} comes from the terminal's gRPC telemetry and is scoped to the terminal-to-PoP segment, so it excludes the rest of the Internet path.

\subsection{Learning Layer: Link-State Forecasting}
\label{sec:learning}

The physics layer provides the covariate stack shared by all three modes; the remaining input is the recent link-state context. Mode~1 reads it from the query terminal itself, Mode~2 must construct it by discovering, localizing, and probing nearby public dishes, and Mode~3 has none. We therefore first describe how \sysname discovers public dishes and measures their recent link state for Mode~2 (\S\ref{sec:discovery}, \S\ref{sec:probes}), and then present the trace-based forecaster for Modes~1 and~2, the covariate-only estimator for Mode~3, and their common predictive-band calibration (\S\ref{sec:forecast}).

\subsubsection{Dish Discovery, Filtering, and Localization}
\label{sec:discovery}

Mode~2 requires a pool of publicly reachable Starlink dishes with estimated locations. \sysname builds this pool in three steps: it discovers and filters candidate hosts, estimates their locations, and selects the neighbors most relevant to each query. We elaborate on these steps below.


\noindent\textbf{Step One: Discovery and filtering.}
\sysname scans Starlink's advertised IPv4 ranges with ZMap~\cite{zmap} and identifies hosts responding on port~443. It retains a host only if it passes three additional checks.

\begin{itemize}[leftmargin=*]
    \item First, the host must respond over both HTTP and ICMP: ICMP is required by the probes in \S\ref{sec:probes}, and the HTTP response is what the fingerprinting of the next check inspects.
    \item Second, its service fingerprints must match a customer terminal rather than a middlebox or performance-enhancing proxy~\cite{censys, lzr, gps}.
    \item Third, its RTT from our measurement vantage point must fall between $30$ and $200$\,ms, the range we use to exclude clearly terrestrial or heavily proxied paths. Hosts that pass the scan and all three checks form the candidate pool.
\end{itemize}


\noindent\textbf{Step Two: Localization.} A Starlink dish routes through a known PoP, so IP geolocation often places the dish near its PoP rather than at its physical location. \sysname therefore uses the PoP only as a regional anchor. It obtains a finer estimate from either commercial GeoIP~\cite{maxmind} or the median location of public measurement clients in the dish's \texttt{/24} prefix~\cite{MLab}. The finer estimate is accepted only if it lies within $200$\,km of the PoP anchor; otherwise, \sysname falls back to the PoP anchor. \S\ref{sec:data_misc} describes these data sources.


\noindent\textbf{Step Three: Neighbor selection and aggregation.} Given the localized pool, a query at $(\mathrm{lat},\mathrm{lon})$ is served by the dishes physically nearest it. We take the closest terminals within one beam cell (diameter $d_c\approx24$\,km)~\cite{pan2023measuring}---the ground footprint of one beam, inside which users share a serving beam and the same weather, and within which user-perceived performance is dominated by the terminals competing in that cell~\cite{pan2024satlinks}---and fall back to the globally closest terminals when the cell is empty. Each neighbor is weighted by $1/\max(1,\mathrm{RTT})$, which discourages a dish that GeoIP places nearby but that routes through a distant PoP, without re-localizing it. Their probes are combined on the $\Delta$-second grid by a per-neighbor median within each grid bin, followed by an RTT-weighted mean across neighbors.

\subsubsection{Probes}
\label{sec:probes}
Public dishes do not expose their per-second telemetry to external users~\cite{hitchhiking}. \sysname therefore estimates their recent link state using bounded active probes. It measures RTT, downlink capacity, and uplink capacity.


\noindent$\bullet$ \textbf{RTT.} A \texttt{scamper}-based ICMP echo probe runs once every $5$\,s~\cite{Luckie2010Scamper}. The returned samples provide the recent RTT history from the measurement vantage point to the dish.

\noindent$\bullet$ \textbf{Downlink capacity.} Downlink capacity is estimated with a \emph{count-varied} packet-train probe~\cite{Dovrolis2004Dispersion, Prasad2003BwEst}, run single-ended off the replies elicited from a terminal we do not control~\cite{Saroiu2002SProbe, Dischinger2007Residential}.
A ladder of short back-to-back trains of increasing length ($m\in\{2,4,8,16\}$ packets) is fired at the dish; each train queues behind the downlink bottleneck and its per-packet replies return spread in time, so the capacity is read from how that inter-reply spread grows with $m$, while the coefficient of variation of the gaps gives a contention indicator. The estimator needs only one prompt reply per probe, and that reply must be of fixed small size so the uplink is not measured by accident. We elicit it with a TTL-scoped ICMP echo, answered by an ICMP time-exceeded; a TCP segment to a closed port (answered by an RST) or a UDP datagram to a closed port (answered by an ICMP port-unreachable) carry the measurement equally well.
\begin{figure}[t]
  \centering
\includegraphics[width=0.95\columnwidth]{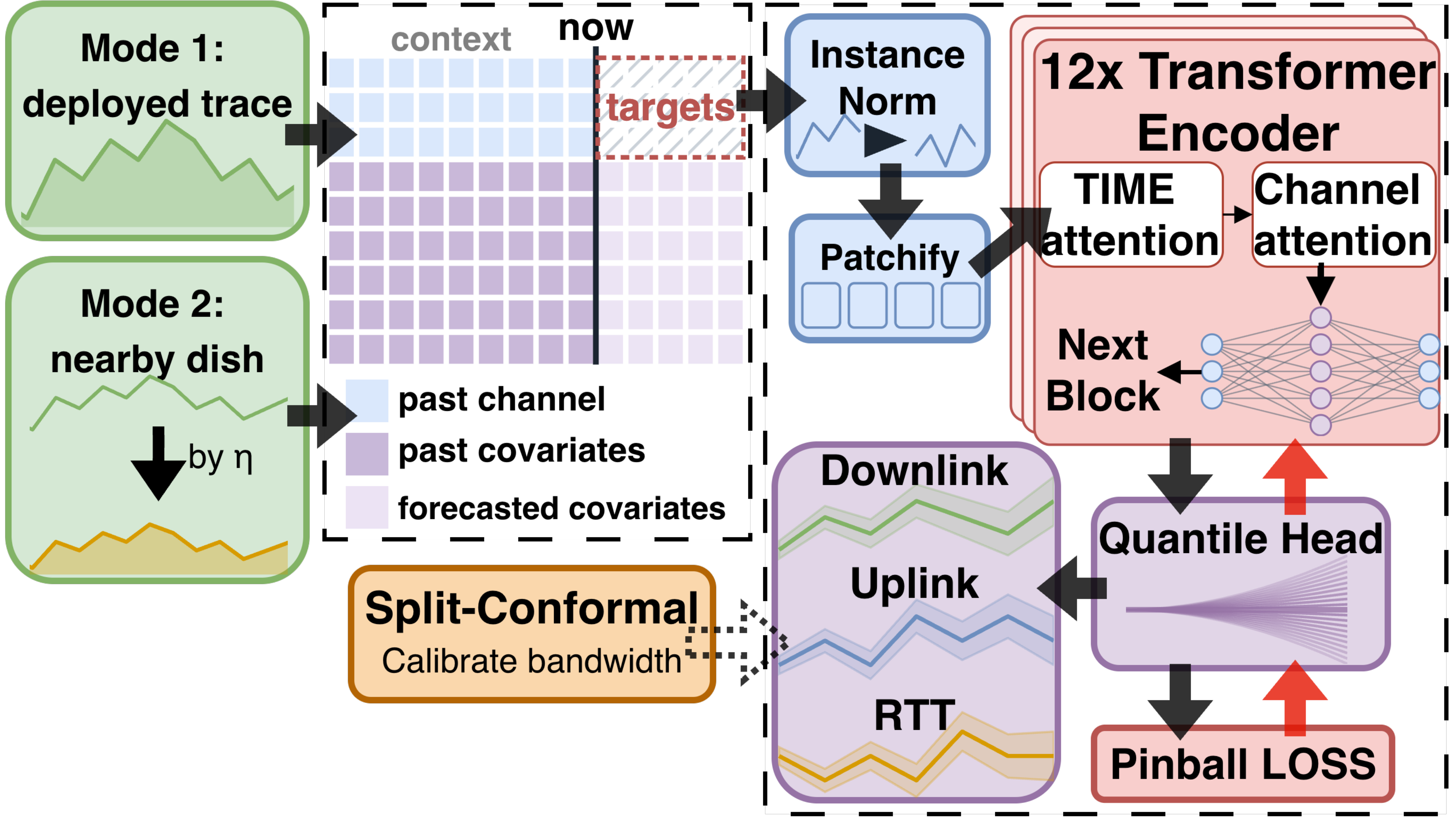}
  \caption{\textbf{Time-series model and fine-tuning.} \textnormal{Modes~1 and~2 use the same time-series stack; covariates continue past \emph{now}. The window is instance-normalized, patchified, and passed through time-series encoder blocks attending over time and across channels.}}
  \label{fig:learning-layer}
  \Description{The time-series model: one window, instance normalization, patchify, twelve encoder blocks, quantile head, conformal calibration.}
\end{figure}

\noindent$\bullet$ \textbf{Uplink capacity.} Uplink is measured symmetrically, with short bounded packets whose replies must drain back through the dish's uplink bottleneck, since the LEO uplink is lower-capacity by design. A plain ICMP echo suffices, and the drain rate gives the uplink capacity.

\noindent$\bullet$ \textbf{Measurement ethics.}
The probes use fixed, bounded packet trains and run at low rates: once every $5$\,s for RTT and once every $30$\,s for each capacity probe. They are designed to avoid sustained saturation of the user's connection and of the shared satellite link. The vantage point hosts a public information and opt-out page, following established Internet-measurement practice~\cite{zmap} and prior LEO measurement work~\cite{hitchhiking}. Dishes whose owners opt out are excluded from all subsequent probes. \S\ref{app:ethics} reports the complete protocol and traffic budget.


\subsubsection{Forecasting model}
\label{sec:forecast}

We use two separate models to forecast link state: one shared by Modes~1 and~2, and another for Mode~3.
Modes~1 and~2 use a fine-tuned foundation model on a GPU, while Mode~3 runs on a CPU. \S\ref{sec:impl} reports model sizes and inference times.

\noindent \textbf{Trace-Based Forecaster: Modes~1 and~2}.
For each query, the trace-based forecaster is given $T_{\mathrm{ctx}}$ seconds of context and returns a predictive distribution at every step $h=1,\ldots,H/\Delta$ of a $\Delta$-second grid over an $H$-second horizon. It jointly forecasts three user-perceived variables: downlink throughput \texttt{dl} in Mbps, uplink throughput \texttt{ul} in Mbps, and RTT \texttt{rtt} in milliseconds. Each output is represented by the quantile grid $\mathcal{Q}=\{0.1,0.2,\ldots,0.9\}$ rather than by a single point estimate.

Each forecasting window contains the three link-state variables over the context interval. Their future values are masked over the forecast horizon. The window also contains eleven covariates defined over both intervals: four directional attenuation estimates from \S\ref{sec:fourleg}; the number of co-visible satellites $N$, their probability-weighted mean elevation $\bar\theta_{\mathrm{el}}$, and the serving satellite's hardware generation (\S\ref{sec:geometry}); and four calendar-phase features encoding daily and weekly demand patterns. Hardware generation is read from the public satellite catalog described in \S\ref{sec:data_geometry}. Future weather forecasts and orbital propagation provide the covariates over the horizon, allowing the model to anticipate a change in link conditions.
\begin{figure}[t]
  \centering
  \includegraphics[width=\columnwidth]{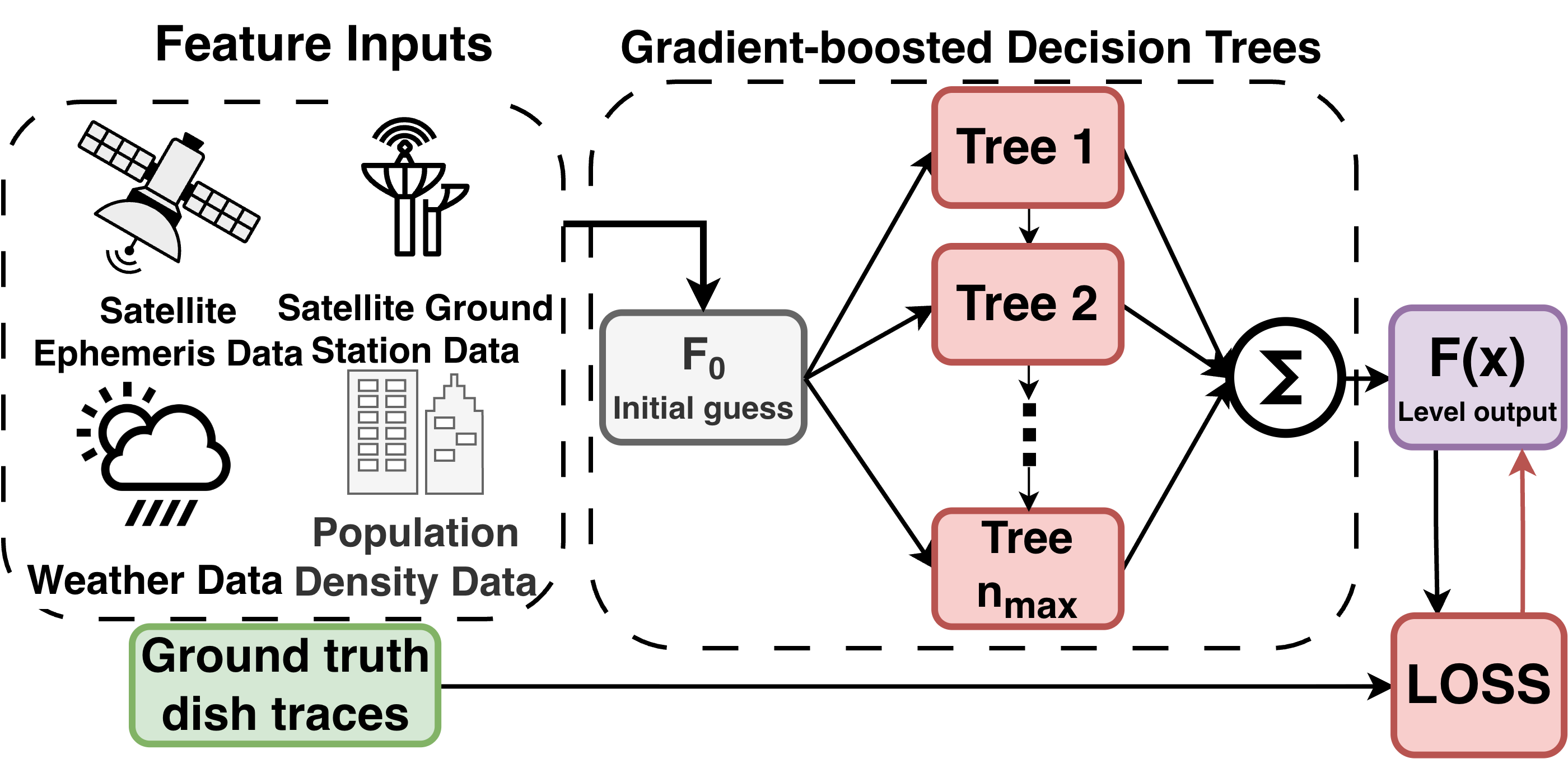}
\caption{\textbf{Mode~3.} \textnormal{We use physics-derived covariates to forecast LEO link performance when no link history is available. A gradient-boosted decision-tree model maps these features to future link conditions.}}
  \label{fig:mode3}
  \Description{Mode~3: the covariates of a window are reduced to one feature vector and mapped by boosted trees to a level.}
\end{figure}

Mode~1 and Mode~2 use the same model and input layout, shown in Fig.~\ref{fig:learning-layer}. Mode~1 fills the context channels with the query terminal's measured throughput and RTT. Mode~2 fills them with the aggregated neighbor estimates.
We use Chronos-2~\cite{chronos2}, a pretrained time-series foundation model that accepts a multivariate history together with known future covariates. We adopt this pretrained model because our labeled data are limited and come from a small number of instrumented sites. We fine-tune the model with a masked pinball loss over the three targets and nine quantile levels, as shown in Fig.~\ref{fig:learning-layer}. The mask handles missing labels and the different sampling rates of throughput and RTT. Throughput values are transformed with $\log(1+x)$ before training and mapped back to Mbps with $\operatorname{expm1}$; RTT remains in its original domain. \S\ref{app:model} details model configuration and training.

Mode~2 requires an additional level correction. Its probes estimate link capacity, while the training labels measure application throughput. For each affected channel $c$, \sysname estimates a global scale factor during training,
\[
\eta_c =
\frac{\operatorname{median}(y_c)}
{\operatorname{median}(x_c)},
\]
where $x_c$ is the neighbor-probe history and $y_c$ is the corresponding label. The factor is applied before normalization and model inference. This removes a systematic capacity-to-throughput offset while leaving the model to learn temporal variation. \S\ref{ss:reconcil} gives the estimation procedure.

\noindent \textbf{Covariate-Only Estimator: Mode~3}.
Mode~3 applies when no informative link trace is available (Fig.~\ref{fig:mode3}). Over a short forecast window, its weather, geometry, and demand covariates change more slowly than the link itself. They can identify the expected link-state level, but they cannot recover the fast fluctuations caused by scheduling, contention, and terminal-specific conditions. Mode~3 therefore predicts one level per channel and applies it across the forecast horizon.

Each Mode~3 window contains twelve inputs. Eleven are shared with the trace-based modes while the twelfth is a population-based contention feature
\[
\log_{10}(1+P_{\mathrm{pop}}),
\]
where $P_{\mathrm{pop}}$ is the population within a $12$\,km radius of the query coordinate. The radius approximates one $24$\,km-diameter user beam. Population is obtained from the $1$\,km WorldPop raster~\cite{worldpop}.
Mode~3 uses a compact gradient-boosted regression-tree ensemble~\cite{friedman2001}. The estimator follows the histogram-based, regularized design used by LightGBM and XGBoost~\cite{lightgbm,xgboost}, with at most $n_{\max}=200$ trees (\S\ref{ss:traincfg}), as shown in Fig.~\ref{fig:mode3}. It is trained under the same quantile objective as the trace-based model and returns the same target and quantile interface. Unlike Mode~1 and Mode~2, it requires no GPU and can train and serve efficiently on a single CPU core.

\noindent\textbf{Predictive-Band Calibration and Serving}.
The median $q_{0.5}$ is the point forecast. The remaining quantiles define predictive bands for planners with different risk tolerances. Because fine-tuning on a small dataset may produce bands that are too narrow or too wide, \sysname calibrates them using split-conformal prediction~\cite{conformal}.

Specifically, on a calibration set held out from fine-tuning, we score each window by how far the truth falls outside its band, $E=\max\{\hat q_{\text{lo}}-y,\;y-\hat q_{\text{hi}}\}$, and widen both edges by the empirical quantile of $E$ at the target level, leaving the median point forecast unchanged. For example, if the nominal $90\%$ throughput band covers only $78\%$ of unseen data, the offset added to each edge is the amount that restores $90\%$; a band that is already over-wide gets a negative offset and is tightened. A reported $q$-level band therefore attains $q$ coverage on average over the calibration distribution, regardless of how tight or loose the fine-tuned quantiles happen to be. The guarantee is marginal and assumes that calibration and test data are exchangeable, which is why \S\ref{sec:eval} measures coverage directly on unseen sites rather than resting on it. The uncertainty the planner consumes thus stays trustworthy even when fine-tuning on the small label set leaves the backbone over- or under-confident.

\begin{figure}[t]
  \centering
  \includegraphics[width=\columnwidth]{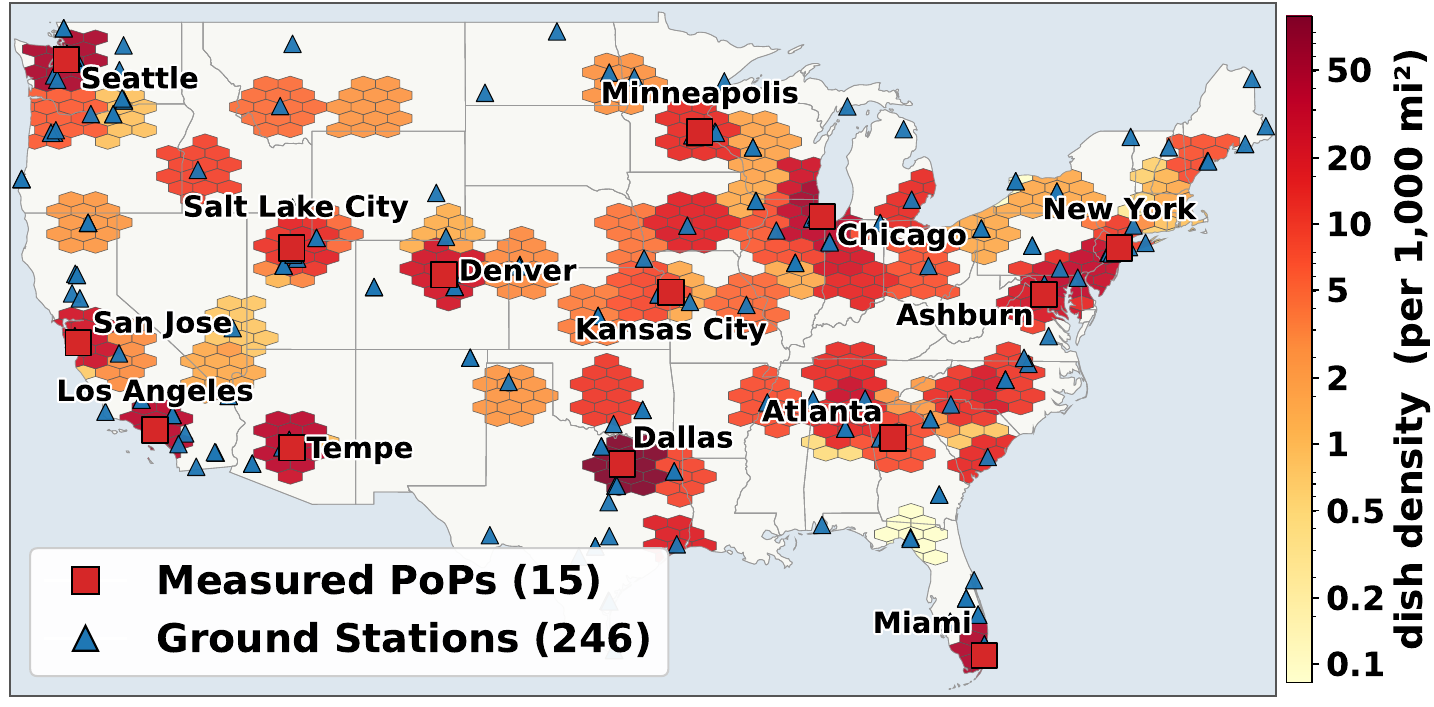}
  \caption{\textbf{Exposed-dish coverage.} \textnormal{Hexagons show the density of publicly reachable Starlink dishes as of June 2026, with each hexagon covering a $100$\,km radius. Red squares and blue triangles indicate our measured PoPs and known ground stations, respectively.}}
  \label{fig:coverage}
  \Description{Map of the United States shaded by the density of usable public terminals.}
\end{figure}

\section{Implementation}
\label{sec:impl}

We run two measurement campaigns. The over-the-air captures that validate the physics layer come from five locations in the eastern U.S. over three days. The link-state traces that train and test the learning layer come from nine sites across five states in the eastern and midwestern U.S. over sixteen days, totaling $8{,}260$ minutes of continuous $1$\,Hz capture under clear skies, cloud cover, and rain.

For the physics layer, the receive chain is in Fig.~\ref{fig:hardware-real}. A Pasternack PEWAN075-15 horn ($15$\,dBi)~\cite{pasternack_pewan075} feeds an Orbital~5400X low-noise block downconverter (LNB)~\cite{Orbital_5400X_LNB} into a USRP~X310~\cite{ettus_x310}, whose two Rx channels are stitched to cover the $11.575$\,GHz downlink at $240$\,Msps---one full Starlink user-downlink channel~\cite{Humphreys2023StarlinkStructure}. A splitter and bias tee carry RF to the radio while passing DC and a Leo Bodnar GPS reference~\cite{leobodnar_gpsclock} to the LNB, keeping its oscillator coherent. The horn rides a Yaesu G-5500DC rotator~\cite{Yaesu_G5500DC} driven by a self-contained tracker~\cite{CSN_SAT_Manual} and aimed from two-line element (TLE) ephemeris, capturing whichever beam falls into its main lobe.

For the link-state traces, a Starlink user terminal~\cite{starlink_user_terminal} in bypass mode hands our on-site vantage a public address, and the vantage records at $1$\,Hz: a bidirectional \texttt{iperf3} transfer over twenty parallel streams to a Lenovo SR650 V3 for throughput, a ping for round-trip time, and the terminal's gRPC telemetry for boresight, obstruction fraction, and PoP-ping statistics. Everywhere else the context comes from third-party dishes, probed from a single NVIDIA Jetson AGX Thor T5000 (\S\ref{sec:discovery}, \S\ref{sec:probes}).

Both layers are implemented in Python. The physics layer takes the inputs of \S\ref{sec:dataset}, propagates element sets with SGP4 to resolve co-visibility and the serving satellite, and ray-marches the four bent-pipe legs through the weather voxels, accumulating gaseous, cloud, rain, frozen-hydrometeor, and ionospheric terms into one attenuation per leg; with the geometry and calendar channels this is the covariate stack. The learning layer assembles windows on a $\Delta$-second grid, keeps only those whose target history is complete, fine-tunes the backbone for Modes~1 and~2, fits the Mode~3 estimator, and applies the conformal offset at serving time.

Modes~1 and~2 fine-tune all $119.5$\,M parameters of the Chronos-2 backbone; Mode~3 uses \texttt{scikit-learn} boosted trees~\cite{sklearn} (\S\ref{app:model} gives both configurations). Fine-tuning one configuration takes a median of four minutes and a three-channel forecast costs $27$\,ms per query, while Mode~3 trains in $11$\,s and serves in $0.5$\,ms on CPU. In lines of Python and shell, the implementation is $4{,}149$ lines for the physics layer, $4{,}815$ for its measurement, $4{,}814$ for probing, and $1{,}211$ for window assembly and training.

\section{Dataset}
\label{sec:dataset}

This section documents the \emph{public} data sources the system uses to construct the covariate stack. The measured time-series stack we collect is described in \S\ref{sec:impl} and \S\ref{sec:eval}.

\noindent $\bullet$ \textbf{Weather.}
\label{sec:data_weather}
Multi-Radar\slash Multi-Sensor (MRMS) CONUS mosaics ($1$\,km grid; $2$--$5$\,min refresh) supply precipitation intensity and type, rain-layer (echo-top) geometry, dual-polarization reflectivity, and melting-layer data~\cite{MRMS_NSSL,MRMS_BAMS,MRMS_Radar24}, which set the rain and snow path lengths, the rain rate $R$, and the mixed-phase handling (\S\ref{app:rain}). The NOAA HRRR Big Data Program (BDP) bucket ($3$\,km grid; $15$--$45$\,min refresh; $40$ pressure levels over $1013.2$--$50$\,hPa)~\cite{hrrr_grib2_prs, HRRR_BDP} supplies cloud base and top, from which the diagnosed layer thickness becomes a slant path for cloud and dry-snow attenuation (\S\ref{app:cloud}). NOAA SWPC operational feeds~\cite{SWPC_ProductsData,SWPC_DataService} give solar-wind plasma and interplanetary magnetic field, GOES X-ray and proton flux, the planetary $K_p$ index, and gridded electron content, which drive the dispersive-delay and scintillation terms (\S\ref{app:iono}).

\noindent $\bullet$  \textbf{Geometry.}
\label{sec:data_geometry}
The geometry inputs determine which bent-pipe link serves a query.

\noindent \textit{(1) Satellite ephemeris.} We take the two-line element set~\cite{spacetrack, celestrak} nearest a query time and propagate it with SGP4~\cite{Vallado2006SGP4}, giving the co-visibility set: visible-satellite count, serving candidates, and the serving satellite's \emph{hardware generation}.


\noindent \textit{(2) Ground stations and PoPs.} Ground-station locations, antenna counts, dish diameters, licensed frequencies, and maximum equivalent isotropically radiated power (EIRP) are scraped from FCC earth-station filings~\cite{FCC_ES}, giving 246 licensed sites in current use. For PoPs,
a PTR lookup on a customer address returns \texttt{customer.\allowbreak\{location\}.\allowbreak pop.\allowbreak starlinkisp.net}, which we cross-check against Starlink's published IP-geolocation feed, as in HitchHiking~\cite{hitchhiking}.

\noindent $\bullet$ \textbf{Miscellaneous.}
\label{sec:data_misc}

\noindent\textit{(1) Exposed dish lookup.} Candidate terminals live in Starlink's autonomous system, AS14593, whose prefixes are published in the global routing table; we scan them for exposed services with ZMap~\cite{zmap}, following established Internet-wide scanning practice~\cite{censys, masscan, lzr, gps}.

\noindent\textit{(2) Geolocation.} For finer localization (\S\ref{sec:discovery}) we use MaxMind GeoLite2-City~\cite{maxmind}, which resolves a dish to a town\slash city\slash region, and public M-Lab NDT measurements~\cite{MLab}, whose median speed-test-client location within the dish's \texttt{/24} prefix hints at city and region. Coverage after filtering and geolocation is in Fig.~\ref{fig:coverage}.

\noindent \textit{(3) Population.} Contention is invisible to Mode~3, so we estimate it from the WorldPop unconstrained $1$\,km global mosaic~\cite{worldpop}, summarized as the population within $12$\,km of the query coordinate.

\section{Evaluation}
\label{sec:eval}

In this section, we first validate the physics layer by comparing its predicted slant-path attenuation against attenuation measured directly from a Starlink downlink. We then evaluate the learning layer as a forecaster of user-perceived link state, and finally check the forecasts' effect downstream.

\subsection{Methodology}
\label{sec:eval_method}
\subsubsection{Physics Layer}
\label{sec:hardware}

We evaluate the physics layer on its own terms, as a link-quality predictor. Built from the ITU-R P-series models, it produces an attenuation for every leg of the bent pipe, yet only one leg is directly measurable: the satellite-to-user Ku downlink is the only channel we can capture over the air, while the satellite-to-ground-station legs hinge on the operator's private beam-sharing and alignment and cannot be observed end to end. We therefore validate the model against that one channel, comparing its predicted slant-path attenuation with the attenuation measured directly from the Starlink Ku downlink, and then use the validated model to provide all four legs' link quality to the forecaster.
We capture the downlink with the Ku-band receive chain of \S\ref{sec:impl} (Fig.~\ref{fig:hardware-real}).

\begin{figure}[t]
  \centering
  \includegraphics[width=\columnwidth]{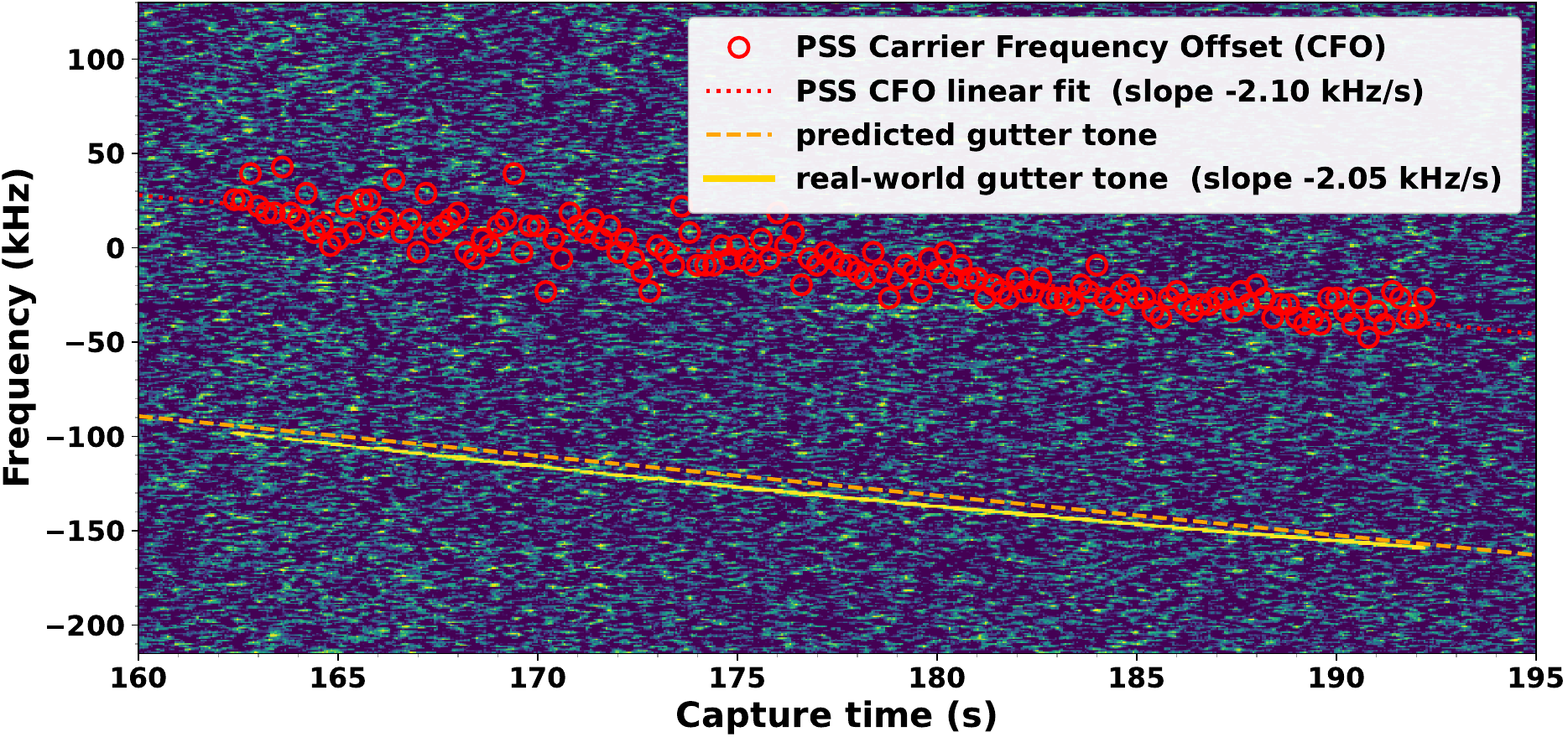}\vspace{-1mm}
  \caption{\textbf{PSS match.} \textnormal{Spectrogram of the captured Starlink Ku-band downlink signal, where the carrier-frequency offset (CFO) estimated from detected PSS symbols closely follows the drift of the real-world gutter tone.}}
  \label{fig:psscfo}
  \Description{Spectrogram of the Ku-band downlink, with the detected synchronization sequence and the matching leakage tone.}
\end{figure}

We read the excess loss from the downlink's Primary Synchronization Sequence (PSS). The frame layout that makes this possible---the channel, the frame period, and the PSS symbol---is not published by the operator; we take it from Humphreys et al.~\cite{Humphreys2023StarlinkStructure}, who recovered it from over-the-air captures. The part of that structure our measurement depends on is redrawn in Fig.~\ref{fig:layout} (\S\ref{app:pss}). On top of it we build Alg.~\ref{alg:pss}, which correlates the captured baseband against a PSS replica over a time--Doppler grid and stacks the per-frame scores across a short window; a matched trace is our unit of analysis. Fig.~\ref{fig:psscfo} shows one such match. The PSS appears at the channel's center frequency $F_{c}$, which sits $F/2$ above the channel midpoint, where $F=\hat F_s/\hat N\approx234$\,kHz is the subcarrier spacing (Table~\ref{tab:pss_params}); the midpoint itself lies in the mid-channel gutter, whose subcarriers are nominally vacant, so any tone we see there is local-oscillator leakage rather than signal~\cite{Humphreys2023StarlinkStructure}. On the plot, the solid yellow line is the real-world gutter tone captured at that time and the dashed orange line is its location estimated from the detected PSS (red circles); the Doppler rate and the frequency of the two agree closely.

To isolate the atmospheric attenuation from the received power, we account for every term of the link budget:
\begin{equation}
P_{\mathrm{rx}} = P_{\mathrm{tx}} + G_{\mathrm{tx}} - \mathrm{FSPL} + G_{\mathrm{rx}} - A_{\mathrm{IF}} - A_{\mathrm{atm}}
\end{equation}
From SpaceX's own technical filing~\cite{spacex2018mod} and the FCC authorization that adopts it~\cite{fcc2022gen2}, we take the satellite's EIRP, which covers both $P_{\mathrm{tx}}$ and $G_{\mathrm{tx}}$: a Ku-band user-downlink EIRP density of $-56.22$\,dBW/Hz at nadir, rising to $-51.60$\,dBW/Hz at maximum slant, for the $550$\,km shell our sites are served by~\cite{spacex2018mod}; over the $240$\,MHz channel these correspond to about $27.6$ and $32.2$\,dBW. The satellite adjusts transmit power with the beam's steering angle so that the power flux density at the ground stays constant~\cite{spacex2018mod}. The free-space path loss (FSPL) follows from the geometry, and the attenuation of the whole IF chain $A_{\mathrm{IF}}$ is measured with a vector network analyzer. The receive gain $G_{\mathrm{rx}}$ is known from the SDR and the rest of the chain, and the absolute scale of the measured $P_{\mathrm{rx}}$ is fixed by calibrating the receive chain against clear-sky captures at each site. The atmospheric attenuation is then obtained from Eq.~\eqref{eq:atm} and compared with the model's prediction:
\begin{equation}
\label{eq:atm}
A_{\mathrm{atm}} = \mathrm{EIRP} + G_{\mathrm{rx}} - \mathrm{FSPL} - P_{\mathrm{rx}} - A_{\mathrm{IF}}
\end{equation}

\begin{algorithm}[t]
\caption{PSS detection: search over time and Doppler. Symbols and parameter values are listed in Table~\ref{tab:pss_params} (\S\ref{app:pss}).}
\label{alg:pss}
\small
\begin{algorithmic}[1]
\Statex \textbf{Input:} received samples $y[\cdot]$, coarse Doppler grid $B$, local PSS replica $p[\cdot]$, window length $T_w$
\Statex \textbf{Output:} Doppler $\hat\beta^{\star}$, frame anchor $\hat\tau_0^{\star}$, lock SNR $\mathrm{snr}_M$
\For{each $\beta \in B$}
    \State $\Delta(\beta) \gets T_f(1-\beta)\hat F_s$, $N_f(\beta) \gets \lfloor T_w\hat F_s/\Delta(\beta)\rfloor$
    \For{$\tau_0 = 0,\dots,\lceil\Delta(\beta)\rceil-1$}
        \For{$k = 0,\dots,N_f(\beta)-1$}
            \State $c_k \gets \Call{PSSCorr}{\tau_0 + k\,\Delta(\beta),\ \beta}$
        \EndFor
        \State $M(\beta,\tau_0) \gets \tfrac{1}{N_f(\beta)}\sum_{k} c_k$
    \EndFor
\EndFor
\State $(\hat\beta^{\star},\hat\tau_0^{\star}) \gets \arg\max_{\beta,\tau_0} M(\beta,\tau_0)$
\State $\mathrm{snr}_M \gets M(\hat\beta^{\star},\hat\tau_0^{\star})\,/\,\mathrm{median}(M)$
\Statex
\Function{PSSCorr}{$n,\beta$}
    \State $L \gets \hat N + \hat N_g$
    \State $t_{\mathrm{in}} \gets [0,\dots,L-1]/\hat F_s$, $t_{\mathrm{out}} \gets (1-\beta)\,t_{\mathrm{in}}$
    \For{$i = 0,\dots,L-1$}
        \State $\tilde p(i) \gets \mathrm{InterpLinear}\bigl(p[\cdot],\, t_{\mathrm{in}},\, t_{\mathrm{out}}(i)\bigr)$
        \State $p_\beta(i) \gets \tilde p(i)\,e^{-j2\pi\beta F_c\, i/\hat F_s}$
    \EndFor
    \State $z \gets \sum_{i} y(n+i)\, p_\beta^{*}(i)$
    \State $R_y \gets \sum_{i} |y(n+i)|^2$, $R_p \gets \sum_{i} |p_\beta(i)|^2$
    \State \Return $|z|^2 / (R_y R_p + \varepsilon)$
\EndFunction
\end{algorithmic}
\end{algorithm}

\subsubsection{Learning Layer}

\begin{figure}[t]
  \centering
  \includegraphics[width=\columnwidth]{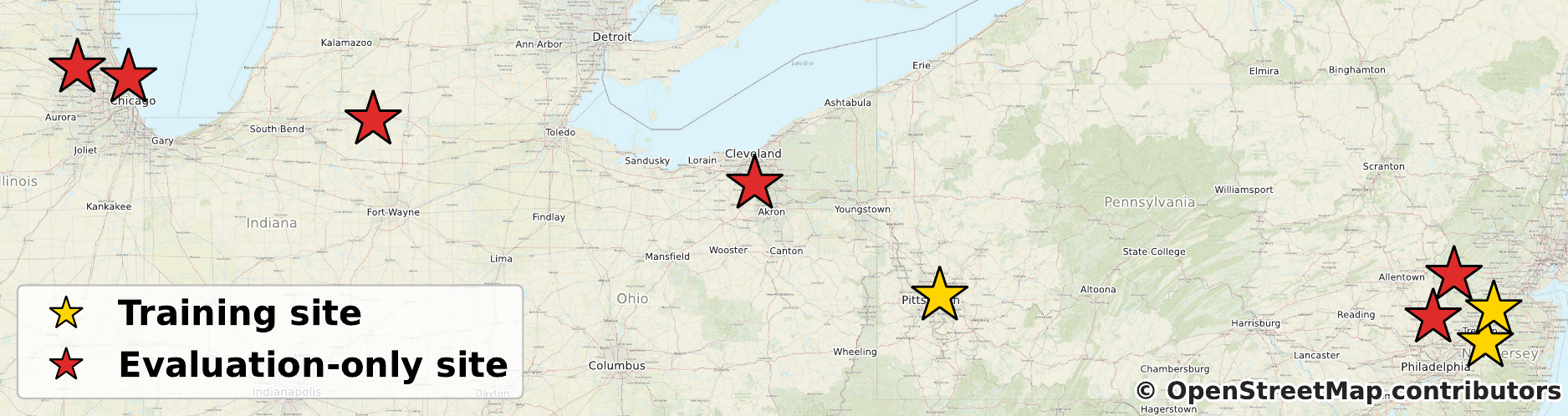}
  \caption{\textbf{Instrumented sites.} \textnormal{Yellow stars denote training sites; red stars denote test-only sites.}}
  \label{fig:sites}
  \Description{Map of the nine instrumented measurement sites.}
\end{figure}

We evaluate the learning layer on real user-perceived link state recorded by the deployed vantage of \S\ref{sec:impl} at nine instrumented locations. The campaign spans $8{,}260$ minutes of measured link state, and no two sites fall within one beam-cell diameter (\S\ref{sec:discovery}), so each site samples a distinct serving beam. To test generalizability, the evaluation covers only sites unseen during training and validation: we train on three sites and evaluate on six others that share no measurement session and no serving beam (Fig.~\ref{fig:sites}).

A window pairs $T_{\mathrm{ctx}}$ seconds of target history with an $H$-second forecast horizon, both sampled on a $\Delta$-second grid, and the model runs at many $(\Delta,T_{\mathrm{ctx}},H)$ settings (\S\ref{app:combos} reports the full sweep; \S\ref{app:model} gives the window details). Accuracy is the mean absolute error (MAE) over every predicted step at the evaluation sites. The downlink and uplink labels come from \texttt{iperf3} bursts and the RTT labels from \texttt{ping} (\S\ref{sec:impl}); all are bin-averaged where the ground-truth cadence ($1$\,Hz) is finer than the prediction cadence.

Within the three training sites, training and validation windows follow a 90/10 split made within each session by time. For baselines, we compare against published forecasters for LEO/Starlink link state: StarNet~\cite{starnet}, a recurrent sequence-to-sequence predictor over its own history and satellite/weather covariates; T3P~\cite{t3p}, a boosted-tree throughput model with a recurrent latency model; Horizon~\cite{horizon}, a tree ensemble that forecasts from speed-test results at unseen locations; and BG-CFQS~\cite{bgcfqs}, a budget-selected conservative-quantile predictor. Each is trained following its authors' described setup and evaluated zero-shot on our unseen traces---the same generalization question we ask of our own model. The three history-driven baselines anchor to our dish's trace scale by their own mechanisms: StarNet rescales its inputs and target with the min-max statistics of the training corpus, while T3P and BG-CFQS anchor on the recent history they consume. Horizon takes no history at all, so it does not anchor. All four predict downlink throughput, and all but BG-CFQS predict round-trip time; none forecasts the uplink.

\subsection{Attenuation Model Validation}
\label{sec:model_val}

\begin{figure*}[t]
  \centering
  \begin{minipage}[t]{0.30\textwidth}
    \centering
    \includegraphics[width=\linewidth]{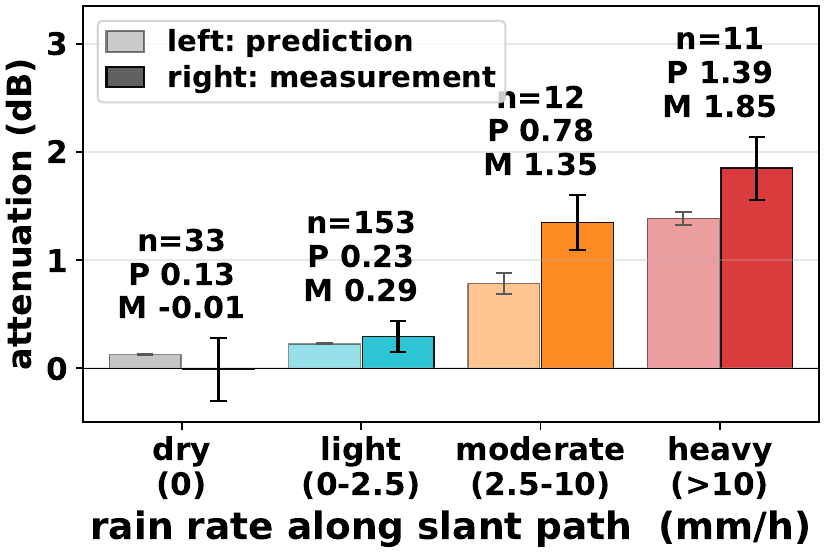}
    \captionof{figure}{\textbf{Ku-downlink validation.} \textnormal{Measured vs.\ predicted attenuation by rain-rate bin; $n$ counts the $15$\,s traces in each bin (P: predicted mean, M: measured mean).}}
    \label{fig:pssval}
    \Description{Predicted against measured Ku attenuation by rain-rate bin.}
  \end{minipage}\hfill
  \begin{minipage}[t]{0.68\textwidth}
    \centering
    \includegraphics[width=\linewidth]{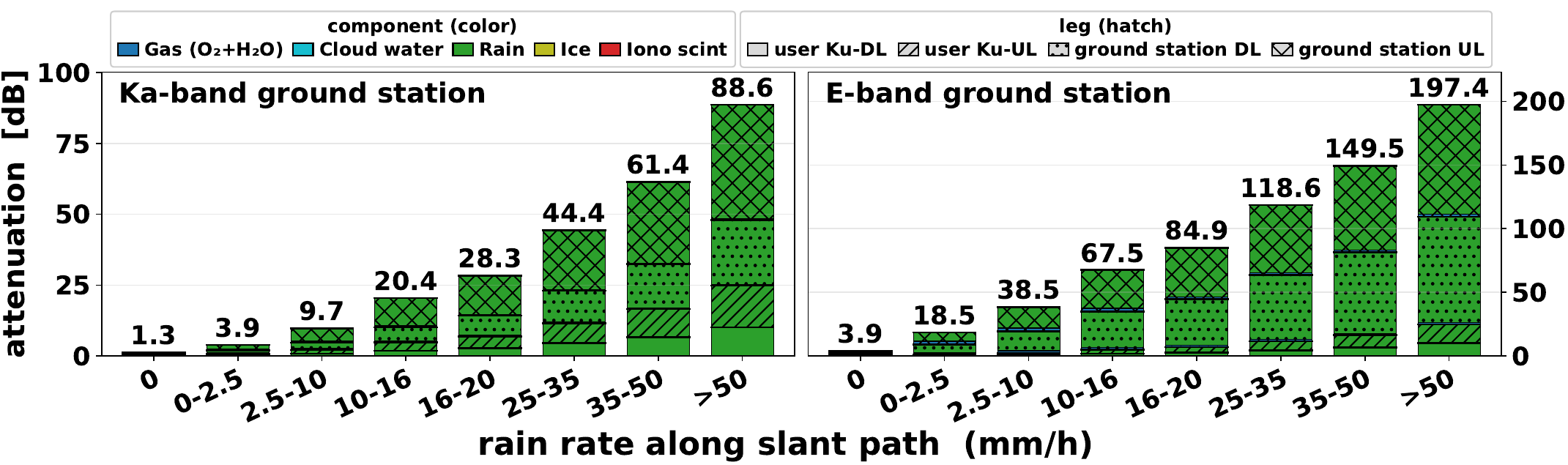}
    \captionof{figure}{\textbf{Four-leg attenuation breakdown.} \textnormal{Predicted four-leg bent-pipe attenuation across rain-rate bins for a Ka-band ground station (left) and an E-band ground station (right); color is the atmospheric component and hatch is the leg.}}
    \label{fig:legbreak}
    \Description{Predicted four-leg attenuation by rain-rate bin, split by atmospheric component and leg, for Ka-band and E-band ground stations.}
  \end{minipage}
\end{figure*}

Fig.~\ref{fig:pssval} shows the predicted attenuation tracking the radar rain rate closely, with the measured attenuation following the prediction in the mean and rising monotonically from dry conditions to heavy rain; the residual between them stays within ${\approx}0.6$\,dB. Since this is the only leg we can validate directly (\S\ref{sec:hardware}), and the model is built from the ITU-R P-series recommendations, which are parameterized by frequency rather than fitted at Ku, agreement on this leg is supporting, though indirect, evidence that the same construction carries to the higher-frequency legs. Fig.~\ref{fig:legbreak} applies it to all four: the higher-frequency ground-station links dominate the total, reaching tens of decibels at Ka band and up to about $200$\,dB at E band under very heavy rain.

\subsection{Forecasting Accuracy}
\label{sec:forecast_acc}

\begin{table}[t]
  \centering\footnotesize
  \setlength{\tabcolsep}{1pt}
  \begin{tabular*}{\columnwidth}{@{\extracolsep{\fill}}lllccc@{}}
    \toprule
    Method & Inputs & Outputs & DL & UL & RTT \\
    \midrule
    \textbf{Mode~1}                 & own trace      & time series & \textbf{12.68} & \textbf{7.68}  & \textbf{5.69} \\
    StarNet~\cite{starnet} & own trace      & time series & 15.2  & ---   & 6.4  \\
    BG-CFQS~\cite{bgcfqs}  & own trace      & time series & 16.0  & ---   & ---  \\
    T3P~\cite{t3p}         & own trace      & time series & 38.2  & ---   & 12.5 \\
    \midrule
   \textbf{Mode~2}                & neighbor trace & time series & \textbf{17.70} & \textbf{10.87} & \textbf{6.37} \\
    \midrule
    \textbf{Mode~3}                 & covariates     & level       & \textbf{20.83} & \textbf{10.29} & \textbf{6.48} \\
    Horizon~\cite{horizon} & covariates     & level       & 27.6  & ---   & 30.7 \\
    \bottomrule
  \end{tabular*}
  \caption{Per-second forecasting error (MAE) at $75$\,s context, $15$\,s horizon; DL, UL in Mbps, RTT in ms.}\vspace{-6mm}
  \label{tab:ladder}
\end{table}

We first compare against the baselines under a single fixed configuration---a $75$\,s context, a $15$\,s horizon, and one prediction per second---so that every method uses the same windows and can forecast at this configuration. The methods differ in what they take as input, so we read each mode against the baselines of its own input--output class.

\noindent$\bullet$ \textbf{Own history (Mode~1).} A terminal's own recent trace is Mode~1's input class, shared by StarNet, T3P, and BG-CFQS (Table~\ref{tab:ladder}, top block). Mode~1 has the lowest MAE in the block---and across every block---on all channels: $12.68$\,Mbps downlink against StarNet's $15.2$, BG-CFQS's $16.0$, and T3P's $38.2$, and $5.69$\,ms RTT against StarNet's $6.4$ and T3P's $12.5$.

\noindent$\bullet$ \textbf{Nearest public dish (Mode~2).} No existing forecaster runs without an on-site deployment while still taking a recent trace as input, so Mode~2 has no external counterpart. We therefore compare it against the same time-series-output forecasters. Its downlink MAE is slightly worse than the on-site-deployment baselines ($17.70$ vs.\ $15.2$ and $16.0$\,Mbps)---good enough for a method that deploys nothing on site---and far better than T3P ($38.2$\,Mbps). Its RTT MAE is on par with StarNet's ($6.37$ vs.\ $6.4$\,ms).

\noindent$\bullet$ \textbf{Covariates only (Mode~3).} Horizon is the one published forecaster in Mode~3's output class, so the two are compared on downlink and RTT. Mode~3 is better on both: $20.83$ vs.\ Horizon's $27.6$\,Mbps downlink, and $6.48$ vs.\ $30.7$\,ms RTT.

No baseline forecasts uplink, so we read its error against the channel's own scale rather than a competitor. On the unseen sites, measured uplink averages $35$\,Mbps (range $17$--$57$\,Mbps between the $5$th and $95$th percentiles). Mode~1's $7.68$\,Mbps error is about a fifth of a typical uplink, and Mode~2's $10.87$ and Mode~3's $10.29$ stay under a third---usable accuracy on a channel no prior LEO forecaster reports at all.

Moreover, our model serves many $(\Delta,T_{\mathrm{ctx}},H)$ configurations; the full sweep is in \S\ref{app:combos}, whose Table~\ref{tab:combos} shows the general trend across the three cadence categories: fine ($\Delta\le15$\,s), moderate ($30$--$300$\,s), and coarse ($\ge600$\,s). Coarsening the cadence bins the targets over longer windows, so errors generally fall. Mode~1's mean RTT error falls from $4.6$ to $1.3$\,ms and its downlink error from $13.9$ to $8.4$\,Mbps between the fine and coarse regimes, and Modes~2 and~3 follow the same pattern.

\begin{table}[t]
  \centering
  \footnotesize
  \setlength{\tabcolsep}{3pt}
  \begin{tabular*}{\columnwidth}{@{\extracolsep{\fill}}llccc@{}}
    \toprule
    Mode & Chan. & $80\%$ & $90\%$ & $P_{0.1}$ \\
    \midrule
    \multirow{3}{*}{Mode~1} & DL  & $0.77\!\to\!0.81$ & $0.87\!\to\!0.90$ & $0.88\!\to\!0.90$ \\
                            & UL  & $0.70\!\to\!0.82$ & $0.82\!\to\!0.91$ & $0.85\!\to\!0.91$ \\
                            & RTT & $0.80\!\to\!0.80$ & $0.91\!\to\!0.90$ & $0.91\!\to\!0.90$ \\
    \midrule
    \multirow{3}{*}{Mode~2} & DL  & $0.47\!\to\!0.80$ & $0.64\!\to\!0.90$ & $0.59\!\to\!0.90$ \\
                            & UL  & $0.64\!\to\!0.81$ & $0.79\!\to\!0.90$ & $0.76\!\to\!0.90$ \\
                            & RTT & $0.56\!\to\!0.80$ & $0.76\!\to\!0.90$ & $0.70\!\to\!0.90$ \\
    \midrule
    \multirow{3}{*}{Mode~3} & DL  & $0.46\!\to\!0.80$ & $0.60\!\to\!0.90$ & $0.65\!\to\!0.90$ \\
                            & UL  & $0.62\!\to\!0.80$ & $0.75\!\to\!0.90$ & $0.74\!\to\!0.90$ \\
                            & RTT & $0.64\!\to\!0.80$ & $0.74\!\to\!0.90$ & $0.88\!\to\!0.90$ \\
    \bottomrule
  \end{tabular*}
  \caption{Predictive-band coverage on the unseen sites, per mode and channel, as raw\,$\to$\,calibrated.}\vspace{-10mm}
  \label{tab:coverage}
\end{table}

\subsection{Predictive Band}

The results so far give the median prediction, but the forecaster also provides a predictive band (the $0.1$--$0.9$ quantiles), so a planner can read a conservative bound directly rather than a point guess. A nominal $80\%$ band should contain the truth about $80\%$ of the time. As noted in \S\ref{sec:intro}, no prior LEO forecaster reports a predictive band whose coverage is calibrated and measured, though probabilistic forecasting for network decisions is well established elsewhere~\cite{sprout, cs2p, kairos}. We therefore measure, on the test set, the fraction of observations each band actually contains.

Table~\ref{tab:coverage} shows each band as raw\,$\to$\,calibrated. The raw figure is the coverage of the model's own quantiles on the test set. The calibrated figure adds a light split-conformal step~\cite{conformal}: from test-set residuals we learn how far the truth tends to fall outside the raw band, then shift each edge outward by that fixed amount (or inward when the raw band is already too wide). The widening is fit on one split of the unseen sites while coverage is evaluated on a disjoint split it never saw, and the amount is a single number per mode and channel, computed once, so it adds nothing to inference. We report $80\%$ and $90\%$ bands and $P_{0.1}=P(y\ge\hat q_{0.1})$, the fraction of truths above the $80\%$ band's lower edge; the nominal values are $0.80$, $0.90$, and $0.90$, respectively.

Mode~1's raw bands are already close to nominal on sites it never saw (within $0.03$ on downlink and RTT, and $0.10$ low on the uplink $80\%$ band). Mode~2's and Mode~3's raw bands fall well short, covering only about two-thirds of the nominal fraction on downlink and between $70\%$ and nominal on uplink and RTT. The conformal step then closes most of that gap, bringing each mode back to roughly the correct coverage; these calibrated bands are what a downstream planner may consume. The one caveat is that the widening is fit on unseen-site labels, so a brand-new site with no local data to calibrate against would see the raw column, not the calibrated one.

\subsection{Transferability to Existing Forecasters}
\label{sec:transfer}

Our physics-informed link-quality predictor also transfers to other systems, and does more for them than raw weather data does. The state-of-the-art LEO forecasting work before us is StarNet~\cite{starnet}. Its feature set carries three raw near-ground weather channels---cloud cover, pressure, and humidity. We replace its raw weather features with our ray-traced four-leg link quality inside StarNet's own recurrent architecture. Here StarNet is retrained on our measurements and evaluated on unseen sites. Its downlink error falls from $18.4\pm2.5$\,Mbps with its own raw weather channels to $15.1\pm0.6$\,Mbps with ours, and becomes far steadier from run to run (Table~\ref{tab:transfer}). Because the lift appears inside an architecture that is not ours, the link-quality covariates can transfer to other models, not only ours.

\begin{table}[t]
  \centering\small
  \setlength{\tabcolsep}{3pt}
  \begin{tabular*}{\columnwidth}{@{\extracolsep{\fill}}lccc@{}}
    \toprule
    Model & Raw weather & Ours & $\Delta$ \\
    \midrule
    StarNet~\cite{starnet} & $18.4\pm2.5$ & $15.1\pm0.6$ & $-18\%$ \\
    \bottomrule
  \end{tabular*}
  \caption{Throughput MAE (Mbps) when StarNet's raw weather features are replaced with our ray-traced four-leg link quality and the model is retrained.}
  \label{tab:transfer}\vspace{-10mm}
\end{table}

\subsection{Downstream Task Applications}
\label{sec:downstream}
\begin{figure*}[t]
  \centering
  \includegraphics[width=0.58\textwidth]{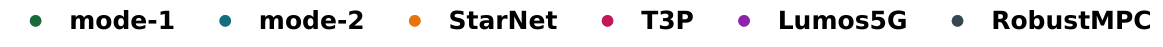}\\[1pt]
  \begin{subfigure}[t]{0.245\textwidth}\centering\includegraphics[width=\linewidth]{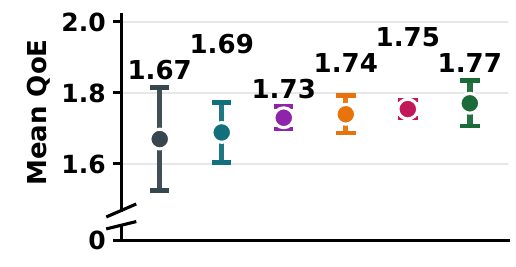}\caption{\textbf{Video-streaming QoE}}\label{fig:abr-vodqoe}\end{subfigure}\hfill
  \begin{subfigure}[t]{0.245\textwidth}\centering\includegraphics[width=\linewidth]{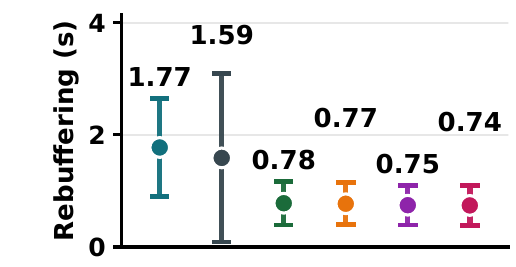}\caption{\textbf{Video rebuffering}}\label{fig:abr-vodrebuf}\end{subfigure}\hfill
  \begin{subfigure}[t]{0.245\textwidth}\centering\includegraphics[width=\linewidth]{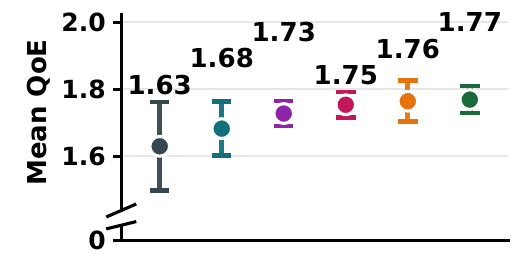}\caption{\textbf{Live-streaming QoE}}\label{fig:abr-liveqoe}\end{subfigure}\hfill
  \begin{subfigure}[t]{0.245\textwidth}\centering\includegraphics[width=\linewidth]{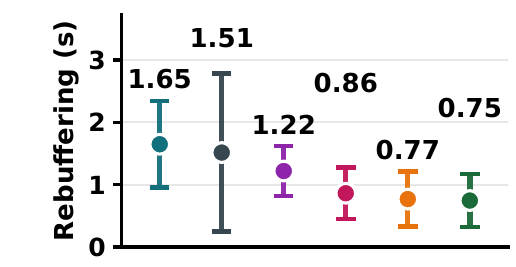}\caption{\textbf{Live rebuffering}}\label{fig:abr-liverebuf}\end{subfigure}
  \vspace{5pt}
  \caption{\textbf{Downstream ABR.} \textnormal{Experiments use real TCP with $30$ parallel flows; higher QoE (a,\,c) and lower rebuffering (b,\,d) are better.}}
  \label{fig:abr}
  \Description{Mean quality of experience and mean rebuffering for six predictors, under video on demand and live streaming.}
\end{figure*}

\noindent$\bullet$ \textbf{Setting.} We also investigate whether \sysname's throughput forecasts help a downstream application make better decisions. We run the standard DASH adaptive-bitrate (ABR) benchmark over emulated real transport. Each measured link trace is replayed packet by packet through mahimahi's \texttt{mm-link}, wrapped in \texttt{mm-delay} for an $80$\,ms round trip~\cite{mahimahi}, and a client running inside the emulated link downloads real video chunks over HTTP. The content is the $193$\,s EnvivioDash3 clip~\cite{enviviodash3}, in the encoding packaged by Pensieve~\cite{pensieve}: $48$ chunks of about four seconds each, every one available at six bitrates ($0.3$, $0.75$, $1.2$, $1.85$, $2.85$, and $4.3$\,Mbps). The bitrate controller is RobustMPC~\cite{mpc}, planning over the next $4$ chunks; each predictor supplies only the throughput estimate it consumes, so the controller is held fixed and the forecast is the variable. We also split each trace among $30$ parallel flows that fairly share the link, a moderate contention scenario reflecting multiple devices that typically share one Starlink Wi-Fi network. Quality of experience (QoE) is RobustMPC's linear objective~\cite{mpc}---delivered bitrate minus a rebuffering penalty minus a bitrate-switching penalty---reported per chunk. We stream every unseen test-set trace under each predictor and buffer size.

\noindent$\bullet$ \textbf{Results.} The baselines are that same RobustMPC controller driven by other throughput sources: T3P~\cite{t3p}, StarNet~\cite{starnet}, Lumos5G~\cite{lumos5g} (a cellular throughput predictor), and RobustMPC's own harmonic-mean-of-recent-chunks estimate. Fig.~\ref{fig:abr} reports both scenarios. For video on demand (a $60$\,s playout buffer), Mode~1 gives the highest QoE of any arm ($1.77$), just ahead of T3P ($1.75$), StarNet ($1.74$), Lumos5G ($1.73$), and RobustMPC ($1.67$); Mode~1 and the three baselines all rebuffer near $0.75$\,s, about half of RobustMPC's $1.59$\,s. Mode~2, built from neighbor public-dish traces, reaches a QoE of $1.69$---above RobustMPC's $1.67$ but a little below the other baselines---and gives ground mainly in rebuffering ($1.77$\,s). For live streaming (a tight $16$\,s buffer), Mode~1 is best on both metrics---$1.77$ QoE at the lowest stall of any arm ($0.75$\,s); the margin is modest, which we attribute mainly to the limited range of link conditions in our current traces, and we expect it to widen with more, and more varied, data. Mode~2 again reaches competitive quality ($1.68$, above RobustMPC's $1.63$) but trails in rebuffering ($1.65$\,s); the cause is the sampling rate of Mode~2's input: our measurement-ethics constraint caps probing of a stranger's dish at roughly one sample per $30$\,s (\S\ref{sec:probes}), whereas the own-terminal telemetry used by Mode~1 and the other baselines, like this per-second ABR loop, runs at $1$\,s, so the neighbor trace is too coarse for the second-to-second swings a small buffer lives on---a gap a denser probe could close.

\section{Related Work}
\label{sec:related}

\noindent\textbf{Measuring Starlink and LEO.} Beyond the profiling studies noted in \S\ref{sec:intro} and a recent packet-level dissection of Starlink's scheduling and queuing~\cite{cech2026dissecting}, a growing body of work has mapped how the link responds to conditions: degradation under weather~\cite{ullah2025starlink, lottermoser2026weather, ehsani2026storm}, under severe space weather~\cite{10.1145/3788084}, and under mobility~\cite{Laniewski2024StarlinkRoad}. Reaching terminals one does not own has driven two enabling threads: reverse-engineering the Ku-band waveform and its synchronization structure, for Starlink~\cite{Humphreys2023StarlinkStructure, kozhaya2025starlink} and for LEO downlinks more broadly~\cite{Komodromos2025OneWebPNT}, and probing publicly exposed dishes under a strict budget, as in the HitchHiking methodology~\cite{hitchhiking}, alongside shared testbeds that open LEO links to third-party experimentation~\cite{10.1145/3750832.3750835}. \sysname{} draws on both: it reads the waveform directly to validate the physics layer, and probes public dishes to assemble Mode~2's context. The difference is one of purpose---these studies establish what a link did, while \sysname{} predicts what it will do next.

\noindent\textbf{Network link-state prediction.} Predicting access-link capacity has a long history outside satellite access: CS2P~\cite{cs2p} predicts CDN video throughput from cross-session similarity, Oboe~\cite{oboe} adapts bitrate control to the prevailing network regime, Lumos5G~\cite{lumos5g} predicts mmWave-5G throughput from user-equipment context, and Sprout~\cite{sprout} forecasts cellular link rate with an explicit uncertainty cushion inside the transport loop---the closest prior instance of serving an interval rather than a point. The LEO forecasters of \S\ref{sec:intro} share one shape: each reads a terminal's own history (or, for Horizon, crowdsourced speed tests) and forecasts downlink and round-trip time only, which is why Table~\ref{tab:ladder} compares on those two channels and why uplink has no external baseline.

\noindent\textbf{Calibrated uncertainty.} The machinery for turning a forecast into an interval with a coverage guarantee is well established in statistics: quantile regression~\cite{koenker1978}, conformal prediction and in particular conformalized quantile regression~\cite{conformal}, and adaptive variants that relax the exchangeability assumption for time series~\cite{aci, enbpi}. Apart from BG-CFQS's one-sided risk-calibrated quantile~\cite{bgcfqs}, it has not been applied to a LEO link's state---the gap \S\ref{sec:eval} closes, with coverage measured in Table~\ref{tab:coverage} and decision value in \S\ref{sec:downstream}.

\noindent\textbf{Time-series foundation models.} A recent line of work carries the foundation-model recipe---pretrain one large model on a broad corpus, then reuse it across tasks with little task-specific training---from language and vision into time-series forecasting. Chronos-2~\cite{chronos2}, TimesFM~\cite{timesfm}, and Moirai~\cite{moirai} are pretrained on large, heterogeneous corpora of real series and then forecast unfamiliar ones, often rivaling or beating models trained for a single dataset on standard benchmarks~\cite{gifteval}; lightweight linear baselines had earlier shown that such an advantage is not automatic for deep forecasters~\cite{dlinear}. Their appeal here is data efficiency: a backbone that already encodes generic temporal structure can be specialized from a modest number of examples, which is what makes our limited labeled set workable at all (\S\ref{app:model}).

\section{Conclusion}
\label{sec:conclusion}

We present \sysname as a step toward predictive satellite networking: understanding and anticipating how the physical world translates into user-perceived network performance. By linking atmospheric conditions and orbital geometry with network measurements, \sysname establishes a general foundation for studying LEO links across space and time. As more measurements become available, this foundation can support increasingly accurate models of satellite connectivity and enable a wide range of applications, from network and deployment planning to adaptive communication, resource management, and proactive network control.

\begin{acks}
This work was supported in part by the National Science Foundation under award No.~2433914, No.~2433915 and No.~2554332.
\end{acks}

\bibliographystyle{ACM-Reference-Format}
\bibliography{reference}

\appendix

\section{Ethics}
\label{app:ethics}

This work raises no ethical concerns with human subjects, but it does measure third-party equipment over the public Internet, so we describe our safeguards in full.

Our measurements are of network infrastructure, not people. Every active probe is exchanged with a terminal's firmware at the network layer---no human sees, answers, or is interrupted by it---and we only use a public IP address and per-probe timing (round-trip time and inter-arrival gaps); we cannot observe user traffic, payloads, or any human-meaningful data. Because Starlink terminals sit behind carrier-grade NAT, a public address localizes a customer at most to a regional point of presence, not a household. On these grounds the study is not human-subjects research under the U.S. Common Rule---a non-human-subjects determination granted by our institution's review board---and it follows the reasoning of the HitchHiking measurement it builds on~\cite{hitchhiking}. We otherwise adhere to the ethical norms of the Internet-measurement community~\cite{zmap}, and IP addresses are hashed at rest, then dropped, and are never published or shared for any external use.

We keep our load on any third-party terminal far below a level that could affect its service. Probes are low-rate and non-saturating: each cycle sends a few tens of kilobytes---less than a single web-page load---and never floods the shared satellite radio link. This yields a closed-form per-dish budget of under $1$--$2$\,MB per hour. A terminal that stops responding is dropped, and a permanent, never-resurrected blocklist is applied as the first filter of every measurement round; it absorbs every complaint, takedown, and opt-out request.

We host a public opt-out page on the measurement vantage: a request appends the terminal's address to the blocklist and permanently excludes it, with no further interaction required. We have received one opt-out request so far, and that terminal is permanently excluded from every future measurement. Our active methods extend the latency-only HitchHiking envelope to active downlink and uplink estimation, a setting without prior art, but inspired by other network domains' techniques~\cite{Downey1999Pathchar}; we therefore rely on the conservative per-dish budget, the blocklist, and the opt-out to stay within the community's non-degradation norm. We did not coordinate with the operator in advance; a larger or longer deployment should.

\section{Forecasting Model: Architecture and Training}
\label{app:model}

This appendix gives the backbone, the domain correction, and the training objective behind the learned block of \S\ref{sec:forecast}. Fig.~\ref{fig:learning-layer} shows the whole path from one assembled window to a calibrated band. Labels and covariates are joined on a per-minute grid into per-window tensors that carry the target channels over the context and the covariate channels over both the context and the horizon, the latter taken from forecast products. Window construction is where the operating mode is applied: the same assembly code produces a Mode~1, Mode~2, or Mode~3 window by changing only what fills the target-context channels, which is what lets one trained architecture serve all three deployment settings. Windows whose target history is incomplete are dropped.

\subsection{Why a Pretrained Backbone}

We build on a pretrained time-series foundation model rather than train a forecaster from scratch, for two reasons. Beyond the label scarcity noted in \S\ref{sec:forecast}, pretraining on billions of observations supplies the generic temporal structure --- trend, seasonality, burst, mean reversion --- so fine-tuning spends our scarce labels only on what is specific to satellite access. The second reason is generality across deployment settings. Useful operating points range from per-second to per-hour decisions, so a fixed-shape architecture would have to be re-trained for each combination, which is precisely the scaling limit of the prior work we compare against~\cite{starnet}, whereas a pretrained model consumes variable context and horizon lengths natively. Of the current time-series foundation models --- MOIRAI, TimesFM, Chronos --- we use Chronos-2~\cite{chronos2} for representational fit: it is the only one we could deploy whose input format separates an observed past from a known future, which is exactly the shape of this problem.

\subsection{Reconciling Capacity with Throughput}
\label{ss:reconcil}

Inside the time-series stack, the throughput and latency history is the dish-measured throughput for Mode~1 and a neighbor trace for Mode~2. For Mode~2 the two do not live in the same domain, because a probe recovers capacity rather than throughput (\S\ref{sec:forecast}). To resolve this, for each channel $c$, a per-channel factor is estimated during training only,
\begin{equation}
\eta_c \;=\;
\frac{\operatorname{med}\{\,y_c \;:\; \text{label future}\,\}}
     {\operatorname{med}\{\,x_c \;:\; \text{model-input history}\,\}},
c\in\{\texttt{dl},\texttt{ul},\texttt{rtt}\}.
\label{eq:eta}
\end{equation}
Applying $\eta$ before instance normalization keeps the domain correction outside the network, so the fine-tune focuses on satellite link dynamics rather than on absorbing a fixed level shift. At inference, a single global $\eta$ per channel is used.

\subsection{Training Objective}

Training minimizes the quantile, or pinball, loss~\cite{koenker1978, gneiting2007}.
For a prediction $\hat{y}_{q,h}$ of level $q$ at horizon step $h$ against label $y_h$,
\begin{equation}
\mathcal{L}=\frac{1}{|\mathcal{P}|}\sum_{c}\;\sum_{h=1}^{H}\;\sum_{p\in\mathcal{P}}
m_{c,h}\,\rho_p\!\left(y_{c,h}-\hat{y}_{p,c,h}\right),
\label{eq:pinball}
\end{equation}
where $c$ ranges over the two throughput channels and the latency channel, $\mathcal{P}$ is the grid of nine quantile levels we train and serve (the backbone itself is pretrained on a finer $21$-level grid~\cite{chronos2}), and $m_{c,h}$ masks steps at which no label exists. The asymmetric check function $\rho_p$ is the quantile-regression loss~\cite{koenker1978}, and averaging it over a grid of levels is a proper scoring rule~\cite{gneiting2007}, consistent for each quantile it contains and approximating the continuous ranked probability score that the true conditional distribution minimizes uniquely. It is also the native loss of the backbone~\cite{chronos2}, which lets the fine-tune continue in the same structure the pretrained weights were learned in.

Downlink and uplink are sampled every $30$\,s while latency is sampled every second, so the throughput channels are far sparser in time. Without $m_{c,h}$ the denser channel would dominate the gradient by count. The loss is computed on instance-normalized targets. The log transform matters most for the two throughput channels, whose distributions are heavy-tailed; in log space a fixed absolute error is a fixed relative error, so the objective is not dominated by the high-rate tail. Uplink also concentrates near zero with intermittent bursts, and the same transform keeps that mass from being modeled as noise around a large mean. Round-trip time is smooth and needs neither treatment, but shares the same objective without harm.

\subsection{Training Configuration}
\label{ss:traincfg}

Modes~1 and~2 fine-tune the full backbone with AdamW at a learning rate of $2\times10^{-5}$, batch size $32$, and early stopping on a validation split, under a budget of $8000$ steps. Mode~3 uses $n_{\max}{=}200$ trees, a learning rate of $0.05$, and at most $15$ leaves per tree. Both settings come from an empirical hyper-parameter sweep.

\section{Primary-Synchronization-Sequence Measurement}
\label{app:pss}

\begin{figure}[t]
  \centering
  \includegraphics[width=\columnwidth]{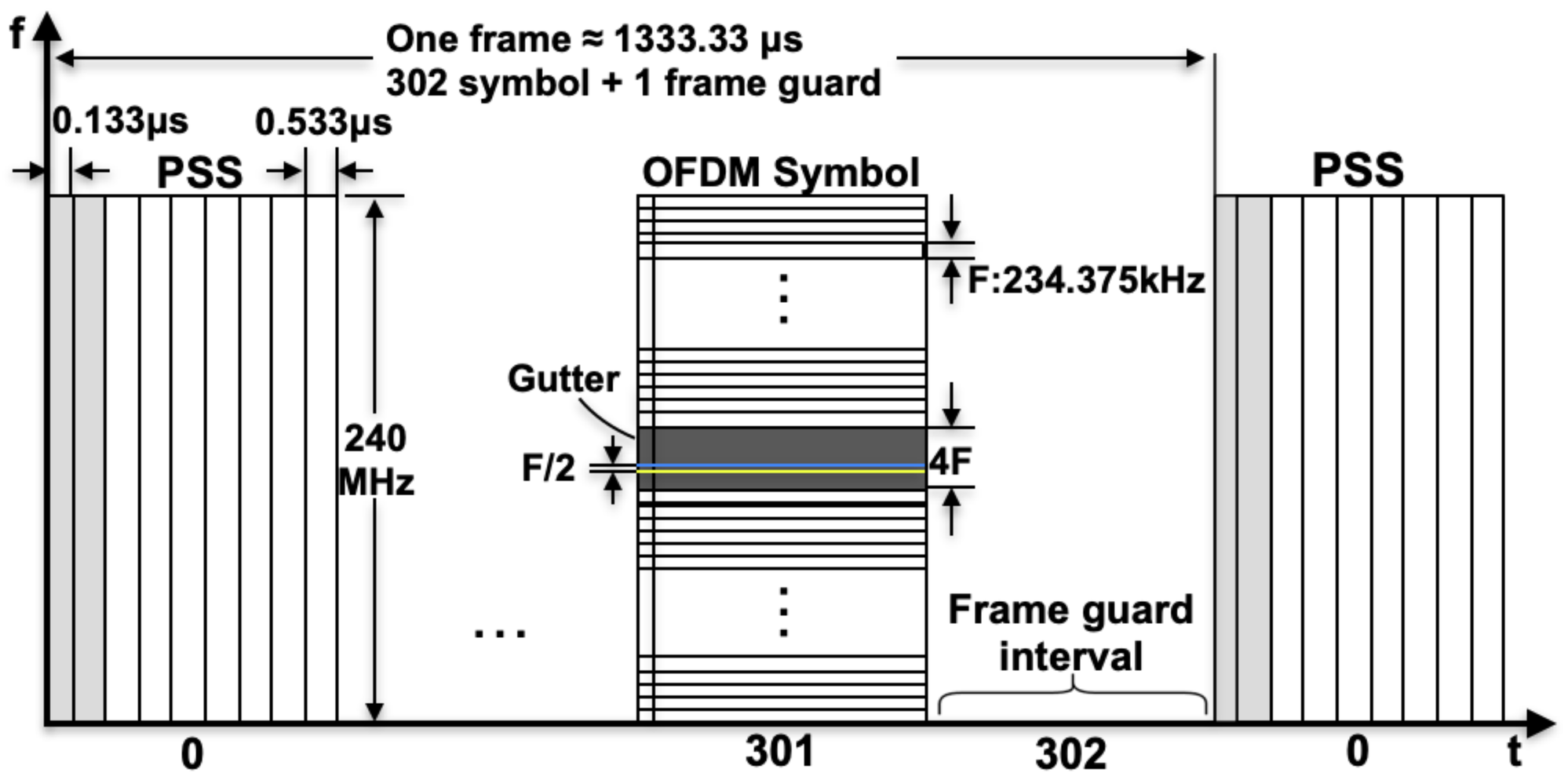}
  \caption{\textbf{Starlink frame layout.} \textnormal{The Ku-band downlink structure in time and frequency that our measurement relies on: the channel, the frame period, and the PSS symbol we correlate against. The operator does not publish this structure; we redraw it from findings of Humphreys et al.~\cite{Humphreys2023StarlinkStructure}.}}
  \label{fig:layout}
  \Description{Ku-band downlink frame layout in time and frequency, marking the PSS symbol.}
\end{figure}

\begin{table}[t]
\centering\small
\setlength{\tabcolsep}{3pt}%
\begin{tabular}{>{\raggedright\arraybackslash}p{1.35cm}
                >{\raggedright\arraybackslash}p{4.05cm}
                >{\raggedright\arraybackslash}p{2.35cm}}
\toprule
\textbf{Symbol} & \textbf{Meaning} & \textbf{Value} \\
\midrule
$y[\cdot]$ & received baseband after channel selection & --- \\
$p[\cdot]$ & local observed-band PSS replica & --- \\
$\beta$ & trial Doppler (time-scaling) & grid over $B$ \\
$B$ & searched $\beta$ set & $[-1.5,\,1.5]\times10^{-5}$, step $3\times10^{-6}$ \\
$\hat F_s$ & working sample rate & $240\,\mathrm{MHz}$ \\
$\hat N$ & OFDM FFT size & $1024$ \\
$\hat N_g$ & cyclic-prefix length & $32$ \\
$L$ & full PSS interval & $\hat N+\hat N_g=1056$ \\
$T_f$ & frame period & $1/750\,\mathrm{s}\approx1.333\,\mathrm{ms}$ \\
$F_c$ & RF channel center & $11.575\,\mathrm{GHz}$ \\
$z$ & matched-filter output & $\sum\tilde y(i)\,p^*(i)$ \\
$R_y,R_p$ & received / replica energy & --- \\
$c$ & normalized score for one $(n,\beta)$ & $|z|^2/(R_yR_p+\varepsilon)$ \\
$\varepsilon$ & numerical floor & $10^{-18}$ \\
$T_w$ & multi-PSS observation window & $\approx50\,\mathrm{ms}$ \\
$\Delta(\beta)$ & inter-frame stride at $\beta$ & $T_f(1-\beta)\hat F_s$ \\
$M(\beta,\tau_0)$ & frame-stacked normalized score & mean of $c$ over frames \\
\bottomrule
\end{tabular}
\caption{Symbols and parameters for full-PSS correlation.}
\label{tab:pss_params}
\end{table}

This appendix details the measurement used to validate the attenuation
model in \S\ref{sec:model_val}. Excess slant-path loss is read from the Starlink Ku downlink's
Primary Synchronization Sequence (PSS), a known length-$1056$ OFDM preamble repeated at the
$750$\,Hz frame rate~\cite{Humphreys2023StarlinkStructure}. The receiver correlates the
captured baseband against a local PSS replica over a joint time-Doppler grid: at each
candidate frame start $n$ and trial Doppler $\beta$ the replica is resampled and
carrier-compensated and a \emph{normalized} correlation score is formed, and the per-frame scores are stacked across a short observation window to reject noise (Alg.~\ref{alg:pss}). A window is accepted as a lock, and split into strong and weak grades, by its multi-frame periodicity support and the consistency of its Doppler. Symbols and parameter values in the algorithm are collected in Table~\ref{tab:pss_params}; the received PSS power of the link budget of
\S\ref{sec:model_val} is the matched-filter magnitude at the accepted $(\hat\beta,\hat\tau_0)$.

\section{Attenuation Model: Derivations and Data Fields}
\label{app:atten}

This appendix collects the per-mechanism attenuation models behind the four-leg attenuation of \S\ref{sec:fourleg}, together with the public-product fields each mechanism reads (\S\ref{sec:data_weather} states what those sources contribute). Frequency $f$ is in GHz, distance in km, pressure in hPa, and attenuation in dB.
\begin{center}
\begin{tabularx}{\linewidth}{@{}p{0.16\linewidth}X@{}}
\toprule
\textbf{Symbol} & \textbf{Definition} \\
\midrule
$f$ 
& Frequency (GHz); wavelength $\lambda=c/(f\cdot10^9)$ (m). \\

$\theta_{\rm el}$ 
& Elevation from the local horizon; azimuth $\psi$ (deg). \\

$p,e$ 
& Dry-air and water-vapor partial pressure (hPa). \\

$T$ 
& Temperature (K); ITU ratio $\tau\triangleq300/T$. \\

$q$ 
& Specific humidity (kg/kg); $q_c,q_i,q_r,q_s,q_g$ per species. \\

$h_s$ 
& Station altitude above mean sea level (km). \\

$\Delta s_j$ 
& Slant thickness of layer $j$ (km); vertical thickness $\Delta h_j$. \\

$D,x$ 
& Particle diameter (m); size parameter $x=\pi D/\lambda$. \\

$\phi_w$ 
& Liquid volume fraction of a mixed-phase particle. \\

$N(D)$ 
& Particle size distribution. \\
\bottomrule
\end{tabularx}
\end{center}


\noindent For $\theta_{\rm el}\gtrsim10^\circ$ a vertical layer thickness maps to slant thickness as $\Delta s_j\approx\Delta h_j/\sin\theta_{\rm el}$; rain uses the exact slant geometry of ITU-R~P.618~\cite{ITURP618}.

\subsection{Propagation and System Background}
\label{app:background}

Each mechanism in the slant-path sum behaves differently. Atmospheric gases, mainly oxygen and water vapor, absorb through molecular resonances and their continua; even at Ku band, away from the strongest lines, the line wings and continua still contribute a humidity- and temperature-dependent loss~\cite{ITURP676, Liebe1989MPM, Rosenkranz1998}. Hydrometeors, from cloud droplets and rain to snow, graupel, and hail, add scattering and extinction whose regime depends on particle size relative to wavelength, from a Rayleigh approximation for small droplets to Mie and non-spherical treatments for large or icy particles~\cite{Chen1975RAND_R1694, BeardChuang1987RaindropShape, Leinonen2014PyTMatrix}, so different weather produces qualitatively different attenuation at the same carrier. Refraction bends and lengthens the path, and the ionosphere adds dispersive delay and scintillation, but at Ku band both are second-order beside the tropospheric losses~\cite{Saakian2020RadioWavePropagationFundamentals, Zhao2025RadioWavePropagationSatelliteSystems}.

The links we study are commercial broadband LEO systems; we measure Starlink, and the same propagation and bent-pipe structure applies to other Ku/Ka constellations such as OneWeb. Their low altitude buys strong signal and low latency, but also fast apparent motion, large Doppler, and rapidly changing elevation and slant geometry~\cite{Zhao2025RadioWavePropagationSatelliteSystems}. Traffic reaches the Internet either through the bent pipe of \S\ref{sec:geometry} or through laser inter-satellite links~\cite{delPortillo2019LEOComparison, mohan2024}. We focus on the bent-pipe path, since it dominates where ground stations are dense: the predominant terminal-to-gateway distance is under $1200$\,km, roughly the ground footprint of a single satellite~\cite{mohan2024}, which is the case across the continental United States we measured. It is also what makes the link measurable: these Ku-band downlinks are wideband and highly structured, and prior work has recovered the Starlink frame layout and its synchronization sequences~\cite{Humphreys2023StarlinkStructure}, with similar recurring sequences reported in OneWeb~\cite{Komodromos2025OneWebPNT}. We take that structure as given and build on it in \S\ref{sec:hardware}.

\subsection{Gaseous Absorption}
\label{app:gas}

Gaseous attenuation is summed layer by layer along the slant path, following the ITU line-by-line structure~\cite{ITURP676}: oxygen lines, water-vapor lines, and the dry-air continuum, with the Van Vleck--Weisskopf line shape~\cite{VanVleck1945}, measured oxygen broadening and line-mixing coefficients~\cite{O2Tretyakov2005}, and continuum and collision-induced terms~\cite{Mlawer2012MTCKD, Richard2012CIA}. For layer $j$ with state $(p_j,T_j,q_j)$ we convert specific humidity to vapor pressure by $e_j=q_jP_j/(\varepsilon+(1-\varepsilon)q_j)$ with $\varepsilon=0.622$ and $P_j=p_j+e_j$, and form the layer specific attenuation
\begin{equation}
\gamma_j(f)=0.1820\,f\bigl(N_{\mathrm O_2,j}''(f)+N_{\mathrm{H_2O},j}''(f)\bigr),
\label{eq:gamma_layer_appB}
\end{equation}
where each imaginary refractivity sums line strengths $S_i$ against the Van Vleck--Weisskopf shape $F_i$, with the dry-air continuum $N_D''$ added for oxygen. Writing the shape in factored form with half-width $\Delta f_i$ and interference term $\delta_i$,
\begin{equation}
\begin{aligned}
\Phi_i(x)&\triangleq\frac{\Delta f_i-\delta_i x}{x^2+(\Delta f_i)^2},\\
F_i(f)&=\frac{f}{f_i}\bigl(\Phi_i(f_i-f)+\Phi_i(f_i+f)\bigr).
\end{aligned}
\label{eq:Phi_def_appB}
\end{equation}
with $\tau\triangleq300/T$ and $p,e$ in hPa, the line strength, half-width, and interference term are
\begin{equation}
S_i=\begin{cases} a_1\times10^{-7}\,p\,\tau^{3}\exp[a_2(1-\tau)], & \mathrm{O_2},\\[0.15em]
b_1\times10^{-1}\,e\,\tau^{3.5}\exp[b_2(1-\tau)], & \mathrm{H_2O},\end{cases}
\label{eq:Si_appB}
\end{equation}
\begin{equation}
\Delta f_{p,i}=\begin{cases} a_3\times10^{-4}\bigl(p\,\tau^{0.8-a_4}+1.1\,e\,\tau\bigr), & \mathrm{O_2},\\[0.15em]
b_3\times10^{-4}\bigl(p\,\tau^{b_4}+b_5\,e\,\tau^{b_6}\bigr), & \mathrm{H_2O},\end{cases}
\label{eq:Dfp_appB}
\end{equation}
with the Zeeman correction $\Delta f_i=\sqrt{(\Delta f_{p,i})^2+2.25\times10^{-6}}$ for oxygen, and for water the Doppler form $\Delta f_i=0.535\,\Delta f_{p,i}+\sqrt{Z_i}$ where $Z_i=0.217(\Delta f_{p,i})^2+2.1316\times10^{-12}f_i^2/\tau$. The interference term is $\delta_i=(a_5+a_6\tau)\times10^{-4}(p+e)\tau^{0.8}$ for oxygen and $\delta_i=0$ for water. The dry-air continuum uses the Debye width $d=5.6\times10^{-4}(p+e)\tau^{0.8}$,
\begin{equation}
N_D''(f)=f\,p\,\tau^{2}\!\left[\frac{6.14\times10^{-5}}{d\bigl(1+(f/d)^2\bigr)}+\frac{1.4\times10^{-12}\,p\,\tau^{1.5}}{1+1.9\times10^{-5}f^{1.5}}\right].
\label{eq:dry_cont_appB}
\end{equation}
The per-line center frequencies $f_i$ and coefficients $a_{1..6},b_{1..6}$ are the ITU-R~P.676 line list~\cite{ITURP676}, too long to reproduce. The slant-path total is
\begin{equation}
A_{\rm gas}(f,\theta_{\rm el})=\sum_{j=1}^N \gamma_j(f)\,\Delta s_j\approx\sum_{j=1}^N \gamma_j(f)\,\frac{\Delta h_j}{\sin\theta_{\rm el}}.
\label{eq:gas_slant_sum_appB}
\end{equation}

\subsection{Cloud and Fog Scattering}
\label{app:cloud}

For cloud and fog droplets ($D\lesssim0.01$\,cm) the Rayleigh approximation holds and specific attenuation is linear in liquid water content, $\gamma_c(f,T)=K_\ell(f,T)\,\rho_\ell$ with $\rho_\ell$ in g/m$^3$. Writing the complex permittivity of water as $\varepsilon(f)=\varepsilon'(f)-j\varepsilon''(f)$ and $\eta(f)=(2+\varepsilon'(f))/\varepsilon''(f)$,
\begin{equation}
K_\ell(f,T)=\frac{0.819\,f}{\varepsilon''(f)\bigl(1+\eta(f)^2\bigr)}.
\label{eq:Kl_appC}
\end{equation}
The permittivity follows the ITU double-Debye water model~\cite{ITURP840},
\label{app:cloud_debye}
\begin{equation}
\begin{aligned}
\varepsilon''(f)&=\frac{f(\varepsilon_0-\varepsilon_1)}{f_p\left[1+(f/f_p)^2\right]}+\frac{f(\varepsilon_1-\varepsilon_2)}{f_s\left[1+(f/f_s)^2\right]},\\
\varepsilon'(f)&=\frac{\varepsilon_0-\varepsilon_1}{1+(f/f_p)^2}+\frac{\varepsilon_1-\varepsilon_2}{1+(f/f_s)^2}+\varepsilon_2,
\end{aligned}
\label{eq:double_debye_appC}
\end{equation}
with $\varepsilon_0=77.66+103.3(300/T-1)$, $\varepsilon_1=0.0671\varepsilon_0$, $\varepsilon_2=3.52$, and relaxation frequencies $f_p=20.20-146(300/T-1)+316(300/T-1)^2$ and $f_s=39.8f_p$ in GHz.
Liquid water content follows from the cloud-liquid mixing ratio, $\rho_{\ell,j}=1000\,\rho_{{\rm air},j}\,q_{c,j}$, with moist-air density $\rho_{\rm air}=(P-e)/(R_dT)+e/(R_vT)$, and the path total is $A_{\rm cloud}=\sum_j K_\ell(f,T_j)\rho_{\ell,j}\Delta s_j$.

\noindent\textit{Product fields.} Cloud base, top, and ceiling come from Unified Post Processor diagnostics (\emph{cloudBase}, \emph{cloudTop}, \emph{cloudCeiling}) computed from the hydrometeor mixing ratios \emph{CLWMR}, \emph{ICMR}, \emph{RWMR}, \emph{SNMR}, and \emph{GRLE}, using the NCEP banding low $p>642$\,hPa, mid $642$--$350$\,hPa, high $p<350$\,hPa from 3D pressure-level data~\cite{RAP_Var_Diag,UPP_Guide,UPP_Git}. The diagnosed layer thickness converts to slant path $L=h_{\text{cloud}}/\sin\theta$ for cloud and dry-snow attenuation.

\subsection{Rain Scattering}
\label{app:rain}

Rain specific attenuation is the ITU-R~P.838 power law~\cite{ITURP838},
\begin{equation}
\gamma_R(f;R)=k(f,\theta_{\rm el},\tau_p)\,R^{\alpha(f,\theta_{\rm el},\tau_p)},
\label{eq:rain_gamma}
\end{equation}
with $R$ in mm/h and $\tau_p$ the polarization tilt ($45^\circ$ for circular). With $x=\log_{10}f$, the horizontal and vertical coefficients are log-frequency Gaussian fits
\begin{equation}
\begin{aligned}
\log_{10}k_{H,V}&=\sum_{j=1}^{4}a_j\exp\!\left(-\frac{(x-b_j)^2}{c_j}\right)+m_kx+c_k,\\
\alpha_{H,V}&=\sum_{j=1}^{5}a_j\exp\!\left(-\frac{(x-b_j)^2}{c_j}\right)+m_\alpha x+c_\alpha,
\end{aligned}
\label{eq:k_alpha_fit}
\end{equation}
whose coefficients are collected in Table~\ref{tab:p838}. They combine over polarization and elevation as
\begin{equation}
k=\frac{k_H+k_V+(k_H-k_V)\cos^2\theta_{\rm el}\cos2\tau_p}{2},
\label{eq:k_pol}
\end{equation}
\begin{equation}
\alpha=\frac{k_H\alpha_H+k_V\alpha_V+(k_H\alpha_H-k_V\alpha_V)\cos^2\theta_{\rm el}\cos2\tau_p}{2k}.
\label{eq:alpha_pol}
\end{equation}

\begin{table*}[t]
\centering\footnotesize
\setlength{\tabcolsep}{4pt}
\begin{tabular*}{\textwidth}{@{\extracolsep{\fill}}c rrr rrr rrr rrr@{}}
\toprule
& \multicolumn{3}{c}{$k_H$} & \multicolumn{3}{c}{$k_V$} & \multicolumn{3}{c}{$\alpha_H$} & \multicolumn{3}{c}{$\alpha_V$}\\
\cmidrule(lr){2-4}\cmidrule(lr){5-7}\cmidrule(lr){8-10}\cmidrule(lr){11-13}
$j$ & $a_j$ & $b_j$ & $c_j$ & $a_j$ & $b_j$ & $c_j$ & $a_j$ & $b_j$ & $c_j$ & $a_j$ & $b_j$ & $c_j$\\
\midrule
1 & $-5.33980$ & $-0.10008$ & $1.13098$ & $-3.80595$ & $0.56934$ & $0.81061$ & $-0.14318$ & $1.82442$ & $-0.55187$ & $-0.07771$ & $2.33840$ & $-0.76284$\\
2 & $-0.35351$ & $1.26970$ & $0.45400$ & $-3.44965$ & $-0.22911$ & $0.51059$ & $0.29591$ & $0.77564$ & $0.19822$ & $0.56727$ & $0.95545$ & $0.54039$\\
3 & $-0.23789$ & $0.86036$ & $0.15354$ & $-0.39902$ & $0.73042$ & $0.11899$ & $0.32177$ & $0.63773$ & $0.13164$ & $-0.20238$ & $1.14520$ & $0.26809$\\
4 & $-0.94158$ & $0.64552$ & $0.16817$ & $0.50167$ & $1.07319$ & $0.27195$ & $-5.37610$ & $-0.96230$ & $1.47828$ & $-48.2991$ & $0.791669$ & $0.116226$\\
5 & --- & --- & --- & --- & --- & --- & $16.1721$ & $-3.29980$ & $3.43990$ & $48.5833$ & $0.791459$ & $0.116479$\\
\midrule
$m$ & \multicolumn{3}{c}{$-0.18961$} & \multicolumn{3}{c}{$-0.16398$} & \multicolumn{3}{c}{$0.67849$} & \multicolumn{3}{c}{$-0.053739$}\\
$c$ & \multicolumn{3}{c}{$0.71147$} & \multicolumn{3}{c}{$0.63297$} & \multicolumn{3}{c}{$-1.95537$} & \multicolumn{3}{c}{$0.83433$}\\
\bottomrule
\end{tabular*}
\caption{ITU-R~P.838 coefficients for rain power law $\gamma_R=kR^{\alpha}$~\cite{ITURP838}, used in \eqref{eq:k_alpha_fit}. The $k$ fits use four Gaussian terms, the $\alpha$ fits five.}
\label{tab:p838}
\end{table*}

We treat the rain column as vertically uniform up to the MRMS echo top $h_{\rm top}$, so the instantaneous loss is a ray integral through the 2.5D field,
\begin{equation}
A_{\rm rain}(t)=\int \gamma_R\!\bigl(f;R(\mathbf{r}(s),t)\bigr)\,\mathbf{1}\!\left[h(s)\le h_{\rm top}(\mathbf{r}(s),t)\right]ds.
\label{eq:rain_ray_integral}
\end{equation}
Marching in ground range $g$ on a local tangent plane gives $h(g)=h_s+g\tan\theta_{\rm el}$ and $ds=dg/\cos\theta_{\rm el}$. With a step $\Delta g$ matched to the $1$\,km MRMS grid, sampling $R_m=R(\mathbf{r}_m,t)$ and $h_{{\rm top},m}=h_{\rm top}(\mathbf{r}_m,t)$ at $g_m=m\Delta g$,
\begin{equation}
\begin{aligned}
A_{\rm rain}(t)&\approx\sum_{m=0}^{M-1}\gamma_R(f;R_m)\,w_m\,\frac{\Delta g}{\cos\theta_{\rm el}},\\
w_m&=\mathbf{1}\!\left[h(g_m)\le h_{{\rm top},m}\right].
\end{aligned}
\label{eq:rain_sum_basic}
\end{equation}
A hard cutoff at the echo top biases the last cell, so when the ray crosses it inside a step ($h_m\le h_{{\rm top},m}<h_{m+1}$) we include the partial fraction
\begin{equation}
w_m^\star=\frac{h_{{\rm top},m}-h_m}{h_{m+1}-h_m}\in[0,1],
\label{eq:partial_step}
\end{equation}
and stop marching once $R_m$ is negligible for several consecutive cells and the ray clears the local echo top by a margin.

\noindent\textit{Product fields.} Precipitation intensity and type from \emph{PrecipRate} and \emph{PrecipFlag}~\cite{MRMS_NSSL,MRMS_BAMS}; vertical echo geometry from \emph{EchoTop-18} and \emph{EchoTop-30}, with quality-controlled and dual-polarization slices \emph{ReflectivityQC}, \emph{Z\textsubscript{dr}}, \emph{RhoHV}~\cite{MRMS_EchoTop_Prod,NWS_DualPol_Products}; melting-layer diagnostics \emph{BrightBandTop} and \emph{BrightBandBottom} with the \emph{WarmRainProbability} diagnostic~\cite{MRMS_BrightBand_Prod,MRMS_WarmRain_Prod}. These drive rain and snow path lengths, the rain rate $R$ for $kR^\alpha$, and mixed-phase handling.

\subsection{Snow, Ice, Graupel, and Hail}
\label{app:snow}

Frozen hydrometeors are modeled through extinction of the coherent beam, $A_{\rm snow/ice}(f,t)\approx\sum_n \gamma_{{\rm snow/ice},n}(f,t)\,\Delta s_n$, with frozen water content $\mathrm{FWC}_n=1000\,\rho_{{\rm air},n}q_{f,n}$ from the frozen-species mixing ratio.

For partially melted or water-coated particles we first mix dielectrics. Given ice and liquid permittivities $\varepsilon_i,\varepsilon_w$ and liquid volume fraction $\phi_{w,n}$, the Bruggeman rule~\cite{Bruggeman1935, Tjaden2016Bruggeman} defines $\varepsilon_{{\rm eff},n}$ through $\phi_{w,n}(\varepsilon_w-\varepsilon_{\rm eff})/(\varepsilon_w+2\varepsilon_{\rm eff})+(1-\phi_{w,n})(\varepsilon_i-\varepsilon_{\rm eff})/(\varepsilon_i+2\varepsilon_{\rm eff})=0$, and the effective refractive index is
\begin{equation}
m_{{\rm eff},n}=\sqrt{\varepsilon_{{\rm eff},n}}.
\label{eq:meff_from_eps}
\end{equation}
Liquid permittivity reuses the double-Debye model of \S\ref{app:cloud}; ice uses a temperature-dependent microwave model~\cite{Matzler2006Ice} with $\varepsilon'_{\rm ice}(T_C)\approx3.1884+9.1\times10^{-4}T_C$.

When the size parameter $x(D,f)=\pi D/\lambda\ll1$ the local attenuation is linear in mass concentration. For pure ice, $\gamma_{{\rm ice},n}\approx K_{\rm ice}(f,T_n)\,\mathrm{IWC}_n$ with $\mathrm{IWC}_n=1000\,\rho_{{\rm air},n}q_{i,n}$; for mixed particles that still satisfy the small-particle condition, $\gamma_{{\rm mix},n}\approx K_{\rm mix}(f,T_n)\,C_n$ on the mixed-particle mass concentration $C_n$, where
\begin{equation}
\begin{aligned}
K_{\rm ice}&=0.819\,f\,\Im\!\left\{\frac{\varepsilon_{\rm ice}-1}{\varepsilon_{\rm ice}+2}\right\},\\
K_{\rm mix}&=0.819\,f\,\Im\!\left\{\frac{\varepsilon_{{\rm eff},n}-1}{\varepsilon_{{\rm eff},n}+2}\right\}.
\end{aligned}
\label{eq:Kice_Kmix}
\end{equation}
The two differ only in which permittivity enters, so a particle's melting state propagates directly into its extinction. When $x\gtrsim1$, as for graupel, hail, and large aggregates, the Rayleigh form is no longer accurate and we compute extinction with a Mie solver for spheres or a T-matrix solver for non-spherical particles~\cite{Leinonen2014PyTMatrix}, using
\begin{equation}
\sigma_{{\rm ext},n}(D;f,T_n)=\frac{\pi D^2}{4}\,Q_{\rm ext}\!\bigl(m_n(f,T_n),x(D,f)\bigr),
\label{eq:sigma_ext}
\end{equation}
with $m_n=\sqrt{\varepsilon_{\rm ice}(f,T_n)}$ for dry ice and $m_n=m_{{\rm eff},n}$ for mixed particles, where the extinction efficiency is the Mie series $Q_{\rm ext}(m,x)=\tfrac{2}{x^2}\sum_{k\ge1}(2k+1)\Re(a_k+b_k)$ in the Riccati--Bessel coefficients $a_k,b_k$, evaluated numerically. This gives
\begin{equation}
\gamma_{{\rm snow/ice},n}(f)=4.343\times10^{3}\int_0^\infty \sigma_{{\rm ext},n}(D;f,T_n)\,N_n(D)\,dD.
\label{eq:gamma_mie_local}
\end{equation}
The Mie branch needs a size distribution. With only bulk mass available we take an exponential PSD $N_n(D)=N_{0,n}e^{-\Lambda_nD}$ with $\Lambda_n=1/D_{c,n}$, and match it to the bulk mass concentration $M_n$ with material density $\rho_p$ to give $N_{0,n}=M_n\Lambda_n^4/(\pi\rho_p)$.

\subsection{Ionosphere: Dispersive Delay and Scintillation}
\label{app:iono}

At Ku band ionospheric absorption is small, but the ionosphere still affects the link through dispersive delay and scintillation fading. With slant total electron content $\mathrm{TEC}=\int N_e(s)\,ds$, the phase and group refractive indices are $n_{ph}\approx1-40.3N_e/f^2$ and $n_g\approx1+40.3N_e/f^2$, so the ionosphere produces a group delay $\Delta\tau=40.3\,\mathrm{TEC}/(cf^2)$ together with a carrier phase advance of magnitude $2\pi\cdot40.3\,\mathrm{TEC}/(cf)$ --- both proportional to TEC and inversely proportional to $f^2$.

Scintillation enters as a fading term. From the amplitude scintillation index $S_4=\sqrt{(\langle I^2\rangle-\langle I\rangle^2)/\langle I\rangle^2}$ on the received intensity $I(t)=|r(t)|^2$, modeling intensity as lognormal for weak-to-moderate scintillation gives
\begin{equation}
\sigma_{\rm iono,dB}=\frac{10}{\ln 10}\sqrt{\ln(1+S_4^2)}\approx4.343\sqrt{\ln(1+S_4^2)},
\label{eq:sigma_iono}
\end{equation}
which reduces to $\sigma_{\rm iono,dB}\approx4.343\,S_4$ for small $S_4$. Where only a reference level is available we scale it as $S_4\approx S_{4,\mathrm{ref}}(f/f_{\mathrm{ref}})^{-v}\csc\theta_{\rm el}$ with a regime-dependent exponent $v$.

\subsection{Hydrometeor Type, Size, and Shape Inference}
\label{app:hydro_retrieval}

The scattering models above need a hydrometeor class, a characteristic size, a wet fraction, and a dielectric. We retrieve $\chi_n=(c_n,D_{c,n},\phi_{w,n},m_{{\rm eff},n},N_n(D))$ at each ray sample from the available operational fields, in two modes.

\noindent\textbf{Direct mode.} When dual-polarization variables are available we screen for uniform liquid rain with $\rho_{hv,n}\ge\rho_{\rm th}$, taking $\rho_{\rm th}$ in $0.97$--$0.99$, which suppresses bins dominated by mixed phase, hail, or highly variable populations~\cite{NWS_DualPol_Products}. Differential reflectivity $Z_{DR}$ is then the primary shape and mean-size cue: positive values indicate horizontally broader particles such as oblate rain drops~\cite{BeardChuang1987RaindropShape}, values near zero near-spherical targets, and negative values ice-dominated orientations~\cite{SeligaBringi1976ZDR}. The hydrometeor-class product supplies a first class assignment where present, and characteristic size follows a calibrated polarimetric retrieval in which $Z_{DR}$ carries mean size while $Z_H$ and $K_{DP}$ refine concentration~\cite{Gorgucci2002PolarimetricDSD}.

\noindent\textbf{Fallback mode.} Without dual-polarization inputs we infer the phase regime at sample altitude $z_n$ from bright-band structure and temperature,
\begin{align}
z_n<h_{{\rm BB,bot},n},\ T_n>0^\circ\mathrm{C} &\Rightarrow \text{liquid},\nonumber\\
h_{{\rm BB,bot},n}\le z_n\le h_{{\rm BB,top},n} &\Rightarrow \text{melting / mixed},\label{eq:regimes_appG}\\
z_n>h_{{\rm BB,top},n},\ T_n\le0^\circ\mathrm{C} &\Rightarrow \text{frozen},\nonumber
\end{align}
and within the bright band take the wet fraction as a clipped linear profile between its bounds,
\begin{equation}
\phi_{w,n}=\mathrm{clip}\!\left(\frac{h_{{\rm BB,top},n}-z_n}{h_{{\rm BB,top},n}-h_{{\rm BB,bot},n}},\,0,\,1\right),
\label{eq:phiw_appG}
\end{equation}
so a particle melts continuously from dry at the top of the band to fully wet at its base. Class is taken as the dominant bulk mass among rain, snow, and graupel, $c_n=\arg\max_c \rho_{{\rm air},n}q_{c,n}$, and characteristic size from class-dependent intensity proxies: increasing with rain rate for liquid, with bulk snow mass for aggregates, and with graupel mass, VIL, and echo-top height for graupel and hail~\cite{MRMS_BAMS,MRMS_VIL_Prod,MRMS_EchoTop_Prod,MRMS_BrightBand_Prod}. The retrieved $(c_n,D_{c,n},\phi_{w,n})$ then instantiate the exponential PSD and, for wet or melting particles, the Bruggeman effective permittivity~\cite{MeneghiniLiao2000MixedPhaseDielectric}, and $\chi_n$ passes to the matching branch: Rayleigh for small ice, Mie for graupel and hail, and the liquid-rain model for rain bins.

\clearpage
\onecolumn
\section{All-Combination Tables for the Model}
\label{app:combos}

Every row is one $(\Delta, T_{\mathrm{ctx}}, H)$ operating point---cadence, context, and horizon, all in seconds---and the values are mean absolute error aggregated over the test set, per mode ($1$/$2$/$3$) and channel (downlink and uplink in Mbps, RTT in ms). Each mode is scored only on the windows its own input makes eligible---Mode~1 needs a fully measured own-history context, Mode~2 a neighbor trace, and Mode~3 neither but still enough windows to fit its tree---so a dash ($-$) marks a mode with no scorable window at that point.

{\footnotesize
\renewcommand{\arraystretch}{1.20}
\setlength{\tabcolsep}{1pt}
\begin{longtable}{@{\extracolsep{\fill}}rrrrrrrrrrrr@{\hspace{0.5em}}rrrrrrrrrrrr}
\toprule
 &  &  & \multicolumn{3}{c}{DL (Mbps)} & \multicolumn{3}{c}{UL (Mbps)} & \multicolumn{3}{c}{RTT (ms)} &  &  &  & \multicolumn{3}{c}{DL (Mbps)} & \multicolumn{3}{c}{UL (Mbps)} & \multicolumn{3}{c}{RTT (ms)}\\
\cmidrule(lr){4-6}\cmidrule(lr){7-9}\cmidrule(lr){10-12}\cmidrule(lr){16-18}\cmidrule(lr){19-21}\cmidrule(lr){22-24}
$\Delta$ & $T_{\mathrm{ctx}}$ & $H$ & M1 & M2 & M3 & M1 & M2 & M3 & M1 & M2 & M3 & $\Delta$ & $T_{\mathrm{ctx}}$ & $H$ & M1 & M2 & M3 & M1 & M2 & M3 & M1 & M2 & M3\\
\midrule
\endfirsthead
\toprule
 &  &  & \multicolumn{3}{c}{DL (Mbps)} & \multicolumn{3}{c}{UL (Mbps)} & \multicolumn{3}{c}{RTT (ms)} &  &  &  & \multicolumn{3}{c}{DL (Mbps)} & \multicolumn{3}{c}{UL (Mbps)} & \multicolumn{3}{c}{RTT (ms)}\\
\cmidrule(lr){4-6}\cmidrule(lr){7-9}\cmidrule(lr){10-12}\cmidrule(lr){16-18}\cmidrule(lr){19-21}\cmidrule(lr){22-24}
$\Delta$ & $T_{\mathrm{ctx}}$ & $H$ & M1 & M2 & M3 & M1 & M2 & M3 & M1 & M2 & M3 & $\Delta$ & $T_{\mathrm{ctx}}$ & $H$ & M1 & M2 & M3 & M1 & M2 & M3 & M1 & M2 & M3\\
\midrule
\endhead
\midrule
\endfoot
\bottomrule
\caption{Forecasting error (MAE) at every measured operating point, over the six unseen sites. Each row is one $(\Delta,T_{\mathrm{ctx}},H)$ point: cadence, context, and horizon, all in seconds. A dash marks a mode with no scorable window there. The points are shown in two side-by-side blocks: read the left block top to bottom, then the right.}
\label{tab:combos}\\
\endlastfoot
  1 & 30 & 5 & 9.96 & 20.18 & 21.23 & 5.89 & 16.84 & 10.21 & 5.35 & 7.34 & 6.24 & 15 & 120 & 60 & 14.08 & $-$ & 16.33 & 8.29 & $-$ & 8.92 & 4.07 & $-$ & 4.02 \\
  1 & 30 & 10 & 11.85 & 19.57 & 20.00 & 6.99 & 14.88 & 10.04 & 5.62 & 7.37 & 6.40 & 15 & 150 & 15 & 12.20 & $-$ & 14.33 & 7.33 & $-$ & 8.56 & 3.76 & $-$ & 3.92 \\
  1 & 30 & 15 & 13.11 & 18.99 & 19.79 & 7.76 & 14.77 & 9.95 & 5.84 & 7.44 & 6.37 & 15 & 150 & 30 & 13.23 & $-$ & 15.74 & 7.95 & $-$ & 8.27 & 3.90 & $-$ & 3.92 \\
  1 & 30 & 20 & 14.35 & 19.20 & 20.71 & 8.51 & 14.50 & 10.07 & 5.96 & 7.24 & 6.44 & 15 & 150 & 45 & 13.75 & $-$ & 16.68 & 8.10 & $-$ & 9.17 & 3.97 & $-$ & 4.17 \\
  1 & 30 & 30 & 15.36 & 19.43 & 20.41 & 9.16 & 13.73 & 9.98 & 6.10 & 7.17 & 6.45 & 15 & 150 & 60 & 14.09 & $-$ & 15.64 & 8.30 & $-$ & 8.09 & 4.05 & $-$ & 4.34 \\
  1 & 30 & 45 & 16.01 & 18.90 & 19.39 & 9.63 & 13.98 & 10.24 & 6.22 & 7.03 & 6.57 & 15 & 180 & 15 & 12.13 & $-$ & 16.95 & 7.34 & $-$ & 10.05 & 3.74 & $-$ & 4.55 \\
  1 & 30 & 60 & 16.39 & 19.27 & 18.92 & 9.88 & 14.30 & 10.23 & 6.31 & 7.23 & 6.57 & 15 & 180 & 30 & 13.07 & $-$ & 14.93 & 7.91 & $-$ & 8.27 & 3.86 & $-$ & 4.59 \\
  1 & 45 & 5 & 9.88 & 20.01 & 23.22 & 5.89 & 16.20 & 9.80 & 5.37 & 7.32 & 6.36 & 15 & 180 & 45 & 13.50 & $-$ & 16.84 & 8.03 & $-$ & 8.34 & 3.93 & $-$ & 4.75 \\
  1 & 45 & 10 & 11.78 & 19.59 & 20.23 & 6.97 & 14.64 & 9.99 & 5.66 & 7.35 & 6.28 & 15 & 180 & 60 & 13.79 & $-$ & 15.85 & 8.20 & $-$ & 8.72 & 4.00 & $-$ & 4.63 \\
  1 & 45 & 15 & 12.91 & 19.34 & 20.89 & 7.70 & 14.17 & 9.86 & 5.76 & 7.45 & 6.43 & 15 & 240 & 15 & 11.84 & $-$ & 14.89 & 7.14 & $-$ & 8.46 & 3.68 & $-$ & 4.66 \\
  1 & 45 & 20 & 14.02 & 18.88 & 20.99 & 8.34 & 14.57 & 9.92 & 5.91 & 7.21 & 6.46 & 15 & 240 & 30 & 12.72 & $-$ & 14.10 & 7.69 & $-$ & 8.56 & 3.80 & $-$ & 4.32 \\
  1 & 45 & 30 & 14.93 & 18.85 & 21.00 & 8.96 & 15.15 & 9.89 & 6.01 & 7.07 & 6.48 & 15 & 240 & 45 & 13.10 & $-$ & 16.39 & 7.80 & $-$ & 8.58 & 3.87 & $-$ & 4.01 \\
  1 & 45 & 45 & 15.67 & 18.41 & 22.19 & 9.46 & 14.95 & 10.14 & 6.10 & 6.97 & 6.52 & 15 & 240 & 60 & 13.33 & $-$ & 16.73 & 7.98 & $-$ & 8.95 & 3.95 & $-$ & 4.06 \\
  1 & 45 & 60 & 16.01 & 19.32 & 20.65 & 9.59 & 16.48 & 9.63 & 6.19 & 7.20 & 6.53 & 15 & 300 & 60 & 13.24 & $-$ & 17.41 & 7.94 & $-$ & 8.39 & 3.95 & $-$ & 4.31 \\
  1 & 60 & 5 & 9.96 & 19.22 & 19.66 & 5.93 & 13.70 & 9.93 & 5.37 & 7.40 & 6.20 & 15 & 300 & 300 & 13.66 & $-$ & 16.14 & 8.28 & $-$ & 7.95 & 4.04 & $-$ & 4.41 \\
  1 & 60 & 10 & 11.24 & 17.70 & 18.62 & 6.81 & 14.21 & 9.76 & 5.58 & 7.50 & 6.29 & 15 & 900 & 60 & 12.71 & $-$ & 24.08 & 7.61 & $-$ & 8.03 & 3.62 & $-$ & 3.97 \\
  1 & 60 & 15 & 12.89 & 17.52 & 19.43 & 7.67 & 14.15 & 9.96 & 5.69 & 7.52 & 6.18 & 15 & 900 & 300 & 13.44 & $-$ & 25.03 & 8.02 & $-$ & 9.03 & 3.67 & $-$ & 4.04 \\
  1 & 60 & 20 & 13.78 & 18.29 & 19.23 & 8.23 & 14.71 & 10.15 & 5.81 & 7.39 & 6.31 & 15 & 900 & 900 & 14.37 & $-$ & 24.76 & 9.04 & $-$ & 8.39 & 3.62 & $-$ & 4.01 \\
  1 & 60 & 30 & 14.78 & 18.21 & 19.17 & 8.88 & 13.95 & 10.37 & 5.91 & 7.14 & 6.43 & 15 & 900 & 1800 & 14.61 & $-$ & 25.05 & 9.19 & $-$ & 8.70 & 3.65 & $-$ & 3.83 \\
  1 & 60 & 45 & 15.48 & 17.32 & 19.13 & 9.31 & 13.86 & 10.71 & 6.02 & 6.92 & 6.40 & 15 & 1800 & 60 & 13.48 & $-$ & 19.86 & 8.08 & $-$ & 7.23 & 3.52 & $-$ & 3.73 \\
  1 & 60 & 60 & 15.88 & 17.61 & 19.18 & 9.62 & 14.62 & 10.38 & 6.10 & 7.03 & 6.57 & 15 & 1800 & 300 & 13.75 & $-$ & 20.54 & 8.22 & $-$ & 7.62 & 3.59 & $-$ & 3.89 \\
  1 & 75 & 5 & 9.49 & 18.81 & 19.50 & 5.75 & 13.85 & 10.03 & 5.33 & 7.43 & 6.32 & 15 & 1800 & 900 & 14.50 & $-$ & 21.11 & 9.17 & $-$ & 7.83 & 3.59 & $-$ & 3.85 \\
  1 & 75 & 10 & 11.23 & 17.26 & 19.93 & 6.76 & 19.26 & 10.02 & 5.49 & 7.85 & 6.32 & 15 & 1800 & 1800 & 14.95 & $-$ & 21.99 & 9.97 & $-$ & 9.23 & 3.62 & $-$ & 3.69 \\
  1 & 75 & 15 & 12.54 & 17.70 & 20.83 & 7.58 & 10.87 & 10.29 & 5.66 & 6.37 & 6.48 & 15 & 1800 & 2700 & 13.88 & $-$ & 26.05 & 9.56 & $-$ & 9.53 & 3.63 & $-$ & 3.58 \\
  1 & 75 & 20 & 13.73 & 17.42 & 20.23 & 8.28 & 14.84 & 10.29 & 5.79 & 7.38 & 6.41 & 15 & 1800 & 3600 & $-$ & $-$ & $-$ & $-$ & $-$ & $-$ & 3.67 & $-$ & 3.50 \\
  1 & 75 & 30 & 14.67 & 17.80 & 20.35 & 8.82 & 13.48 & 10.44 & 5.89 & 7.18 & 6.49 & 15 & 3600 & 60 & $-$ & $-$ & $-$ & $-$ & $-$ & $-$ & 3.54 & $-$ & 3.48 \\
  1 & 75 & 45 & 15.34 & 16.99 & 20.40 & 9.26 & 13.91 & 10.99 & 5.99 & 6.91 & 6.58 & 15 & 3600 & 300 & $-$ & $-$ & $-$ & $-$ & $-$ & $-$ & 3.53 & $-$ & 3.61 \\
  1 & 75 & 60 & 15.71 & 17.63 & 21.11 & 9.52 & 13.91 & 10.48 & 6.08 & 7.10 & 6.57 & 15 & 3600 & 900 & $-$ & $-$ & $-$ & $-$ & $-$ & $-$ & 3.41 & $-$ & 3.49 \\
  1 & 90 & 5 & 9.56 & $-$ & 18.94 & 5.79 & $-$ & 10.57 & 5.25 & $-$ & 6.02 & 15 & 3600 & 1800 & $-$ & $-$ & $-$ & $-$ & $-$ & $-$ & 3.33 & $-$ & 3.28 \\
  1 & 90 & 10 & 11.34 & $-$ & 18.90 & 6.81 & $-$ & 10.06 & 5.50 & $-$ & 6.33 & 15 & 3600 & 2700 & $-$ & $-$ & $-$ & $-$ & $-$ & $-$ & 3.26 & $-$ & 3.22 \\
  1 & 90 & 15 & 12.68 & $-$ & 19.28 & 7.59 & $-$ & 10.44 & 5.62 & $-$ & 6.24 & 15 & 3600 & 3600 & $-$ & $-$ & $-$ & $-$ & $-$ & $-$ & 3.22 & $-$ & 3.20 \\
  1 & 90 & 20 & 13.77 & $-$ & 19.01 & 8.27 & $-$ & 10.35 & 5.77 & $-$ & 6.39 & 15 & 3600 & 5400 & $-$ & $-$ & $-$ & $-$ & $-$ & $-$ & 3.19 & $-$ & $-$ \\
  1 & 90 & 30 & 14.66 & $-$ & 20.20 & 8.83 & $-$ & 11.17 & 5.88 & $-$ & 6.47 & 15 & 7200 & 60 & $-$ & $-$ & $-$ & $-$ & $-$ & $-$ & 2.66 & $-$ & $-$ \\
  1 & 90 & 45 & 15.35 & $-$ & 21.02 & 9.27 & $-$ & 10.78 & 5.98 & $-$ & 6.44 & 15 & 7200 & 300 & $-$ & $-$ & $-$ & $-$ & $-$ & $-$ & 2.67 & $-$ & $-$ \\
  1 & 90 & 60 & 15.68 & $-$ & 22.12 & 9.47 & $-$ & 10.24 & 6.06 & $-$ & 6.56 & 15 & 7200 & 900 & $-$ & $-$ & $-$ & $-$ & $-$ & $-$ & 2.73 & $-$ & $-$ \\
  1 & 120 & 5 & 9.65 & $-$ & 18.86 & 5.81 & $-$ & 10.87 & 5.33 & $-$ & 6.05 & 15 & 7200 & 1800 & $-$ & $-$ & $-$ & $-$ & $-$ & $-$ & 2.73 & $-$ & $-$ \\
  1 & 120 & 10 & 11.31 & $-$ & 18.46 & 6.73 & $-$ & 10.04 & 5.55 & $-$ & 6.26 & 30 & 300 & 60 & 11.84 & 13.81 & 12.48 & 6.78 & 9.41 & 8.44 & 3.38 & 5.14 & 3.60 \\
  1 & 120 & 15 & 12.62 & $-$ & 20.94 & 7.54 & $-$ & 10.21 & 5.66 & $-$ & 6.25 & 30 & 300 & 300 & 12.95 & 14.29 & 13.72 & 7.59 & 10.88 & 7.73 & 3.52 & 5.37 & 3.86 \\
  1 & 120 & 20 & 13.76 & $-$ & 20.60 & 8.11 & $-$ & 10.35 & 5.80 & $-$ & 6.40 & 30 & 900 & 60 & 10.78 & 13.18 & 13.88 & 6.49 & 7.83 & 7.07 & 3.05 & 5.13 & 3.33 \\
  1 & 120 & 30 & 14.69 & $-$ & 19.73 & 8.72 & $-$ & 10.73 & 5.91 & $-$ & 6.41 & 30 & 900 & 300 & 11.50 & 17.14 & 21.87 & 7.25 & 14.06 & 8.03 & 3.14 & 5.49 & 3.55 \\
  1 & 120 & 45 & 15.23 & $-$ & 18.88 & 9.09 & $-$ & 10.73 & 6.03 & $-$ & 6.47 & 30 & 900 & 900 & 12.72 & 15.14 & 19.32 & 7.68 & 9.95 & 6.72 & 3.12 & 5.24 & 3.53 \\
  1 & 120 & 60 & 15.70 & $-$ & 19.14 & 9.36 & $-$ & 10.23 & 6.13 & $-$ & 6.60 & 30 & 900 & 1800 & 12.88 & 18.16 & 21.19 & 7.94 & 9.95 & 7.23 & 3.19 & 5.15 & 3.38 \\
  1 & 150 & 5 & 9.75 & $-$ & 19.90 & 5.76 & $-$ & 10.62 & 5.38 & $-$ & 6.27 & 30 & 1800 & 60 & 12.02 & 15.55 & 24.03 & 7.18 & 9.69 & 7.86 & 3.01 & 4.89 & 3.07 \\
  1 & 150 & 10 & 11.47 & $-$ & 19.21 & 6.73 & $-$ & 9.96 & 5.60 & $-$ & 6.42 & 30 & 1800 & 300 & 11.99 & 15.77 & 19.05 & 7.63 & 13.97 & 6.35 & 3.11 & 5.47 & 3.37 \\
  1 & 150 & 15 & 12.84 & $-$ & 22.72 & 7.49 & $-$ & 10.62 & 5.74 & $-$ & 6.32 & 30 & 1800 & 900 & 12.60 & 14.07 & 18.19 & 8.05 & 9.20 & 6.69 & 3.14 & 5.30 & 3.35 \\
  1 & 150 & 20 & 13.90 & $-$ & 21.70 & 8.15 & $-$ & 10.74 & 5.88 & $-$ & 6.43 & 30 & 1800 & 1800 & 12.65 & 15.48 & 20.57 & 8.69 & 10.86 & 7.35 & 3.17 & 5.05 & 3.23 \\
  1 & 150 & 30 & 14.73 & $-$ & 21.32 & 8.72 & $-$ & 10.53 & 5.98 & $-$ & 6.51 & 30 & 1800 & 2700 & 11.90 & 15.21 & 24.30 & 8.25 & 9.29 & 8.11 & 3.17 & 4.86 & 3.07 \\
  1 & 150 & 45 & 15.42 & $-$ & 21.94 & 9.05 & $-$ & 10.88 & 6.09 & $-$ & 6.54 & 30 & 1800 & 3600 & $-$ & 15.66 & $-$ & $-$ & 9.02 & $-$ & 3.21 & 4.91 & 2.97 \\
  1 & 150 & 60 & 15.74 & $-$ & 21.58 & 9.23 & $-$ & 10.77 & 6.17 & $-$ & 6.73 & 30 & 3600 & 60 & $-$ & $-$ & $-$ & $-$ & $-$ & $-$ & 3.06 & $-$ & 2.73 \\
  1 & 180 & 5 & 9.72 & $-$ & 22.16 & 5.71 & $-$ & 10.65 & 5.49 & $-$ & 6.29 & 30 & 3600 & 300 & $-$ & $-$ & $-$ & $-$ & $-$ & $-$ & 3.12 & $-$ & 3.15 \\
  1 & 180 & 10 & 11.50 & $-$ & 21.76 & 6.72 & $-$ & 10.02 & 5.70 & $-$ & 6.48 & 30 & 3600 & 900 & $-$ & 14.02 & $-$ & $-$ & 9.89 & $-$ & 3.01 & 4.51 & 3.00 \\
  1 & 180 & 15 & 12.84 & $-$ & 18.90 & 7.53 & $-$ & 10.37 & 5.82 & $-$ & 6.44 & 30 & 3600 & 1800 & $-$ & $-$ & $-$ & $-$ & $-$ & $-$ & 2.90 & $-$ & 2.81 \\
  1 & 180 & 20 & 13.99 & $-$ & 19.49 & 8.11 & $-$ & 10.68 & 5.92 & $-$ & 6.52 & 30 & 3600 & 2700 & $-$ & $-$ & $-$ & $-$ & $-$ & $-$ & 2.80 & $-$ & 2.74 \\
  1 & 180 & 30 & 14.88 & $-$ & 22.02 & 8.60 & $-$ & 10.60 & 6.03 & $-$ & 6.59 & 30 & 3600 & 3600 & $-$ & $-$ & $-$ & $-$ & $-$ & $-$ & 2.76 & $-$ & 2.71 \\
  1 & 180 & 45 & 15.38 & $-$ & 19.29 & 8.97 & $-$ & 10.53 & 6.12 & $-$ & 6.60 & 30 & 3600 & 5400 & $-$ & $-$ & $-$ & $-$ & $-$ & $-$ & 2.77 & $-$ & $-$ \\
  1 & 180 & 60 & 15.72 & $-$ & 18.50 & 9.20 & $-$ & 10.80 & 6.20 & $-$ & 6.64 & 30 & 7200 & 60 & $-$ & $-$ & $-$ & $-$ & $-$ & $-$ & 2.37 & $-$ & $-$ \\
  1 & 240 & 5 & 9.69 & $-$ & 20.35 & 5.71 & $-$ & 11.10 & 5.39 & $-$ & 6.28 & 30 & 7200 & 300 & $-$ & $-$ & $-$ & $-$ & $-$ & $-$ & 2.49 & $-$ & $-$ \\
  1 & 240 & 10 & 11.53 & $-$ & 19.96 & 6.68 & $-$ & 10.69 & 5.65 & $-$ & 6.45 & 30 & 7200 & 900 & $-$ & $-$ & $-$ & $-$ & $-$ & $-$ & 2.55 & $-$ & $-$ \\
  1 & 240 & 15 & 12.84 & $-$ & 20.26 & 7.47 & $-$ & 10.44 & 5.82 & $-$ & 6.39 & 30 & 7200 & 1800 & $-$ & $-$ & $-$ & $-$ & $-$ & $-$ & 2.54 & $-$ & $-$ \\
  1 & 240 & 20 & 13.96 & $-$ & 20.97 & 8.07 & $-$ & 10.65 & 5.94 & $-$ & 6.64 & 45 & 300 & 60 & 10.22 & 11.27 & 14.23 & 5.70 & 7.88 & 6.69 & 3.08 & 5.20 & 2.63 \\
  1 & 240 & 30 & 14.83 & $-$ & 20.73 & 8.59 & $-$ & 10.69 & 6.06 & $-$ & 6.57 & 45 & 300 & 300 & 11.89 & 12.01 & 13.81 & 6.71 & 8.25 & 7.51 & 3.29 & 5.38 & 3.34 \\
  1 & 240 & 45 & 15.40 & $-$ & 18.99 & 8.98 & $-$ & 11.04 & 6.18 & $-$ & 6.62 & 45 & 900 & 60 & 9.49 & 11.28 & 12.71 & 5.59 & 8.10 & 6.65 & 2.77 & 4.99 & 3.58 \\
  1 & 240 & 60 & 15.70 & $-$ & 19.24 & 9.11 & $-$ & 10.73 & 6.27 & $-$ & 6.81 & 45 & 900 & 300 & 10.73 & 11.70 & 16.57 & 6.42 & 8.38 & 5.99 & 2.88 & 5.29 & 3.58 \\
  1 & 300 & 60 & 15.75 & $-$ & 20.97 & 9.04 & $-$ & 10.92 & 6.41 & $-$ & 7.00 & 45 & 900 & 900 & 11.48 & 13.71 & 18.23 & 6.56 & 9.04 & 5.71 & 2.82 & 5.09 & 3.17 \\
  1 & 300 & 300 & 16.93 & $-$ & 21.05 & 9.62 & $-$ & 10.51 & 6.63 & $-$ & 7.00 & 45 & 900 & 1800 & 11.37 & 16.11 & 20.16 & 6.64 & 8.46 & 6.01 & 2.86 & 4.97 & 3.08 \\
  1 & 900 & 60 & 16.05 & $-$ & 29.78 & 8.93 & $-$ & 10.99 & $-$ & $-$ & $-$ & 45 & 1800 & 60 & 9.76 & 11.80 & 23.51 & 5.97 & 8.57 & 5.70 & 2.65 & 4.86 & 3.61 \\
  1 & 900 & 300 & 17.29 & $-$ & 25.24 & 9.55 & $-$ & 10.46 & $-$ & $-$ & $-$ & 45 & 1800 & 300 & 9.84 & 11.50 & 20.12 & 6.28 & 8.33 & 5.11 & 2.78 & 5.21 & 3.34 \\
  1 & 900 & 900 & 18.61 & $-$ & 25.02 & 10.53 & $-$ & 10.52 & $-$ & $-$ & $-$ & 45 & 1800 & 900 & 10.54 & 12.23 & 20.49 & 6.68 & 8.12 & 5.53 & 2.80 & 5.15 & 3.07 \\
  1 & 1800 & 60 & 17.14 & $-$ & 29.56 & 9.98 & $-$ & 10.79 & $-$ & $-$ & $-$ & 45 & 1800 & 1800 & 10.30 & 13.94 & 22.02 & 7.06 & 8.62 & 5.99 & 2.81 & 4.91 & 2.92 \\
  1 & 1800 & 300 & 18.05 & $-$ & 21.94 & 10.57 & $-$ & 10.39 & $-$ & $-$ & $-$ & 45 & 1800 & 2700 & 9.72 & 15.06 & 23.79 & 6.70 & 8.29 & 6.82 & 2.83 & 4.68 & 2.80 \\
  1 & 1800 & 900 & 18.52 & $-$ & 22.68 & 11.29 & $-$ & 10.95 & $-$ & $-$ & $-$ & 45 & 1800 & 3600 & $-$ & 15.58 & $-$ & $-$ & 7.51 & $-$ & 2.88 & 4.50 & 2.71 \\
  5 & 30 & 5 & 9.26 & 17.30 & 21.39 & 5.39 & 16.13 & 9.29 & 4.02 & 6.41 & 4.97 & 45 & 3600 & 60 & $-$ & $-$ & $-$ & $-$ & $-$ & $-$ & 2.80 & $-$ & 2.96 \\
  5 & 30 & 10 & 11.65 & 16.89 & 21.46 & 6.77 & 13.03 & 10.32 & 4.31 & 6.44 & 5.07 & 45 & 3600 & 300 & $-$ & 13.02 & $-$ & $-$ & 9.90 & $-$ & 2.86 & 4.88 & 3.13 \\
  5 & 30 & 15 & 13.56 & 17.44 & 21.66 & 7.78 & 14.72 & 10.72 & 4.55 & 6.15 & 5.13 & 45 & 3600 & 900 & $-$ & 12.91 & $-$ & $-$ & 9.41 & $-$ & 2.75 & 4.39 & 2.73 \\
  5 & 30 & 20 & 14.89 & 17.41 & 20.97 & 8.44 & 14.41 & 10.18 & 4.74 & 6.21 & 5.06 & 45 & 3600 & 1800 & $-$ & $-$ & $-$ & $-$ & $-$ & $-$ & 2.60 & $-$ & 2.57 \\
  5 & 30 & 30 & 16.10 & 17.08 & 20.14 & 9.14 & 13.32 & 9.79 & 4.87 & 5.78 & 5.06 & 45 & 3600 & 2700 & $-$ & $-$ & $-$ & $-$ & $-$ & $-$ & 2.52 & $-$ & 2.53 \\
  5 & 30 & 45 & 16.68 & 16.98 & 19.58 & 9.62 & 13.93 & 9.61 & 4.95 & 5.80 & 5.17 & 45 & 3600 & 3600 & $-$ & $-$ & $-$ & $-$ & $-$ & $-$ & 2.51 & $-$ & 2.56 \\
  5 & 30 & 60 & 16.69 & 17.46 & 20.40 & 9.73 & 13.97 & 9.34 & 5.05 & 5.92 & 5.10 & 45 & 3600 & 5400 & $-$ & $-$ & $-$ & $-$ & $-$ & $-$ & 2.55 & $-$ & $-$ \\
  5 & 45 & 5 & 9.81 & 17.01 & 20.13 & 5.52 & 14.19 & 9.88 & 4.00 & 6.30 & 5.04 & 45 & 7200 & 60 & $-$ & $-$ & $-$ & $-$ & $-$ & $-$ & 2.14 & $-$ & $-$ \\
  5 & 45 & 10 & 11.95 & 17.04 & 19.92 & 6.90 & 12.77 & 9.95 & 4.24 & 6.42 & 5.02 & 45 & 7200 & 300 & $-$ & $-$ & $-$ & $-$ & $-$ & $-$ & 2.08 & $-$ & $-$ \\
  5 & 45 & 15 & 13.88 & 17.81 & 19.76 & 7.83 & 15.50 & 9.99 & 4.49 & 6.18 & 5.16 & 45 & 7200 & 900 & $-$ & $-$ & $-$ & $-$ & $-$ & $-$ & 2.14 & $-$ & $-$ \\
  5 & 45 & 20 & 14.75 & 17.69 & 18.35 & 8.38 & 15.49 & 9.38 & 4.63 & 6.19 & 5.16 & 45 & 7200 & 1800 & $-$ & $-$ & $-$ & $-$ & $-$ & $-$ & 2.08 & $-$ & $-$ \\
  5 & 45 & 30 & 15.94 & 17.12 & 19.61 & 9.12 & 13.85 & 9.13 & 4.77 & 5.77 & 5.08 & 60 & 300 & 60 & 9.11 & 13.89 & 14.25 & 5.25 & 8.61 & 8.98 & 2.95 & 5.33 & 3.64 \\
  5 & 45 & 45 & 16.38 & 16.96 & 19.87 & 9.54 & 14.34 & 9.72 & 4.83 & 5.78 & 4.96 & 60 & 300 & 300 & 10.43 & 12.94 & 13.49 & 6.02 & 7.47 & 8.16 & 3.03 & 5.39 & 3.41 \\
  5 & 45 & 60 & 16.34 & 17.10 & 20.26 & 9.57 & 14.20 & 9.76 & 4.93 & 5.90 & 4.99 & 60 & 900 & 60 & 8.29 & 15.71 & 15.25 & 4.90 & 11.71 & 6.58 & 2.58 & 4.97 & 3.37 \\
  5 & 60 & 5 & 9.94 & 18.62 & 17.12 & 5.70 & 13.03 & 9.22 & 3.95 & 6.44 & 5.11 & 60 & 900 & 300 & 9.44 & 14.15 & 14.09 & 5.80 & 10.67 & 5.84 & 2.67 & 5.15 & 3.17 \\
  5 & 60 & 10 & 11.20 & 15.59 & 17.69 & 6.46 & 12.31 & 9.36 & 4.27 & 6.47 & 4.86 & 60 & 900 & 900 & 10.67 & 14.10 & 20.52 & 5.81 & 10.81 & 5.40 & 2.67 & 5.14 & 3.06 \\
  5 & 60 & 15 & 14.20 & 16.78 & 18.73 & 8.17 & 13.64 & 9.43 & 4.48 & 6.20 & 4.86 & 60 & 900 & 1800 & 10.55 & 17.04 & 20.61 & 5.84 & 10.64 & 5.76 & 2.72 & 4.95 & 2.93 \\
  5 & 60 & 20 & 14.36 & 17.37 & 19.14 & 8.31 & 14.69 & 9.71 & 4.63 & 6.30 & 4.79 & 60 & 1800 & 60 & 8.64 & 17.51 & 24.58 & 5.42 & 9.42 & 4.86 & 2.56 & 5.01 & 2.29 \\
  5 & 60 & 30 & 15.57 & 16.72 & 19.43 & 9.09 & 13.09 & 9.39 & 4.75 & 5.87 & 4.95 & 60 & 1800 & 300 & 8.94 & 15.30 & 19.56 & 5.81 & 10.54 & 4.73 & 2.67 & 5.31 & 2.89 \\
  5 & 60 & 45 & 15.82 & 17.00 & 20.27 & 9.50 & 14.59 & 9.43 & 4.80 & 5.66 & 4.96 & 60 & 1800 & 900 & 9.74 & 14.35 & 21.26 & 6.03 & 9.81 & 5.11 & 2.69 & 5.08 & 2.81 \\
  5 & 60 & 60 & 15.85 & 16.55 & 19.55 & 9.53 & 14.24 & 9.47 & 4.88 & 5.75 & 5.02 & 60 & 1800 & 1800 & 9.35 & 13.74 & 23.15 & 6.24 & 9.11 & 4.96 & 2.69 & 4.89 & 2.75 \\
  5 & 75 & 5 & 9.71 & 17.58 & 18.60 & 5.74 & 13.48 & 9.09 & 3.94 & 6.48 & 4.74 & 60 & 1800 & 2700 & 8.96 & 14.63 & 24.85 & 5.71 & 7.34 & 6.07 & 2.66 & 4.58 & 2.59 \\
  5 & 75 & 10 & 11.83 & 14.88 & 18.45 & 6.92 & 12.75 & 9.29 & 4.19 & 6.43 & 4.82 & 60 & 1800 & 3600 & $-$ & 15.38 & $-$ & $-$ & 7.85 & $-$ & 2.68 & 4.59 & 2.44 \\
  5 & 75 & 15 & 13.12 & 16.26 & 18.54 & 7.69 & 13.90 & 9.65 & 4.40 & 6.21 & 5.05 & 60 & 3600 & 60 & $-$ & 12.70 & $-$ & $-$ & 6.47 & $-$ & 2.59 & 3.99 & 1.72 \\
  5 & 75 & 20 & 14.12 & 16.67 & 18.98 & 8.30 & 13.95 & 10.21 & 4.57 & 6.23 & 5.05 & 60 & 3600 & 300 & $-$ & 13.50 & $-$ & $-$ & 8.86 & $-$ & 2.68 & 4.73 & 2.60 \\
  5 & 75 & 30 & 15.13 & 16.97 & 18.24 & 8.95 & 13.57 & 10.07 & 4.70 & 5.87 & 4.95 & 60 & 3600 & 900 & $-$ & 12.43 & $-$ & $-$ & 7.97 & $-$ & 2.56 & 4.23 & 2.49 \\
  5 & 75 & 45 & 15.35 & 16.15 & 19.06 & 9.32 & 14.10 & 9.56 & 4.77 & 5.58 & 5.12 & 60 & 3600 & 1800 & $-$ & $-$ & $-$ & $-$ & $-$ & $-$ & 2.40 & $-$ & 2.18 \\
  5 & 75 & 60 & 15.34 & 16.39 & 17.60 & 9.40 & 14.63 & 9.75 & 4.85 & 5.72 & 5.08 & 60 & 3600 & 2700 & $-$ & $-$ & $-$ & $-$ & $-$ & $-$ & 2.30 & $-$ & 2.17 \\
  5 & 90 & 5 & 9.90 & $-$ & 18.74 & 5.66 & $-$ & 10.58 & 3.89 & $-$ & 5.03 & 60 & 3600 & 3600 & $-$ & $-$ & $-$ & $-$ & $-$ & $-$ & 2.30 & $-$ & $-$ \\
  5 & 90 & 10 & 11.86 & $-$ & 19.00 & 6.85 & $-$ & 10.61 & 4.25 & $-$ & 4.93 & 60 & 3600 & 5400 & $-$ & $-$ & $-$ & $-$ & $-$ & $-$ & 2.39 & $-$ & $-$ \\
  5 & 90 & 15 & 13.44 & $-$ & 18.90 & 7.76 & $-$ & 10.39 & 4.42 & $-$ & 5.17 & 60 & 7200 & 60 & $-$ & $-$ & $-$ & $-$ & $-$ & $-$ & 2.13 & $-$ & $-$ \\
  5 & 90 & 20 & 14.16 & $-$ & 17.70 & 8.31 & $-$ & 9.79 & 4.57 & $-$ & 5.23 & 60 & 7200 & 300 & $-$ & $-$ & $-$ & $-$ & $-$ & $-$ & 2.14 & $-$ & $-$ \\
  5 & 90 & 30 & 15.03 & $-$ & 17.34 & 9.00 & $-$ & 9.57 & 4.69 & $-$ & 5.12 & 60 & 7200 & 900 & $-$ & $-$ & $-$ & $-$ & $-$ & $-$ & 2.25 & $-$ & $-$ \\
  5 & 90 & 45 & 15.21 & $-$ & 18.48 & 9.34 & $-$ & 9.39 & 4.74 & $-$ & 5.08 & 60 & 7200 & 1800 & $-$ & $-$ & $-$ & $-$ & $-$ & $-$ & 2.28 & $-$ & $-$ \\
  5 & 90 & 60 & 15.29 & $-$ & 18.06 & 9.40 & $-$ & 9.79 & 4.79 & $-$ & 5.09 & 90 & 300 & 300 & 9.61 & 13.01 & 12.91 & 5.49 & 7.03 & 8.78 & 2.83 & 5.47 & 2.73 \\
  5 & 120 & 5 & 9.99 & $-$ & 18.46 & 5.70 & $-$ & 9.52 & 3.99 & $-$ & 4.77 & 90 & 900 & 300 & 9.31 & 12.19 & 10.76 & 5.28 & 7.01 & 5.89 & 2.48 & 5.08 & 3.46 \\
  5 & 120 & 10 & 11.81 & $-$ & 19.02 & 6.78 & $-$ & 9.25 & 4.27 & $-$ & 4.61 & 90 & 900 & 900 & 11.11 & 15.19 & 20.44 & 6.01 & 9.11 & 4.77 & 2.44 & 4.89 & 2.90 \\
  5 & 120 & 15 & 13.29 & $-$ & 20.94 & 7.81 & $-$ & 9.15 & 4.46 & $-$ & 4.78 & 90 & 900 & 1800 & 10.89 & 16.52 & 21.01 & 5.56 & 8.03 & 5.08 & 2.49 & 4.65 & 2.77 \\
  5 & 120 & 20 & 14.06 & $-$ & 20.09 & 8.37 & $-$ & 9.33 & 4.54 & $-$ & 4.87 & 90 & 1800 & 300 & 8.16 & 12.79 & 18.21 & 4.88 & 7.55 & 3.95 & 2.29 & 4.99 & 3.19 \\
  5 & 120 & 30 & 14.76 & $-$ & 19.04 & 8.96 & $-$ & 9.70 & 4.63 & $-$ & 4.93 & 90 & 1800 & 900 & 9.22 & 12.45 & 20.07 & 5.77 & 6.79 & 4.05 & 2.31 & 4.92 & 2.66 \\
  5 & 120 & 45 & 15.03 & $-$ & 18.04 & 9.29 & $-$ & 9.64 & 4.68 & $-$ & 5.03 & 90 & 1800 & 1800 & 9.49 & 13.60 & 23.82 & 5.34 & 7.47 & 4.02 & 2.36 & 4.68 & 2.53 \\
  5 & 120 & 60 & 15.09 & $-$ & 19.23 & 9.37 & $-$ & 9.83 & 4.72 & $-$ & 5.01 & 90 & 1800 & 2700 & 9.15 & 14.61 & $-$ & 5.22 & 6.70 & $-$ & 2.39 & 4.47 & 2.37 \\
  5 & 150 & 5 & 10.07 & $-$ & 17.29 & 5.74 & $-$ & 9.25 & 3.84 & $-$ & 5.06 & 90 & 1800 & 3600 & $-$ & 15.47 & $-$ & $-$ & 6.49 & $-$ & 2.42 & 4.27 & 2.24 \\
  5 & 150 & 10 & 11.97 & $-$ & 16.36 & 6.92 & $-$ & 9.15 & 4.22 & $-$ & 5.23 & 90 & 3600 & 300 & $-$ & 13.21 & $-$ & $-$ & 8.22 & $-$ & 2.34 & 4.66 & 3.65 \\
  5 & 150 & 15 & 13.47 & $-$ & 16.65 & 7.75 & $-$ & 9.59 & 4.37 & $-$ & 5.20 & 90 & 3600 & 900 & $-$ & 11.99 & $-$ & $-$ & 7.98 & $-$ & 2.29 & 4.14 & 2.38 \\
  5 & 150 & 20 & 14.21 & $-$ & 16.67 & 8.30 & $-$ & 9.33 & 4.52 & $-$ & 5.05 & 90 & 3600 & 1800 & $-$ & $-$ & $-$ & $-$ & $-$ & $-$ & 2.22 & $-$ & 2.21 \\
  5 & 150 & 30 & 14.79 & $-$ & 17.92 & 8.91 & $-$ & 10.02 & 4.61 & $-$ & 4.95 & 90 & 3600 & 2700 & $-$ & $-$ & $-$ & $-$ & $-$ & $-$ & 2.13 & $-$ & $-$ \\
  5 & 150 & 45 & 14.96 & $-$ & 18.26 & 9.20 & $-$ & 9.89 & 4.64 & $-$ & 4.96 & 90 & 3600 & 3600 & $-$ & $-$ & $-$ & $-$ & $-$ & $-$ & 2.12 & $-$ & $-$ \\
  5 & 150 & 60 & 15.06 & $-$ & 18.07 & 9.27 & $-$ & 9.76 & 4.68 & $-$ & 5.05 & 90 & 7200 & 300 & $-$ & $-$ & $-$ & $-$ & $-$ & $-$ & 1.97 & $-$ & $-$ \\
  5 & 180 & 5 & 10.14 & $-$ & 17.94 & 5.62 & $-$ & 9.84 & 3.96 & $-$ & 5.12 & 90 & 7200 & 900 & $-$ & $-$ & $-$ & $-$ & $-$ & $-$ & 2.15 & $-$ & $-$ \\
  5 & 180 & 10 & 12.22 & $-$ & 18.96 & 6.93 & $-$ & 10.20 & 4.28 & $-$ & 5.06 & 120 & 900 & 300 & 8.18 & 15.81 & 10.91 & 4.65 & 9.64 & 5.82 & 2.16 & 4.78 & 2.83 \\
  5 & 180 & 15 & 13.57 & $-$ & 20.42 & 7.71 & $-$ & 10.78 & 4.38 & $-$ & 5.16 & 120 & 900 & 900 & 11.00 & 14.21 & 17.63 & 5.44 & 8.09 & 4.94 & 2.16 & 4.85 & 2.73 \\
  5 & 180 & 20 & 14.36 & $-$ & 18.66 & 8.28 & $-$ & 10.02 & 4.48 & $-$ & 5.06 & 120 & 900 & 1800 & 10.48 & 16.24 & 18.36 & 5.20 & 8.30 & 4.91 & 2.22 & 4.61 & 2.57 \\
  5 & 180 & 30 & 14.84 & $-$ & 17.53 & 8.80 & $-$ & 10.14 & 4.54 & $-$ & 5.14 & 120 & 1800 & 300 & 7.94 & 15.52 & 17.04 & 4.64 & 10.84 & 4.15 & 2.05 & 4.99 & 2.29 \\
  5 & 180 & 45 & 14.99 & $-$ & 17.99 & 9.11 & $-$ & 9.65 & 4.59 & $-$ & 5.08 & 120 & 1800 & 900 & 8.81 & 13.58 & 19.72 & 5.20 & 8.85 & 3.68 & 2.08 & 4.87 & 2.69 \\
  5 & 180 & 60 & 15.05 & $-$ & 19.66 & 9.16 & $-$ & 9.55 & 4.61 & $-$ & 5.06 & 120 & 1800 & 1800 & 8.02 & 12.61 & 23.04 & 4.27 & 12.03 & 3.99 & 2.13 & 4.75 & 2.45 \\
  5 & 240 & 5 & 10.15 & $-$ & 18.40 & 5.48 & $-$ & 11.02 & 3.91 & $-$ & 4.98 & 120 & 1800 & 2700 & 7.26 & 14.25 & $-$ & 3.85 & 12.03 & $-$ & 2.16 & 4.56 & 2.12 \\
  5 & 240 & 10 & 12.24 & $-$ & 18.31 & 6.74 & $-$ & 11.10 & 4.19 & $-$ & 5.01 & 120 & 1800 & 3600 & $-$ & 15.69 & $-$ & $-$ & 8.25 & $-$ & 2.20 & 4.38 & 1.97 \\
  5 & 240 & 15 & 13.63 & $-$ & 18.80 & 7.62 & $-$ & 10.53 & 4.30 & $-$ & 5.12 & 120 & 3600 & 300 & $-$ & 15.33 & $-$ & $-$ & 9.64 & $-$ & 2.26 & 4.60 & 2.09 \\
  5 & 240 & 20 & 14.11 & $-$ & 18.69 & 8.06 & $-$ & 9.83 & 4.40 & $-$ & 5.16 & 120 & 3600 & 900 & $-$ & 14.29 & $-$ & $-$ & 8.31 & $-$ & 2.24 & 4.25 & 1.93 \\
  5 & 240 & 30 & 14.85 & $-$ & 16.92 & 8.68 & $-$ & 9.39 & 4.46 & $-$ & 5.07 & 120 & 3600 & 1800 & $-$ & $-$ & $-$ & $-$ & $-$ & $-$ & 2.06 & $-$ & $-$ \\
  5 & 240 & 45 & 14.90 & $-$ & 17.15 & 8.98 & $-$ & 9.77 & 4.52 & $-$ & 4.99 & 120 & 3600 & 2700 & $-$ & $-$ & $-$ & $-$ & $-$ & $-$ & 2.00 & $-$ & $-$ \\
  5 & 240 & 60 & 15.04 & $-$ & 17.76 & 9.04 & $-$ & 10.39 & 4.56 & $-$ & 5.03 & 120 & 3600 & 3600 & $-$ & $-$ & $-$ & $-$ & $-$ & $-$ & 2.00 & $-$ & $-$ \\
  5 & 300 & 60 & 15.01 & $-$ & 19.27 & 8.93 & $-$ & 9.94 & 4.52 & $-$ & 5.04 & 120 & 7200 & 300 & $-$ & $-$ & $-$ & $-$ & $-$ & $-$ & 2.13 & $-$ & $-$ \\
  5 & 300 & 300 & 15.69 & $-$ & 18.68 & 9.21 & $-$ & 9.27 & 4.61 & $-$ & 5.03 & 120 & 7200 & 900 & $-$ & $-$ & $-$ & $-$ & $-$ & $-$ & 2.08 & $-$ & $-$ \\
  5 & 900 & 60 & 15.14 & $-$ & 25.21 & 8.57 & $-$ & 10.02 & 4.22 & $-$ & 4.74 & 150 & 900 & 300 & 8.64 & 12.13 & 11.68 & 4.95 & 7.23 & 6.19 & 2.21 & 4.92 & 3.13 \\
  5 & 900 & 300 & 15.83 & $-$ & 22.46 & 8.86 & $-$ & 9.16 & 4.27 & $-$ & 4.68 & 150 & 900 & 900 & 11.46 & 13.09 & 17.98 & 5.83 & 8.22 & 5.40 & 2.19 & 4.58 & 2.70 \\
  5 & 900 & 900 & 16.69 & $-$ & 23.23 & 9.61 & $-$ & 9.44 & 4.24 & $-$ & 4.70 & 150 & 900 & 1800 & 11.70 & 15.41 & 16.65 & 5.57 & 7.22 & 5.48 & 2.24 & 4.68 & 2.56 \\
  5 & 900 & 1800 & 16.76 & $-$ & 24.25 & 10.10 & $-$ & 9.99 & 4.26 & $-$ & 4.52 & 150 & 1800 & 300 & 7.88 & 12.02 & 17.70 & 4.98 & 7.36 & 4.70 & 2.06 & 4.86 & 2.52 \\
  5 & 1800 & 60 & 15.84 & $-$ & 23.38 & 9.12 & $-$ & 9.42 & 4.16 & $-$ & 4.50 & 150 & 1800 & 900 & 8.54 & 12.28 & 17.89 & 5.87 & 6.66 & 4.96 & 2.11 & 4.71 & 2.50 \\
  5 & 1800 & 300 & 16.22 & $-$ & 22.32 & 9.46 & $-$ & 9.08 & 4.23 & $-$ & 4.62 & 150 & 1800 & 1800 & 7.73 & 12.05 & $-$ & 5.24 & 7.57 & $-$ & 2.16 & 4.35 & 2.41 \\
  5 & 1800 & 900 & 16.62 & $-$ & 22.26 & 10.29 & $-$ & 9.64 & 4.24 & $-$ & 4.51 & 150 & 1800 & 2700 & $-$ & 12.13 & $-$ & $-$ & 11.08 & $-$ & 2.19 & 4.28 & 2.14 \\
  5 & 1800 & 1800 & 16.61 & $-$ & 23.74 & 11.24 & $-$ & 11.00 & 4.26 & $-$ & 4.32 & 150 & 1800 & 3600 & $-$ & 13.64 & $-$ & $-$ & 8.22 & $-$ & 2.22 & 4.07 & $-$ \\
  5 & 1800 & 2700 & 15.56 & $-$ & 25.34 & 10.59 & $-$ & 10.97 & 4.21 & $-$ & 4.19 & 150 & 3600 & 300 & $-$ & 11.43 & $-$ & $-$ & 7.16 & $-$ & 2.21 & 4.32 & $-$ \\
  5 & 1800 & 3600 & $-$ & $-$ & $-$ & $-$ & $-$ & $-$ & 4.19 & $-$ & 4.08 & 150 & 3600 & 900 & $-$ & 10.47 & $-$ & $-$ & 6.73 & $-$ & 2.16 & 4.10 & $-$ \\
  5 & 3600 & 60 & $-$ & $-$ & $-$ & $-$ & $-$ & $-$ & 4.15 & $-$ & 4.25 & 150 & 3600 & 1800 & $-$ & $-$ & $-$ & $-$ & $-$ & $-$ & 2.09 & $-$ & $-$ \\
  5 & 3600 & 300 & $-$ & $-$ & $-$ & $-$ & $-$ & $-$ & 4.13 & $-$ & 4.22 & 150 & 3600 & 2700 & $-$ & $-$ & $-$ & $-$ & $-$ & $-$ & 2.01 & $-$ & $-$ \\
  5 & 3600 & 900 & $-$ & $-$ & $-$ & $-$ & $-$ & $-$ & 4.02 & $-$ & 4.10 & 150 & 3600 & 3600 & $-$ & $-$ & $-$ & $-$ & $-$ & $-$ & 1.98 & $-$ & $-$ \\
  5 & 3600 & 1800 & $-$ & $-$ & $-$ & $-$ & $-$ & $-$ & 3.92 & $-$ & 3.88 & 200 & 900 & 300 & 7.67 & 12.20 & $-$ & 4.94 & 7.59 & $-$ & 2.13 & 4.79 & 3.85 \\
  5 & 3600 & 2700 & $-$ & $-$ & $-$ & $-$ & $-$ & $-$ & 3.88 & $-$ & 3.83 & 200 & 900 & 900 & 9.83 & 12.53 & $-$ & 5.22 & 6.71 & $-$ & 1.99 & 4.78 & 3.03 \\
  5 & 3600 & 3600 & $-$ & $-$ & $-$ & $-$ & $-$ & $-$ & 3.85 & $-$ & 3.83 & 200 & 900 & 1800 & 10.52 & 15.40 & $-$ & 5.46 & 7.01 & $-$ & 2.06 & 4.54 & 2.65 \\
  5 & 7200 & 60 & $-$ & $-$ & $-$ & $-$ & $-$ & $-$ & 3.37 & $-$ & 3.62 & 200 & 1800 & 300 & 6.51 & 12.49 & 6.99 & 4.15 & 7.42 & 3.76 & 1.85 & 4.79 & 2.80 \\
  5 & 7200 & 300 & $-$ & $-$ & $-$ & $-$ & $-$ & $-$ & 3.34 & $-$ & 3.35 & 200 & 1800 & 900 & 6.52 & 13.09 & 7.20 & 4.14 & 8.54 & 3.95 & 1.86 & 4.98 & 2.31 \\
  5 & 7200 & 900 & $-$ & $-$ & $-$ & $-$ & $-$ & $-$ & 3.43 & $-$ & 3.55 & 200 & 1800 & 1800 & 6.67 & 14.55 & 13.09 & 3.56 & 6.64 & 3.54 & 1.91 & 4.48 & 2.18 \\
  5 & 7200 & 1800 & $-$ & $-$ & $-$ & $-$ & $-$ & $-$ & 3.44 & $-$ & 3.58 & 200 & 1800 & 2700 & $-$ & 15.07 & $-$ & $-$ & 6.03 & $-$ & 1.96 & 4.33 & $-$ \\
  10 & 300 & 60 & 14.34 & $-$ & 17.06 & 8.72 & $-$ & 8.54 & 4.16 & $-$ & 4.59 & 200 & 1800 & 3600 & $-$ & 15.47 & $-$ & $-$ & 9.28 & $-$ & 2.02 & 3.81 & $-$ \\
  10 & 300 & 300 & 14.77 & $-$ & 18.26 & 8.98 & $-$ & 8.60 & 4.21 & $-$ & 4.62 & 200 & 3600 & 300 & $-$ & 14.07 & $-$ & $-$ & 8.66 & $-$ & 1.93 & 4.29 & $-$ \\
  10 & 900 & 60 & 14.33 & $-$ & 21.86 & 8.38 & $-$ & 8.65 & 3.77 & $-$ & 4.30 & 200 & 3600 & 900 & $-$ & 11.40 & $-$ & $-$ & 6.68 & $-$ & 1.94 & 3.91 & $-$ \\
  10 & 900 & 300 & 14.84 & $-$ & 24.00 & 8.81 & $-$ & 8.92 & 3.84 & $-$ & 4.30 & 200 & 3600 & 1800 & $-$ & $-$ & $-$ & $-$ & $-$ & $-$ & 1.91 & $-$ & $-$ \\
  10 & 900 & 900 & 15.69 & $-$ & 24.40 & 9.61 & $-$ & 8.95 & 3.82 & $-$ & 4.27 & 200 & 3600 & 2700 & $-$ & $-$ & $-$ & $-$ & $-$ & $-$ & 1.87 & $-$ & $-$ \\
  10 & 900 & 1800 & 15.71 & $-$ & 24.75 & 9.85 & $-$ & 9.45 & 3.84 & $-$ & 4.07 & 200 & 3600 & 3600 & $-$ & $-$ & $-$ & $-$ & $-$ & $-$ & 1.85 & $-$ & $-$ \\
  10 & 1800 & 60 & 15.33 & $-$ & 18.32 & 8.88 & $-$ & 8.22 & 3.70 & $-$ & 4.14 & 250 & 900 & 300 & 6.69 & 11.66 & $-$ & 3.37 & 6.70 & $-$ & 1.95 & 4.64 & 4.10 \\
  10 & 1800 & 300 & 15.46 & $-$ & 20.58 & 9.16 & $-$ & 8.57 & 3.80 & $-$ & 4.16 & 250 & 900 & 900 & 8.01 & 13.92 & $-$ & 3.68 & 7.90 & $-$ & 1.87 & 4.64 & 2.90 \\
  10 & 1800 & 900 & 15.86 & $-$ & 21.24 & 10.02 & $-$ & 9.04 & 3.79 & $-$ & 4.07 & 250 & 900 & 1800 & 8.02 & 14.49 & $-$ & 3.77 & 7.12 & $-$ & 1.93 & 4.32 & 2.73 \\
  10 & 1800 & 1800 & 15.77 & $-$ & 23.54 & 10.80 & $-$ & 10.22 & 3.82 & $-$ & 3.90 & 250 & 1800 & 300 & 5.49 & 11.84 & $-$ & 2.62 & 6.04 & $-$ & 1.69 & 4.65 & 2.25 \\
  10 & 1800 & 2700 & 14.72 & $-$ & 25.09 & 10.17 & $-$ & 10.31 & 3.80 & $-$ & 3.76 & 250 & 1800 & 900 & 5.29 & 11.83 & $-$ & 2.89 & 6.41 & $-$ & 1.72 & 4.65 & 2.47 \\
  10 & 1800 & 3600 & $-$ & $-$ & $-$ & $-$ & $-$ & $-$ & 3.79 & $-$ & 3.65 & 250 & 1800 & 1800 & 4.82 & 13.75 & $-$ & 1.83 & 6.68 & $-$ & 1.84 & 4.40 & 2.13 \\
  10 & 3600 & 60 & $-$ & $-$ & $-$ & $-$ & $-$ & $-$ & 3.70 & $-$ & 3.82 & 250 & 1800 & 2700 & $-$ & 16.18 & $-$ & $-$ & 6.44 & $-$ & 1.89 & 4.28 & $-$ \\
  10 & 3600 & 300 & $-$ & $-$ & $-$ & $-$ & $-$ & $-$ & 3.73 & $-$ & 3.81 & 250 & 1800 & 3600 & $-$ & 16.47 & $-$ & $-$ & 6.23 & $-$ & 1.95 & 3.99 & $-$ \\
  10 & 3600 & 900 & $-$ & $-$ & $-$ & $-$ & $-$ & $-$ & 3.60 & $-$ & 3.66 & 250 & 3600 & 300 & $-$ & 13.08 & $-$ & $-$ & 9.46 & $-$ & 1.96 & 4.38 & $-$ \\
  10 & 3600 & 1800 & $-$ & $-$ & $-$ & $-$ & $-$ & $-$ & 3.51 & $-$ & 3.45 & 250 & 3600 & 900 & $-$ & 10.75 & $-$ & $-$ & 6.38 & $-$ & 2.01 & 3.90 & $-$ \\
  10 & 3600 & 2700 & $-$ & $-$ & $-$ & $-$ & $-$ & $-$ & 3.45 & $-$ & 3.40 & 250 & 3600 & 1800 & $-$ & 13.46 & $-$ & $-$ & 7.13 & $-$ & 1.95 & 3.33 & $-$ \\
  10 & 3600 & 3600 & $-$ & $-$ & $-$ & $-$ & $-$ & $-$ & 3.42 & $-$ & 3.40 & 250 & 3600 & 2700 & $-$ & $-$ & $-$ & $-$ & $-$ & $-$ & 1.89 & $-$ & $-$ \\
  10 & 3600 & 5400 & $-$ & $-$ & $-$ & $-$ & $-$ & $-$ & 3.37 & $-$ & $-$ & 250 & 3600 & 3600 & $-$ & $-$ & $-$ & $-$ & $-$ & $-$ & 1.91 & $-$ & $-$ \\
  10 & 7200 & 60 & $-$ & $-$ & $-$ & $-$ & $-$ & $-$ & 2.65 & $-$ & 2.86 & 300 & 900 & 300 & 9.07 & 10.41 & $-$ & 3.92 & 6.48 & $-$ & 1.85 & 4.59 & 3.28 \\
  10 & 7200 & 300 & $-$ & $-$ & $-$ & $-$ & $-$ & $-$ & 2.68 & $-$ & 2.67 & 300 & 900 & 900 & 11.33 & 11.99 & $-$ & 4.79 & 7.45 & $-$ & 1.80 & 4.50 & 2.83 \\
  10 & 7200 & 900 & $-$ & $-$ & $-$ & $-$ & $-$ & $-$ & 2.77 & $-$ & 2.91 & 300 & 900 & 1800 & 11.40 & 14.42 & $-$ & 5.17 & 7.52 & $-$ & 1.84 & 4.39 & 2.42 \\
  10 & 7200 & 1800 & $-$ & $-$ & $-$ & $-$ & $-$ & $-$ & 2.77 & $-$ & $-$ & 300 & 1800 & 300 & 6.26 & 9.93 & $-$ & 2.83 & 5.45 & $-$ & 1.68 & 4.57 & 2.31 \\
  15 & 45 & 15 & 12.15 & 16.01 & 20.02 & 7.18 & 11.84 & 8.87 & 3.95 & 5.38 & 4.59 & 300 & 1800 & 900 & 7.07 & 11.80 & $-$ & 2.93 & 6.07 & $-$ & 1.68 & 4.53 & $-$ \\
  15 & 45 & 30 & 13.47 & 14.65 & 17.89 & 8.05 & 12.08 & 8.08 & 4.10 & 4.97 & 4.62 & 300 & 1800 & 1800 & 7.54 & 13.24 & $-$ & 2.47 & 6.26 & $-$ & 1.74 & 4.35 & $-$ \\
  15 & 45 & 45 & 13.97 & 15.71 & 18.15 & 8.29 & 13.80 & 8.39 & 4.18 & 5.19 & 4.74 & 300 & 1800 & 2700 & $-$ & 12.20 & $-$ & $-$ & 6.35 & $-$ & 1.76 & 3.98 & $-$ \\
  15 & 45 & 60 & 14.40 & 16.07 & 19.34 & 8.55 & 12.35 & 8.03 & 4.29 & 5.41 & 4.64 & 300 & 1800 & 3600 & $-$ & 17.33 & $-$ & $-$ & 5.92 & $-$ & 1.83 & 3.92 & $-$ \\
  15 & 60 & 15 & 11.99 & 15.24 & 16.17 & 7.09 & 12.88 & 7.57 & 3.90 & 5.26 & 5.07 & 300 & 3600 & 300 & $-$ & 11.16 & $-$ & $-$ & 5.69 & $-$ & 1.90 & 3.85 & $-$ \\
  15 & 60 & 30 & 13.23 & 14.80 & 17.03 & 7.85 & 12.40 & 7.46 & 4.03 & 4.96 & 4.71 & 300 & 3600 & 900 & $-$ & 10.94 & $-$ & $-$ & 6.30 & $-$ & 1.90 & 3.86 & $-$ \\
  15 & 60 & 45 & 13.75 & 15.85 & 18.25 & 8.07 & 13.74 & 7.72 & 4.10 & 5.18 & 4.73 & 300 & 3600 & 1800 & $-$ & 12.72 & $-$ & $-$ & 6.61 & $-$ & 1.80 & 3.06 & $-$ \\
  15 & 60 & 60 & 14.15 & 16.01 & 18.48 & 8.30 & 13.20 & 8.06 & 4.21 & 5.42 & 4.69 & 300 & 3600 & 2700 & $-$ & $-$ & $-$ & $-$ & $-$ & $-$ & 1.73 & $-$ & $-$ \\
  15 & 75 & 15 & 11.83 & 15.13 & 16.95 & 7.10 & 13.00 & 9.25 & 3.84 & 5.42 & 4.71 & 300 & 3600 & 3600 & $-$ & $-$ & $-$ & $-$ & $-$ & $-$ & 1.72 & $-$ & $-$ \\
  15 & 75 & 30 & 13.09 & 14.10 & 14.86 & 7.86 & 9.88 & 8.76 & 3.97 & 5.21 & 4.79 & 600 & 1800 & 900 & 7.87 & 9.25 & $-$ & 5.67 & 6.20 & $-$ & 1.15 & 4.56 & $-$ \\
  15 & 75 & 45 & 13.64 & 14.77 & 14.50 & 8.07 & 12.33 & 8.70 & 4.04 & 5.08 & 4.64 & 600 & 1800 & 1800 & 8.93 & 10.54 & $-$ & 6.17 & 7.04 & $-$ & 1.21 & 4.29 & $-$ \\
  15 & 75 & 60 & 14.01 & 14.29 & 17.81 & 8.32 & 12.67 & 8.96 & 4.14 & 5.31 & 4.42 & 600 & 1800 & 2700 & $-$ & 13.70 & $-$ & $-$ & 5.73 & $-$ & 1.21 & 4.16 & $-$ \\
  15 & 90 & 15 & 12.07 & 15.06 & 15.32 & 7.19 & 12.79 & 8.14 & 3.83 & 5.42 & 4.78 & 600 & 1800 & 3600 & $-$ & 14.52 & $-$ & $-$ & 5.80 & $-$ & 1.31 & 3.78 & $-$ \\
  15 & 90 & 30 & 13.20 & 14.04 & 14.31 & 7.90 & 11.06 & 8.89 & 3.97 & 5.00 & 4.47 & 600 & 3600 & 900 & $-$ & 9.68 & $-$ & $-$ & 6.37 & $-$ & 1.19 & 3.82 & $-$ \\
  15 & 90 & 45 & 13.76 & 14.53 & 16.71 & 8.08 & 12.91 & 9.23 & 4.03 & 5.08 & 4.09 & 600 & 3600 & 1800 & $-$ & 12.60 & $-$ & $-$ & 5.93 & $-$ & 1.12 & 3.30 & $-$ \\
  15 & 90 & 60 & 14.19 & 14.01 & 18.27 & 8.31 & 12.48 & 8.82 & 4.13 & 5.29 & 4.12 & 600 & 3600 & 2700 & $-$ & 10.77 & $-$ & $-$ & 5.69 & $-$ & 1.08 & 3.15 & $-$ \\
  15 & 120 & 15 & 11.97 & $-$ & 18.36 & 7.16 & $-$ & 8.15 & 3.77 & $-$ & 3.63 & 900 & 3600 & 900 & $-$ & 9.84 & $-$ & $-$ & 5.29 & $-$ & 1.68 & 3.71 & $-$ \\
  15 & 120 & 30 & 13.09 & $-$ & 18.75 & 7.86 & $-$ & 8.17 & 3.91 & $-$ & 3.67 & 900 & 3600 & 1800 & $-$ & 10.85 & $-$ & $-$ & 5.42 & $-$ & 1.37 & 3.03 & $-$ \\
  15 & 120 & 45 & 13.69 & $-$ & 16.45 & 8.06 & $-$ & 9.01 & 3.98 & $-$ & 3.79 & 900 & 3600 & 2700 & $-$ & 10.88 & $-$ & $-$ & 4.82 & $-$ & $-$ & 3.16 & $-$ \\
\end{longtable}
}
\twocolumn

\end{document}